\documentclass[]{aa}  
\usepackage{graphicx}
\usepackage{txfonts}
\begin{document}
   \title{Metallicity evolution, metallicity gradients, and gas fractions at z$\sim$3.4
            \thanks{Based on data obtained at the VLT through the ESO programs
			178.B-0838, 075.A-0300 and 076.A-0711.}}

		        
   \author{P.~Troncoso
	  \inst{1,2}
          \and
		    R.~Maiolino\inst{3,4}
		  \and
		V.~Sommariva\inst{1}
		 \and
		  G.~Cresci\inst{5}
		  \and
		  F.~Mannucci\inst{6}
		  \and
		 A.~Marconi\inst{7}
		 \and
		  M.~Meneghetti\inst{6,8}
		\and
		  A.~Grazian\inst{1}
		 \and
		 A.~Cimatti\inst{9}
		  \and
		 A.~Fontana\inst{1}
		 \and
		 T.~Nagao\inst{10,11}
		\and
		 L.~Pentericci\inst{1}
		  }		        
     \institute{INAF - Osservatorio Astronomico di Roma, via di Frascati 33, 00040 Monte Porzio Catone, Italy
         \and
         Astronomisches Institut, Ruhr-Universit\"at Bochum, Universit\"atsstra\ss{}e 150, D-44780, Bochum, Germany
         \and
   Cavendish Laboratory, University of Cambridge, 19 J. J. Thomson Ave., Cambridge CB3 0HE, United Kingdom	
         \and
   Kavli Institute for Cosmology, University of Cambridge, Madingley Road, Cambridge CB3 0HA, United Kingdom	
	\and
	INAF-Osservatorio Astronomico di Bologna, via Ranzani 1, I-40127 Bologna, Italy
   	\and
	INAF - Osservatorio Astrofisico di Arcetri, Largo E. Fermi 5, I-50125 Firenze, Italy
	\and
	Dip. di Fisica e Astronomia, Universit\'a di Firenze, via G. Sansone 1, I-50019, Sesto F.no, Firenze, Italy
	\and
	INFN, Sezione di Bologna, viale Berti Pichat 6/2, 40127 Bologna, Italy 
	\and
	Dipartimento di Fisica e Astronomia, Universit\'a di Bologna, Via Ranzani 1, I-40127 Bologna, Italy
	\and
	The Hakubi Center for Advanced Research, Kyoto University, Kyoto 606-8302, Japan
	\and
	Department of Astronomy, Kyoto University, Kitashirakawa-Oiwake-cho, Sakyo-ku, Kyoto 606-8502, Japan	
   }
     \date{Received ; accepted }
     
 \abstract{ 
  We used near-infrared integral field spectroscopic observations from the AMAZE
  and LSD ESO programs to constrain the metallicity in a sample of 40
  star-forming galaxies at 3$<$z$<$5 (most of which are at z$\sim$3.4).
  We measured metallicities by exploiting strong emission-line diagnostics. 
  We found that a significant fraction of star-forming galaxies at z$\sim$3.4 deviate from the 
  fundamental metallicity relation (FMR), 
  with a metallicity of up to a factor of ten lower than expected according to the FMR.
  This deviation does not correlate with the dynamical properties of the galaxy or with the presence of interactions.
 To investigate the origin of the metallicity deviation in more detail, 
 we also inferred information on the gas content by inverting the
  Schmidt-Kennicutt relation, assuming that the latter does not evolve out to
  z$\sim$3.4.
  In agreement with recent CO observational data, we found that
  in contrast with the steeply rising trend at 0$<$z$<$2, 
  the gas fraction in massive galaxies remains constant,
  with an indication of a marginal decline at 2$<$z$<$3.5.
  When combined with the metallicity information,
  we infer that to explain the low metallicity and gas content in z$\sim$3.4 galaxies,
  both prominent outflows and massive pristine gas inflows are needed.
In ten galaxies we can also spatially resolved the metallicity distribution.
We found that the metallicity generally anticorrelates with the distribution
of star formation and with the gas surface density.
We discuss these findings in terms of pristine gas inflows toward the center, 
and outflows of metal-rich gas from the center toward the external regions.
}

 \maketitle


\section{Introduction} 
\label{sec_intro}

Observations and theory have led during the past few years
to a new scenario in which galaxy evolution is primarily regulated by the gas content 
in galaxies, which is, in turn, regulated by gas inflow and outflow phenomena.
Millimeter observations have revealed large amounts of
molecular gas in high-redshift galaxies \citep{tacconi10,daddi10}.
These gas-rich galaxies are generally massive rotating disks,
already in place at these early epochs,
which obey the same Schmidt-Kennicutt \citep[hereafter S-K law,][]{schmidt59,kennicutt98}
star formation law observed in the local Universe and
whose enhanced star formation rate (SFR) is simply a consequence of their
higher gas content relative to local galaxies and is not driven by
enhanced star formation efficiency \citep{genzel04,bouche07,genzel10,daddi10}. 

Several models can account for these observed trends. The emerging scenario is that
at high redshift, gas from the intergalactic medium
quickly replenishes galaxies through smooth cold flows \citep[e.g.][]{dekel09}.  
As a consequence of the high pressure, this gas is mostly transformed into molecular hydrogen, 
which can effectively be used to feed star formation according to the models by
\cite{or09} and \cite{lagos11a}.
These models provide a detailed description of the evolution of the molecular gas content
in galaxies as a function of redshift and as a function of galaxy (halo) mass.
According to these models, the evolution of the cosmic star formation
rate is simply a consequence of the evolution of the molecular gas content
in galaxies through the Schmidt-Kennicutt relation \citep{bouche10,genel08}.

Gas flows are thought to also be primarily responsible for the variation
of metal contents in galaxies and their redshift evolution. In particular,
the well-known mass-metallicity relation has been primarily ascribed to outflow of metal-rich gas, 
which is preferentially expelled from low-mass galaxies \citep{tremonti04},
although effects associated with the initial mass function (IMF), 
downsizing and gas infalls have also been invoked and several thorough theoretical models
have been proposed 
\citep{koppen07,kobayashi07,brooks07,derossi07,cescutti07,finlator08,arrigoni09,caluramenci09,dave11b,sakstein11,dave12,dayal13}.

More recently, it has been found that the gas metallicity also has a secondary
dependence on the SFR. At a given stellar mass the 
metallicity decreases with increasing SFR
\citep{kewley06,ellison08,mannucci10,laralopez10,cresci12,yates12,andrewsmartini12,lilly13,
perezmontero13}.

This three-dimensional relationship between stellar mass, SFR, and
metallicity has been dubbed fundamental metallicity relation \citep[FMR]{mannucci10}.
This relation is very tight (dispersion 0.05dex), suggesting that it 
reflects a smooth secular interplay between star formation and gas flows.
In particular, one of the basic ideas behind the inverse correlation between SFR and metallicity
is that it is primarily associated with inflow of pristine gas:
the accreted gas on the one hand dilutes the gas metallicity, on the other
hand boosts star formation through the Schmidt-Kennicutt relation.
More detailed models and simulations have been presented by various authors
to interpret the FMR
\citep{dave11b,lagos11a,dave12,yates12,dayal13,forbes13}.

Recently, by using spatially resolved spectroscopy of nearby galaxies,
\cite{sanchez13} have not confirmed the metallicity dependence on the SFR,
in contrast to previous studies. Part of the discrepancy may be associated
with the use of the metallicity at the effective radius, instead of the
metallicity obtained from the integrated line emission within the same
aperture as used to infer the SFR. Indeed, \cite{bothwell13} have shown that
the anticorrelation with the SFR (with low dispersion) is only found if
both metallicity and SFR are extracted from the same aperture (and actually
from the same spectrum integrated within the same aperture), while the
scatter increases largely when the metallicity from the central region (SDSS fiber aperture)
is compared with the total, integrated SFR.
\cite{bothwell13} have also found evidence for a more fundamental relation between
HI gas mass, metallicity, and stellar mass (HI-FMR). They suggested that the classical FMR
(SFR-FMR) is a byproduct of the HI-FMR. However, even if the SFR-FMR were a consequence
of the HI-FMR, the former can be considered as a tool for tracing galaxy evolutionary processes.

One important aspect of the FMR is that it is found not to evolve with
redshift out to z$\sim$2.5, suggesting that the same mechanism of galaxy
formation is at work in most galaxies out to this redshift \citep{mannucci10,laralopez10,richard11,belli13}.
Significant evolution was found at z$\sim$3 by \cite{mannucci10} 
by using an initial subsample of the AMAZE (Assessing the Mass-Abundance redshift[Z] Evolution) 
and LSD (Lyman-break galaxies Stellar population and Dynamics) spectroscopic surveys 
\citep{maiolino08,mannucci09}.
These authors found that galaxies at z$>$3 are anomalously metal poor,
specifically, about 0.6 dex below the FMR.

If the evolution of the FMR is confirmed at z$>$3 with higher statistics,
this may suggest a change in the dominant mode of galaxy formation.
One possibility is that an excess
of gas inflow at early epochs causes an excess of metallicity dilution.

Evidence for such massive inflows at z$\sim$3 has been found
through the analysis of metallicity gradients. By exploiting
integral field spectroscopic data of three galaxies at z$\sim$3 from the
AMAZE sample, \cite{cresci10} found evidence for inverted (positive)
metallicity gradients. 
In particular, the region of lowest metallicity
was found to be coincident with the region of highest star formation,
suggesting that strong infall of pristine gas dilutes the metallicity
and boosts star formation in these galaxies at z$\sim$3.
Studies of resolved metallicities at lower redshift have obtained mixed results,
with samples of galaxies showing positive as well as negative gradients
\citep{jones10,queyrel12,jones12},
possibly hinting
at evolutionary trends and dependence on the dynamical status of galaxies.
The inverted gradients at z$\sim$3 are based on only three galaxies, 
hence requiring a larger sample to achieve a statistically sound result.

In this paper, we present the results from the full sample of z$\sim$3-4 galaxies from AMAZE,
an ESO Large Program that exploits the VLT integral field near-IR spectrograph SINFONI(Spectrograph for INtegral Field Observations in the Near Infrared), 
combined with the parallel program LSD, 
performed with adaptive optics (AO), which are described in
more detail in the next sections.
While the dynamical properties of these samples were discussed in \cite{gnerucci11a},
in this paper we focus on the distribution of the SFR (from which we inferred
the gas fraction) and the metallicity properties (integrated and spatially
resolved), by
extending the results based on the first subsample presented in \cite{maiolino08},
\cite{mannucci09}, and \cite{cresci10}.

\section{AMAZE project: sample and data}\label{amaze}

This paper is primarily based on the AMAZE project,
which uses near-IR integral field spectroscopy of star-forming galaxies
at redshift $3 < z < 5$ with SINFONI \citep{eisenhauer03} at the Very Large Telescope
(VLT). A complete description of the AMAZE programs and of the methods used  for data
analysis and reduction are presented in \cite{maiolino08}.  Here we report a brief
summary of the observing program and of the data obtained.

AMAZE is an ESO large program that was awarded 180 hours of observations
with SINFONI.  The sample consists of $31$ Lyman-break galaxies (LBGs) 
in the redshift range $\rm 3<z<5.2$ (most of which are at z$\sim$3.4), 
with Spitzer/IRAC photometry ($3.6-8\mu m$) required to derive reliable stellar masses. 
Those LBGs that host AGNs were discarded based on either UV (optical rest-frame) spectra, 
hard X-ray data, or on the MIPS 24$\mu$m flux \citep[see discussion in][]{maiolino08}.

SINFONI was used in seeing-limited mode, with the 
$0.125\arcsec\times0.25\arcsec$ pixel scale and the H+K grating, yielding a spectral
resolution $R\sim1500$ over the spectral range $1.45-2.41\mu m$. The typical
seeing during the observations was about 0.6--0.7$''$. 

The main goals of the project are to obtain information on the metal 
enrichment and dynamics of galaxies at z$>$3 by exploiting the optical 
nebular lines of [OII]$\lambda$3727, [NeIII]$\lambda$3869, H$\beta$, 
and [OIII]$\lambda\lambda$4959,5007 redshifted into the H and K bands.

Data were reduced by using the ESO-SINFONI pipeline (version 3.6.1).
The pipeline subtracts the sky from the temporally contiguous frames, 
flat-fields the images, spectrally calibrates each individual slice, 
and then reconstructs the cube. Within the pipeline the
pixels are resampled to a symmetric angular size of $0.125\arcsec \times 0.125\arcsec$.
The atmospheric absorption and instrumental response were taken into account and
corrected for by using a suitable standard star.

In addition to SPITZER data, optical and near-IR photometry is available from the archives,
see \cite{steidel03} and \cite{grazian06} and also from new observations \citep{sommariva13}
obtained with the near-IR camera NICS \citep{baffa01}
mounted in the Telescopio Nazionale Galileo (TNG).
Hubble Space Telescope (HST) optical and near-IR archival images (WFPC3, WFPC2, ACS)
are available for most of the sources located in the GOODS-S survey field
as well as for a few additional AMAZE targets in the archive.
Of these we used primarily the images obtained with the filters F606W, F775W, and
F814W to probe the rest-frame UV emission and, with this, 
study the environment and morphology \citep{sommariva13}.

The full list of galaxies in the AMAZE sample along with
some of their main observational properties is given in Table~\ref{table_sample}.
The AMAZE galaxies were extracted from the \cite{steidel03} and \cite{vanzella06} catalogs. 
The AMAZE sample ($< R > \sim$24) is globally representative of the galaxies in
the parent samples \citep{steidel03,vanzella06}, which peak at $< R > \sim$24.5, 
although they are slightly brighter.
Note that the galaxies were not pre-selected on any prior knowledge of nebular line detectability
(either in the optical or in the near-IR).

In addition, five lensed galaxies were selected (marked with an asterisk in Table~\ref{table_sample}) 
that allowed us to probe galaxies with a lower SFR than that of the bulk of the sample.
Four of these are lensed by the cluster Abell 1689, 
and their magnification factor was obtained by
constructing a lens model of the cluster by using the 
34 multiply imaged systems published in \cite{limousin07}.
Twenty-four of these multiply imaged systems have spectroscopic redshifts, 
which ensures that the model is robustly calibrated.
To derive the cluster magnification maps, 
we used the Lenstool public software \citep{kneib96} and
reproduced the \cite{limousin07} mass model using the parameter file available at http://www.astro.ku.dk/~marceau/model.par.
Then, we sampled the magnification maps to derive the local measurements of the magnification at the galaxy positions.
The data for the fifth lensed source, the ``Cosmic Eye'', were taken from the archive,
and its magnification factor was taken from \cite{smail07}.

Preliminary results on the mass-metallicity relation at z$\sim$3.4 based on a first
subset of nine galaxies are presented in \cite{maiolino08}.
As mentioned, the metallicity gradients for three bright and extended AMAZE galaxies
are presented in \cite{cresci10}.
A study of the kinematics and dynamics of the whole sample is presented in
\cite{gnerucci11a,gnerucci11b}, who found a significant fraction (about one third)
of galaxies to be characterized by regular disk-rotation patterns.
\cite{sommariva12} also obtained stellar metallicities from optical (UV rest-frame) spectra
of a few AMAZE galaxies, finding values broadly consistent with the metallicities inferred
from the optical nebular lines.

  \begin{figure*}
   \centering
   \includegraphics[width=.6\linewidth]{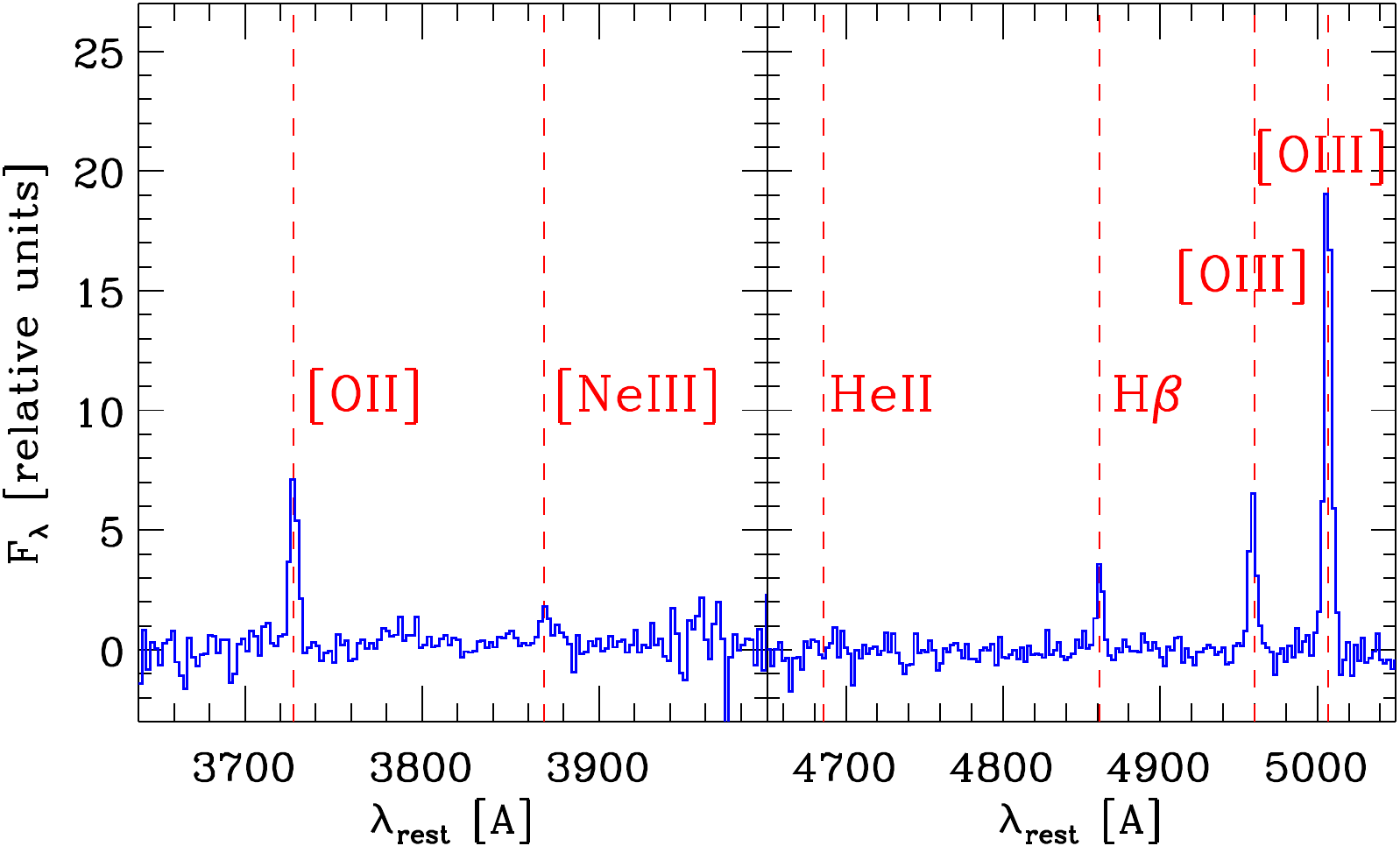}
\caption{Composite spectrum of the AMAZE and LSD spectra at z$\sim$3.4 (34 sources).}
  \label{fig_stack}
\end{figure*}

\section{LSD project}\label{lsd}

The AMAZE data are complemented with the LSD project,
whose data are presented in \cite{mannucci09}.
LSD is a parallel SINFONI project consisting of observations assisted with the AO 
module of nine LBGs at z$\sim$3. Data analysis and reduction are presented
in \cite{mannucci09}. Thanks to the exploitation of the AO module, these data achieve
an angular resolution ($\sim 0.1-0.2''$) much higher than that of the AMAZE data. 

The kinematics for the LSD galaxies is presented in \cite{gnerucci11a}.
HST data of the LSD galaxies is also available through new observations \citep{sommariva13}.

The list of LSD galaxies is given in Table~\ref{table_sample} along with their observational properties.
The LSD sample is representative of the global LBG population selected by \cite{steidel03}, as discussed in \cite{mannucci09}.

\section{Basic observational results}

In this section we provide some basic observational results, such as the integrated
spectra, integrated emission line fluxes, stacked spectra and ancillary data and information.

\subsection{Emission lines}

The spectra were extracted within fixed apertures of 0.75$"$ in diameter (corresponding to
$\sim$ 5.4~kpc projected on sources at $ z\sim 3.4$), which in most cases encloses more 
than 70\% of the emission line flux and generally maximizes the S/N ratio.
The same procedure was applied for the AMAZE and LSD sources previously published.
There are four exceptions, SSA22a-M38, SSA22a-C36, CDFS-12631,
and DFS2237b-C21, for which the line emission extends significantly beyond 0.75$''$.
Only for these four sources an aperture of 1.25$''$ was used.
Figures \ref{fig1sp}, \ref{fig2sp}, \ref{fig3sp}, and \ref{fig4sp} in the appendix
show the spectra extracted for all unlensed sources.

For the three lensed sources at z$\sim$3, LnA1689-1, LnA1689-4 and the Cosmic Eye 
the spectra were extracted manually by choosing high S/N regions.
Their spectra are shown in figure \ref{figlens}.

For all sources the redshift is such that [OII]3727 and [NeIII]3870 is redshifted into the
H band, while H$\beta$ and [OIII]5007 are redshifted into the K band, 
with the exception of galaxies at 4.1$<$z$<$5.2,
for which only [OII] and [NeIII] are observable in the K band.
Five sources are undetected, as noted in Table~\ref{table_sample}, 
two of which are at z$>$4.
CDFS-13497 presents a very poor S/N spectrum (see Fig.\ref{fig2sp}) 
and is excluded from the following analysis.

Fig.\ref{fig_stack} shows the stacked spectrum of all sources at z$<$4.
It is important  to note that the stacked spectrum shows no evidence 
for the HeII nebular line at $\lambda$4686\AA. 
This line is expected if there is contribution from AGNs.
The nondetection of this line additionally supports that 
our selection criteria carefully excluded AGNs from the sample.

The emission lines fluxes were obtained by fitting the emission lines with Gaussian functions
by imposing the same $FWHM$ for all the lines within the same band 
(hence with the same instrumental resolution, which slightly changes with wavelength).
To automatically exclude noise fluctuations or bad pixels we constrained the velocity dispersion 
of the lines to be higher than the instrumental resolution estimated from the sky emission lines. 
The continuum, when detected (marginally), was subtracted with a simple linear slope, and the
uncertainties in the subtraction were included in the final errors on the line fluxes.
In Table \ref{table_flux} the measured emission line fluxes are reported.
As mentioned,
out of the 40 galaxies in the joint AMAZE+LSD sample, 34 have detections of all emission lines
required to infer the metallicity.

As pointed out in \cite{gnerucci11a}, many of the sources in the
seeing-limited AMAZE data and the AO-assisted LSD are spatially resolved.
A detailed study on the morphology and size determination by using different methods of the
AMAZE and LSD galaxies will be published in a companion work (see Sommariva et al. in prep.).
In general, the different methods used to determine the size are inconsistent with each other, 
with differences ranging up to a factor of four or more.
This questions the reliability of the size determination and, even more importantly, its meaning. 
This has implications for the determination of the surface brightness 
of galaxies, as discussed in more detail below (see section \S \ref{s_gas}).

\subsection{Physical properties determined from the broad-band SED} \label{ssection_sedfit}

 \begin{figure}[!ht]
   \centering
   \includegraphics[width=1.\linewidth]{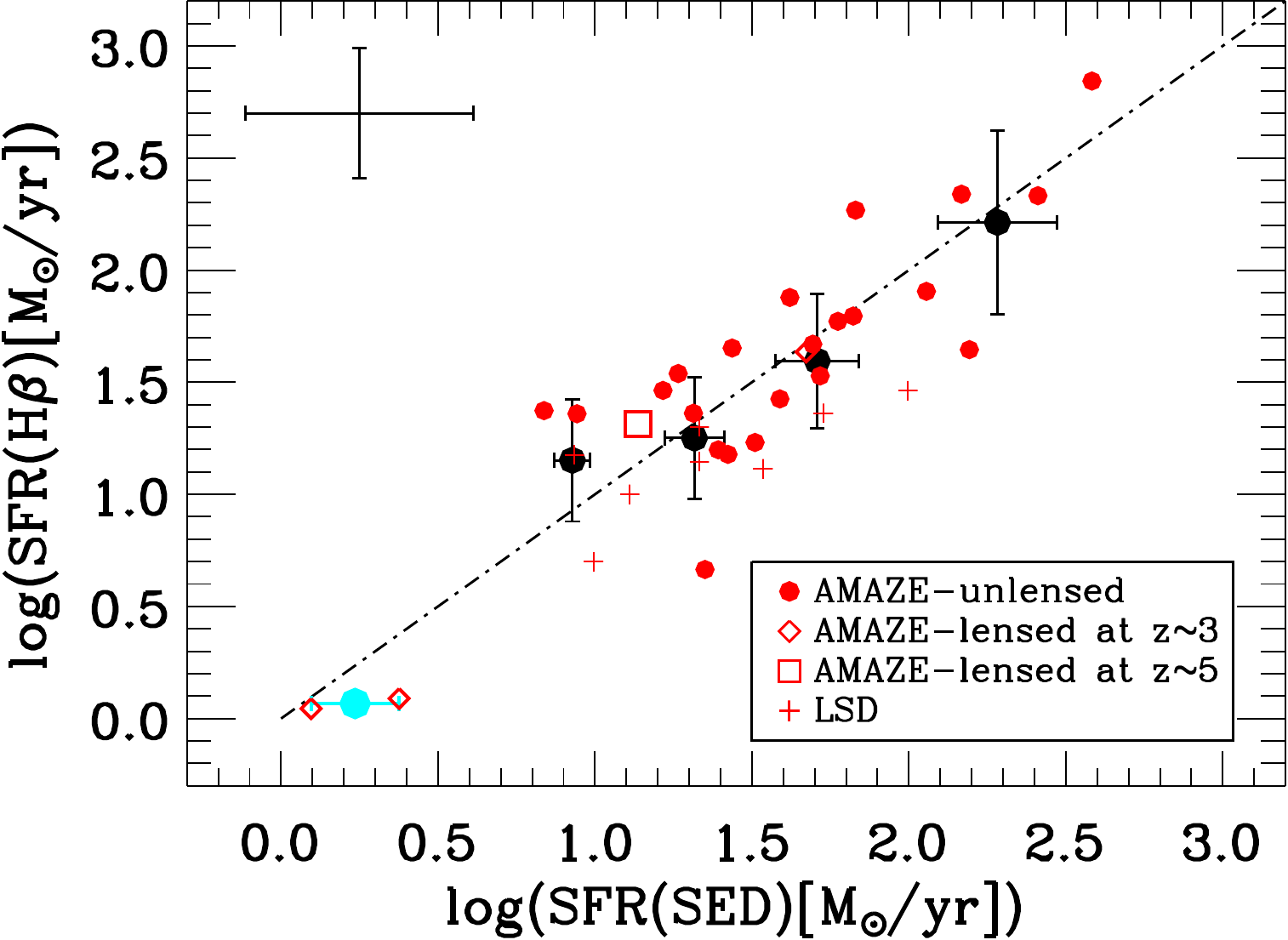}
\caption{Comparison between the SFRs derived from the SED fitting and the SFR based on $H\beta$. 
The dashed line shows the 1:1 relation along which the two estimators agree perfectly. 
The red circles are unlensed AMAZE galaxies at z$\sim$3.4. 
The red diamonds are the three lensed sources at $z\sim$3.4.
The red hollow square is the lensed galaxy at z=4.8
for which H$\beta$ emission has been inferred from [OII]3727
by assuming the [OII]/H$\beta$ ratio appropriate for its metallicity (see Appendix B).
The red crosses show the LSD galaxies. 
Black filled circles show the SFR($H\beta$) averaged within bins of SFR(SED).
The cyan bin is composed of the two lensed galaxies LnA1689-1 and LnA1689-4. 
The error bars on the binned values correspond to the dispersion of the SFR within each bin. 
The error bars on the top-left corner indicate the average uncertainties on
the measurements of SFR(SED) and SFR(H$\beta$) for the individual objects.
}
\label{fig_sfrsedhb}
\end{figure}

For each galaxy in our sample a multiwavelength broad-band spectral energy distribution (SED)
was built with published or archival data and 
was fitted with a set of galaxy spectral templates 
to obtain the stellar mass, SFR, age, and dust reddening.
The broad-band photometric data for the sources in the CDFS were collected from 
the GOODS--MUSIC multiwavelength catalog \citep{grazian06}.
This catalog provides photometric data in 14 spectral bands (from UV to the Spitzer-IRAC bands)
and has recently been updated to include the Spitzer--MIPS data at 24$\mu$m.
For the LBGs extracted from the \cite{steidel03} sample, optical photometric data (U,G,R,I) 
were extracted from the publicly available images \citep{steidel03},
while Spitzer IRAC and MIPS data were obtained from the Spitzer archive
or from new observations \citep{mannucci09}; the photometry extraction 
was performed following the same methods described in \cite{grazian06}.

The galaxy templates and best-fitting technique are the same as those used in previous papers
\citep{grazian06,grazian07}, and similar to those adopted by other groups in the 
literature \citep[e.g.][]{dickinson03,drory04,pozzetti07}.
The galaxy templates were computed with standard spectral synthesis models (\cite{bc03}
hereafter BC03) and were chosen to broadly encompass the variety of 
star formation histories, ages, metallicities, and extinction of real galaxies.
More specifically, we considered an exponentially declining SFR 
with e-folding times ranging from 0.1 to 15 Gyr.
We used the Salpeter IMF ($M_{min}=0.1M_{\odot}$ and $M_{max}=65M_{\odot}$), 
ranging over a set of metallicities (from $Z=0.02 Z_\odot$ to $Z=2.5 Z_\odot$) and 
dust extinction ($0<E(B-V)<1.1$), with a \cite{calzetti00} attenuation curve,
which is generally more appropriate for the stellar component. 
For each model of this grid, the expected magnitudes were computed in the desired filter sets 
(depending on the available public data). 
The best-fitting template was found with a standard $\chi^2$ minimization.
The stellar mass and other physical parameters of the galaxy, such as SFR, age, and dust extinction, 
were fitted simultaneously to the actual SED of the observed galaxy. 
To compare our results with other studies, using the Chabrier IMF,
we applied a correction factor of 1.7 \citep{pozzetti07} to our masses and SFR.
The SED-fitting results are reported in Table \ref{tab_sed}.
For the four lensed sources masses and SFR were corrected
for the magnification factor obtained by us (Ln1689-1, Ln1689-2, Ln1689-4)
or in the literature \citep[Cosmic Eye,][]{smail07}, as reported in Table \ref{tab_sed}.

\subsection{Star formation rates}

The SFR can also be inferred also from the H$\beta$ line flux that, if unreddened,
is proportional to the number of ionizing photons emitted by young hot stars.

We inferred the SFR from H$\beta$ by first correcting for it through the reddening inferred 
from the continuum-fitting, then by applying a stellar-to-nebular differential correction factor of 1/0.44,
as discussed in \cite{calzetti00} (although the applicability of this correction factor
for high redshift galaxies is subject to some debate, as discussed in \cite{reddy12}).
We finally obtained the H$\alpha$ luminosity by assuming the
case-B recombination, that is, $\rm H\alpha/H\beta =2.8$, and by adopting the $\rm H\alpha$ to SFR
conversion factor given by \cite{kennicutt98}.
The errors of the flux measurements presented in Table \ref{table_flux}
and the reddening ($\Delta (E(B-V)) = 0.1$) 
were taken into account to derive the final $SFR(H\beta)$ presented in Table \ref{tab_sed}. 
Uncertainties in $SFR(H\beta)$ are at a one-sigma level.
We note that the correction for the H$\beta$ flux for stellar absorption is negligible. Indeed,
the continuum is not detected, or barely detected, in all of the spectra; therefore,
even assuming that the H$\beta$
stellar absorption on the continuum is as deep as it can be (e.g. post-starburst A-type-like spectra),
this would not change the inferred H$\beta$ flux by more than a few percent.

Fig.~\ref{fig_sfrsedhb} shows the comparison between the SFR inferred from the SED 
with the SFR inferred from H$\beta$, where, for the sake of clarity, 
we averaged the data in bins of SFR(SED) and 
where the error bars give the dispersion of the points in each bin.
Each bin has seven galaxies on average, except for the cyan symbol, which has two galaxies.
The latter shows the average obtained from the two lensed sources LnA1689-1 and LnA1689-4,
which allowed us to explore low SFRs.
The error bars in the top-right corner show the average measurement errors on individual objects.
There is a good general agreement between the two tracers of star formation over
the two orders of magnitude spanned by our data.
A more extensive discussion of the reliability and a comparison of the different SFR tracers
for the galaxies in our sample is given in a companion paper, \cite{castellano13}.
In the following we used (unless stated otherwise) the SFR inferred from H$\beta$ (SFR$_{H\beta}$),
but the results are essentially unchanged if one uses the SFR inferred from the SED.

\section{Metallicity evolution} \label{met_z3}

In this section we investigate the integrated metallicity 
of the LBG galaxies at z$\sim$3-5, its dependence on stellar
mass and SFR, and compare them with lower-redshift galaxies
to infer possible evolutionary effects.

To infer the metallicities we adopted the same method as
in \cite{maiolino08} and in \cite{mannucci09}.
Essentially, the method consists of exploiting the $\rm R_{23}$ parameter (where
$\rm R_{23}=(F([OII]\lambda3727)+F([OIII]\lambda4959)+ F([OIII]\lambda5007))/F(H\beta$), 
which depends on metallicity.
The dependence of this parameter on the metallicity has been calibrated by various authors, 
either by using theoretical models or through empirical calibrations based on other primary
indicators (such as the electronic temperature $\rm T_e$ inferred from auroral lines).
A summary of some of the calibrations is given in \cite{kewley08}. 
\cite{maiolino08} adopted a hybrid calibration method,
in which at low metallicities ($\rm 12+log(O/H) <8.4$, 
where models have problems in reproducing the observed line ratios) 
the $\rm R_{23}$ parameter is calibrated directly through the $\rm T_e$ method, 
while at high metallicities ($\rm 12+log(O/H) > 8.4$, 
where the electron temperature is more difficult to measure and subject
to temperature fluctuations) 
the calibration is obtained by exploiting models. 
The main problem with the $\rm R_{23}$ 
parameter is that it has a double metallicity solution for each value (upper and lower branches). 
\cite{maiolino08} selected the branch through the $[NeIII]\lambda3869/[OII]\lambda3727$ and
$[OIII]\lambda5007/[OII]\lambda3727$ ratios. 
These are tracers of the ionization parameter, 
but also depend monotonically on the metallicity (although with large scatter),
meaning that the ionization parameter correlates, on average, with the metallicity.
In practice, we simultaneously fitted the three diagnostics and found the best solution.
A more detailed discussion on the method adopted to infer the metallicities is given in Appendix B.
Although the use of this method brings with it some caveats, 
it was also adopted in \cite{mannucci10} and thus allows
a direct comparison with the results in that work (FMR).

Finally, for the one lensed AMAZE galaxy at z=4.8,
for which we can only observe [NeIII] and [OII] in the K band, 
we inferred the metallicity from the [NeIII]/[OII] line ratio, 
exploiting the anticorrelation (though with high dispersion)
between this line ratio and the metallicity found in \cite{maiolino08}.
In Table \ref{table_mgas_met} the integrated metallicities inferred for the AMAZE and
LSD galaxies are reported.

\begin{figure*}[!ht]
 \centering
 \includegraphics[width=.7\linewidth]{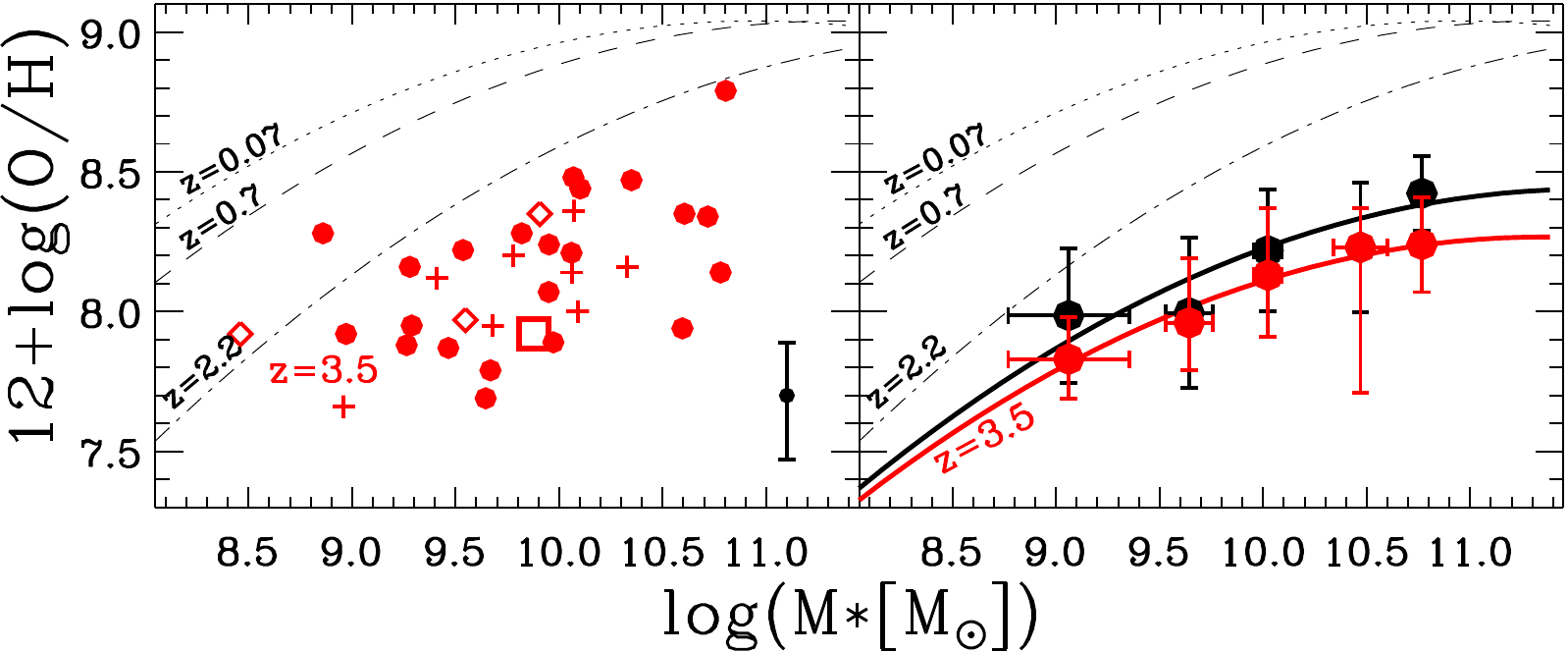}
\caption{Mass-metallicity relation for the galaxies in the AMAZE and LSD samples.
The left panel shows the individual AMAZE and LSD galaxies at z$\sim$3--5 (same
symbol codes as in Fig. \ref{fig_sfrsedhb}).
The overplotted lines (small dot, dashed, dot-dashed, long
dashed) shows the mass-metallicity relation at lower redshifts from previous studies
($z\sim0.07,0.7$, and $2.2$ respectively, see \cite{maiolino08}). 
The average error bar on the metallicity determination is shown in the lower-right corner.
In the right panel the data are binned according to stellar mass.
The black circles show the average metallicities within each stellar mass bin,
while the red circles show the metallicity inferred from the spectra stacked in the same bins.  The black solid line shows
the fit to the average metallicities, while the red solid line shows the fit to
the metallicity of the stacked spectra.} 
\label{fig_mz} 
\end{figure*}

\subsection{Mass--metallicity relation at $z \sim 3.4$ }\label{Mz}

Fig.~\ref{fig_mz} shows the mass-metallicity relation for the full AMAZE+LSD
sample. In the left panel the red symbols show the individual
measurements of the AMAZE and LSD galaxies. 
Red points denote the AMAZE unlensed sources,
diamonds the AMAZE lensed sources at $z\sim3$,
and the red square shows the lensed source at $z\sim 5$.
Red crosses are the LSD sources. In the right panel, 
the full sample at 3$<$z$<$3.7 is binned according to stellar mass in intervals of $0.5$~dex.
The black points show the average metallicities in each stellar mass bin, 
while the red circles show the metallicity inferred from the spectra stacked in the same bins. 
Each bin has six galaxies on average.
The overplotted curves are the best fits of the observed mass-metallicity
relations at different redshifts obtained by various previous surveys \citep[see][]{maiolino08}.
Over the investigated mass range, the metallicity at $z\sim 3.4$ is 
about $0.8$~ dex lower than for local galaxies.
For the best fit to the mass-metallicity relation, 
we adopted the same description of the mass-metallicity relation as in \cite{maiolino08},

\begin{equation}
\label{eq_mz_evol} 
\rm 12+log(O/H) = -0.0864~(log M_* - log M_0)^2 + K_0,
\end{equation}

where $\rm log M_0$ and $\rm K_0$ were determined at each redshift to obtain
the best fit to the observed data points.
By using the complete AMAZE+LSD sample at z$\sim$3.4,
we obtained $\rm log M_0=11.35,11.59$ and $\rm K_0=8.27,8.44$ for the stacked and
average metallicities, respectively\footnote{To avoid the fit to be
dominated by the mass intervals with the largest number of objects, we first
obtained the metallicity average or stacked within mass bins and then
fitted the resulting values with Eq.~\ref{eq_mz_evol}.}.
The inclusion of low-mass systems ($log M_* < 9.5 M\sun$), 
with respect respect to \cite{maiolino08},
results in a flatter relation than at lower redshifts.
However, this putative evolution of the mass-metallicity evolution is still convolved
with the different SFRs probed at different redshifts by the different surveys.
Since the metallicity of galaxies also depends on their SFR, 
as discussed in \cite{mannucci10},
the evolution of the mass-metallicity relation with redshift 
(both in normalization and shape) is largely apparent and mostly due to the higher 
SFR of galaxies observed at high redshift (both because of selection
effects and because of the real intrinsic evolution of the average SFR in galaxies).
Therefore, it is more meaningful to investigate the metallicity evolution relative 
the three-dimensional relation that involves stellar mass as well as SFR, as discussed in the next section.

\subsection{Evolution at z$\sim$3.4 compared with the fundamental metallicity relation}\label{S3e}

As we mentioned in section \S \ref{sec_intro}, 
\cite{mannucci10} have observationally shown that 
galaxies in the redshift range $0 < z < 2.5$ are described 
by a tight three-dimensional relation between stellar mass ($M_*$), 
$SFR$ and gas-phase metallicity.
The homogeneity of the FMR in the redshift range $0< z < 2.5$
suggests that the mechanisms that drive the galaxy evolution 
during this time interval are similar. 
A deviation of galaxies from the FMR at higher redshift (at $z \gtrsim 2.5$) 
would indicate different evolutionary mechanisms at such early epochs.

At z $\sim 3$ \cite{mannucci10} analyzed a subsample of 17 AMAZE and LSD galaxies and
found a deviation of the {\it average} metallicity of these galaxies from the FMR
(at $0 < z < 2.5$) of about $-0.6$ dex. 

In our sample, the metallicity of the full AMAZE+LSD composite spectrum at z$\sim$3.4 (34 galaxies) 
is $8.03^{+0.16}_{-0.18}$, this is $0.43^{+0.16}_{-0.18}~dex$ lower than the expected
metallicity from the FMR ($8.46$) evaluated at the average star formation
($10^{1.5}~M_{\sun} yr^{-1}$) and average stellar mass ($10^{9.8}~M_{\sun}$) of the full sample.
The aim of this section is to investigate the behavior of the full AMAZE+LSD
sample at z$\sim$3.4 (twice the size of that used by \cite{mannucci10}) 
compared with the FMR, by investigating the deviations as a function of SFR and mass
through exploiting the fact that the 34 galaxies of the sample span over two orders of magnitude
in mass and SFR.

  \begin{figure*}[!ht]
   \centering
\includegraphics[width=.7\linewidth]{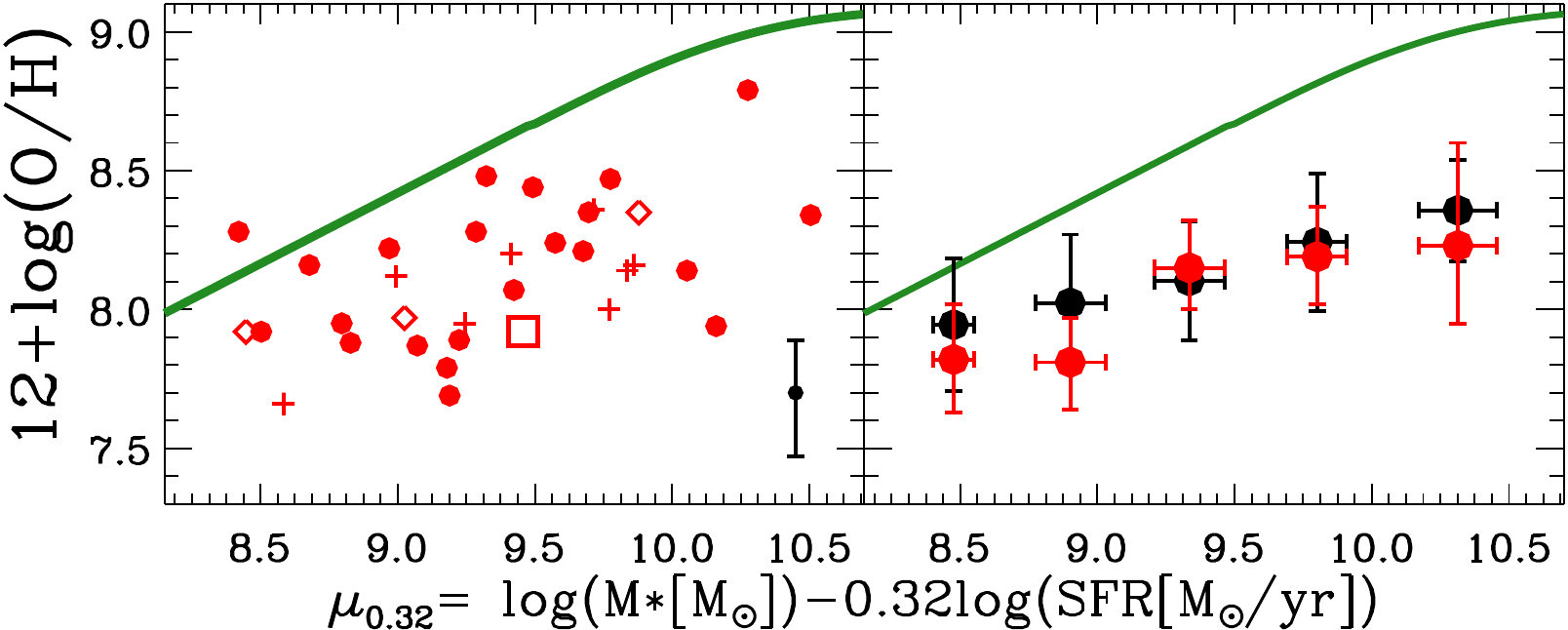}
\caption{Metallicity as a function of the parameter $\mu_{0.32}$. 
The green line plots the FMR projection  as a function of $\mu_{0.32}$ \citep{mannucci11}.
The left panel shows the individual AMAZE and LSD galaxies at z$\sim$3--5 (same
symbol codes as in Fig. \ref{fig_sfrsedhb}).
The average metallicity uncertainty of the sample is shown in the lower-right corner.
In the right panel the data are binned by the $\mu_{0.32}$ parameter.
The black circles show the average metallicities in each $\mu_{0.32}$ bin,
while the red circles show the metallicity of the spectra stacked in the same bins. 
The black error bars show the dispersion of the average metallicity and $\mu_{0.32}$ at each bin. 
The red vertical error bars show the error of the metallicity of the stacked spectrum in each bin.
The red horizontal error bars show the dispersion of the average $\mu_{0.32}$ in each bin. 
}
  \label{fig_ohmu32}
\end{figure*}

To further quantify the deviations of the data at $z\sim 3.4$ with respect to the local/low-z FMR, 
we used the parameter $\rm \mu_{0.32}= \log{(M_*)} - 0.32 \, \log{(SFR)}$,
which provides a projection of the FMR that minimizes the metallicity
scatter of local galaxies \citep{mannucci10} and its extension to the lower masses of \cite{mannucci11}.
This projection of the FMR shows the deviation of z$\sim$3.4 galaxies 
compared with to the local FMR more clearly and more globally.

In figure \ref{fig_ohmu32}, the green line is the best fit of the metallicity distribution of 
local galaxies as a function of $\mu_{0.32}$ \citep[see][]{mannucci11}.
In the left panel, the individual measurements of the full AMAZE+LSD sample 
are shown (same symbol-coding as in Fig.\ref{fig_mz}).
Galaxies at z$\sim$3.4 have a metallicity dispersion of about 0.25 in dex, 
much higher than local galaxies (which have $\sigma \sim 0.05$), 
which probably reflects the mixture of different stages of unsteady chemical 
evolutionary processes at this epoch, in contrast to what is observed at later epochs.

\begin{figure}[!ht]
  \centering
  \includegraphics[width=1.\linewidth]{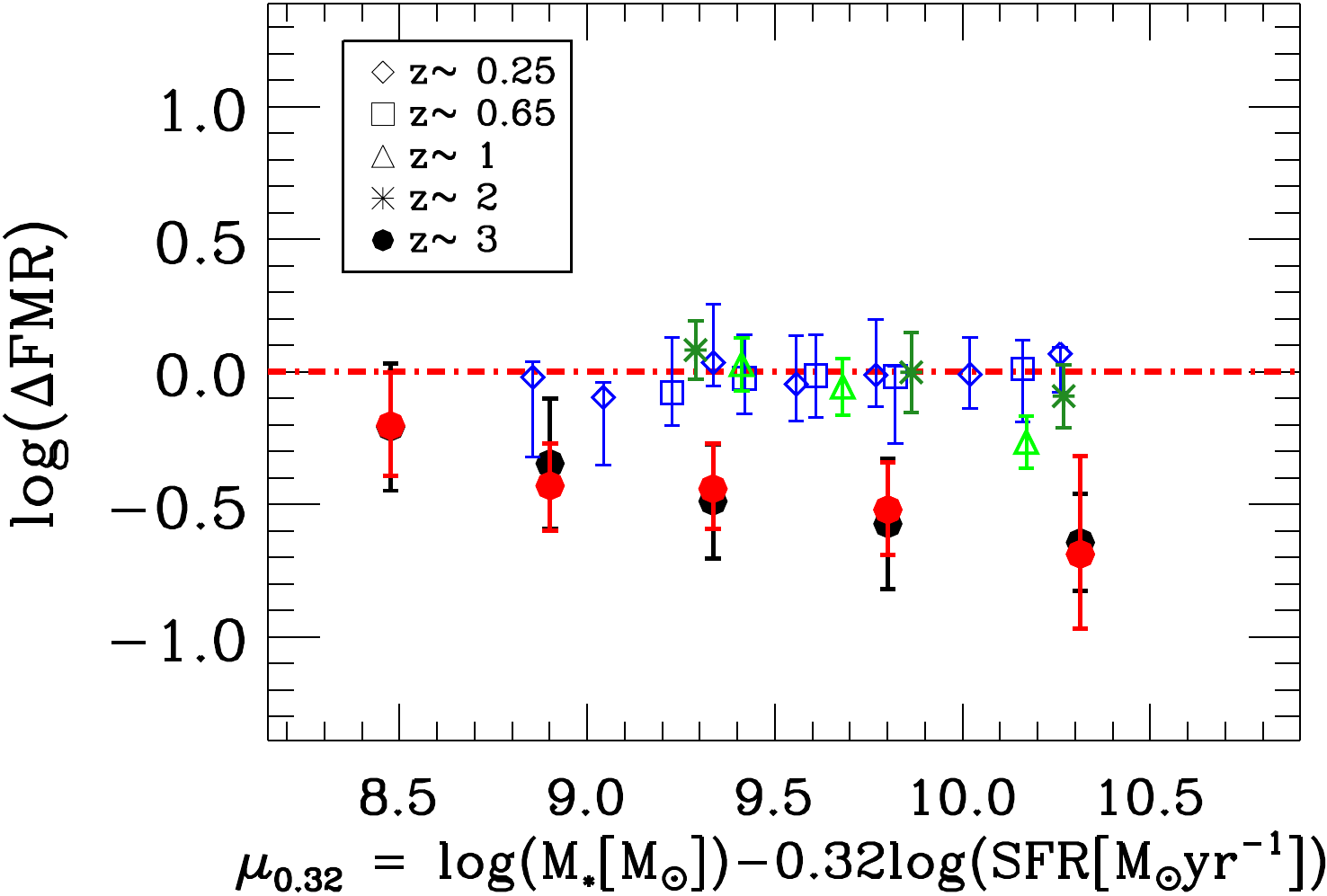} 
  \caption{Deviation of the FMR as a function of the $\mu_{0.32}$ parameter.
  Black and red circles show metallicity differences of the averaged and stacked spectra,
  respectively, within each $\mu _{0.32}$ bin compared with the FMR.
  The black error bars show the dispersion of the average metallicity within each bin.
  The red error bars show the error of the metallicity of the stacked spectrum in each bin.
  Blue diamonds, blue squares, green triangles, and green asterisks 
  show the average metallicity differences compared with the FMR of galaxies 
  at z$\sim$0.25, 0.65, 1, and 2, respectively \citep{cresci12,vergani12,law09,lehnert09}.
  }
  \label{fig_DFMRmu032}
  \end{figure}

In the right panel, the black circles show the average metallicities in $\mu_{0.32}$ bins,
while the red symbols show the metallicity of the stacked spectra in the same bins.
The plot also shows that the deviation from the FMR increases as a function of $\mu_{0.32}$.
This trend is more clearly shown in Fig. \ref{fig_DFMRmu032}, 
where we plot the deviation of z$\sim$3 galaxies 
(averaged and stacked in bins of $\mu_{0.32}$, black and red points, respectively)
as a function of $\mu_{0.32}$. 

The deviations (up to 0.6--1~dex) are clearly much higher than the mean 
scatter of local galaxies on the FMR ($\sim 0.05$~dex). 
For comparison, in Fig.\ref{fig_DFMRmu032} we also show the location of galaxies
at z$\sim$0.25-0.65 \citep{cresci12}, at z$\sim$1 \citep{vergani12}, 
and at z$\sim$2 \citep{law09,lehnert09}, whose average metallicities
are derived with the \cite{maiolino08} calibration for consistency.
Galaxies at z$\sim$0.25-0.65 do not show any deviation compared with the local FMR. 
Galaxies at z$\sim$1--2 do show some marginal deviation
from the FMR at high $\mu _{32}$, but it is much lower than that of galaxies at z$>$3.
This result indicates that the deviation from the FMR is mostly a peculiarity of z$>$3 galaxies, 
which suggests a transition epoch across z$\sim$3 where the main mode of galaxy formation
probably changes.

To better understand the physical parameters that drive the deviations at z$\sim$3.4
from the FMR, we also investigated the dependence of the deviations from FMR as
a function of other galaxy physical parameters.
In particular, we investigated the deviation of z$\sim$3.4 galaxies from
the FMR as a function of stellar mass, SFR, and specific SFR (sSFR=SFR/$M_*$, which is
a proxy of the evolutionary stage of galaxies).
None of these parameters strongly correlate with the deviation from the FMR, 
the main driving quantity of the deviations is the 
$\rm \mu _{32}$ parameter that, as discussed in \cite{mannucci10},
is linked to the star formation efficiency and to the combined effect of
inflows and outflows. 
We investigate these scenarios in more detail in Sect.\ref{sec_models}.

  \begin{figure}[!ht]
  \centering
  \includegraphics[width=1.\linewidth]{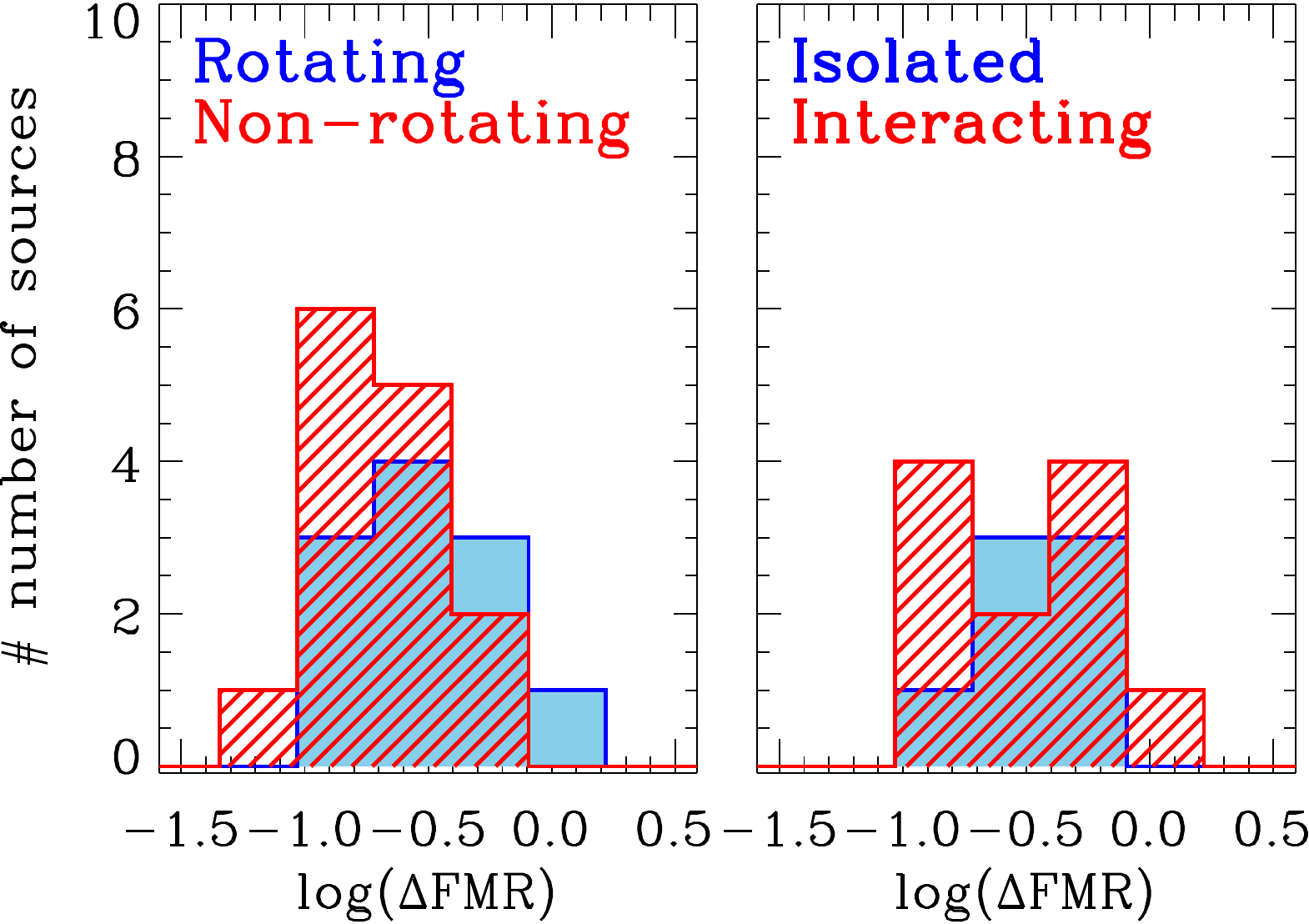}
  \caption{Histogram of the metallicity deviations from the FMR for the AMAZE and LSD galaxies at
  z$\sim$3.4. 
The blue histograms refer to the rotating disks and isolated galaxies. 
The red histograms indicate the nonrotating galaxies and interacting galaxies.
   }
  \label{fig_histodiffFMR}
  \end{figure}

A possible origin of the deviation of galaxies at z$\sim$3.4
from the FMR may be associated with an increased rate of mergers, 
which destabilize low-metallicity gas in galaxy outskirts and drive it toward the central,
active regions, 
as suggested by some models \citep[e.g.][]{rupke10,perez11,pilkington12,torrey12}.

The dynamical properties of the AMAZE+LSD sample has been studied by \cite{gnerucci11a}.
They found a significant fraction ($\sim$30$\%$) of galaxies with regular rotation,
especially among massive systems, indicating that massive rotating disks
(although highly turbulent) were already in place at this early epoch.
Other galaxies show irregular kinematics, suggesting recent or ongoing merging,
while for other galaxies it is not possible to distinguish between rotation or irregular 
kinematics due to low angular resolution or low S/N. 

Fig. \ref{fig_histodiffFMR} shows the distribution of the metallicity difference 
from the FMR of galaxies at z$\sim$3.4.
We divided the sample between rotating and nonrotating systems
\citep[according to][]{gnerucci11a}.
There is no significant difference between rotating
and nonrotating systems in terms of deviations from the FMR;
the average deviation  from the FMR of rotating systems is
$\rm \langle \log{(\Delta FMR)}\rangle = -0.37 _{-0.36}^{+0.27}$,
while the average deviation of nonrotating systems is
$\rm \langle \log{(\Delta FMR)}\rangle = -0.58 _{-0.08}^{+0.23}$.
We also divided the sample into interacting and isolated galaxies 
by visual inspection of the available HST images (including those in the H band). 
There is no significant difference between interacting and isolated systems in terms of 
deviations from the FMR; the average deviation from the FMR of isolated systems is
$\rm \langle \log{(\Delta FMR)}\rangle = -0.29 _{-0.28}^{+0.10}$,
while the average deviation of interacting systems is
$\rm \langle \log{(\Delta FMR)}\rangle = -0.37 _{-0.36}^{+0.26}$.
These results suggests that enhanced merging cannot be the only explanation for the
deviation from the FMR in galaxies at z$\sim$3.4. Moreover, 
there are indications that merging galaxies in the local Universe are following the FMR \citep{cresciprep}.

Another possibility is that the deviation from the FMR at z$>$3 is associated with selection
effects and, in particular, with the color-selection criterion of LBG galaxies.
The color selection tends to avoid dusty galaxies, which are presumably more metal rich.
However, other samples at lower redshifts were color-selected, but do not
show similar strong deviations from the FMR, as discussed above. Moreover,
some of the most deviating galaxies are also the most massive, which should
also be the most dusty (according to the mass-dust relation observed locally
and at high redshift), which constrasts with the putative color bias.
Therefore, the deviation from the FMR seems to be a peculiarity intrinsic to z$>$3 galaxies.

While a discussion of the potential physical scenarios that might explain
the deviations of galaxies at z$>$3 is given in sections \S \ref{sec_models} and \S \ref{sec_metgrad}, 
we note here that gas metallicities are not the only property that indicates a transition in galaxy evolution at z$\sim$3.
Indeed, the finding that galaxies at z$>$3 deviate from the
trends observed at lower redshift has also been obtained in other
studies. \cite{moller13} have found an indication for a transition in the
metal content of DLA at z$\sim$2.6, which they interpreted as a transition from
an epoch dominated by strong gas infall to the more quiescent, later epoch
in terms of gas accretion. A transition at a similar epoch has also been
observed in the reddening properties of Ly$\alpha$ emitters and QSO host
galaxies \citep{fynbo13,nilssonmoller09,nilssonmoller11,bongiovanni10}.

Even more interesting, \cite{prochaskawolfe09} have shown that
the HI content in galaxies does not evolve in the redshift range 0$<$z$<$2.5,
while it does evolve at higher redshifts. They associated the transition phase
with the conversion of HI into H$_2$ at z$<$2.5, which in turn regulates star
formation and metal production. Although additional modeling is required,
it is clear that our result suggesting a transition of galaxy evolution 
occurring at z$\sim$2.5-3.0 is in line with various other observational results.

There is yet another possibility to explain the deviation 
of our sample at z$\sim$3.4 from the FMR: LBGs at high-z migth belong
to a different class of objects that, at any redshift, do not obey the FMR.
\cite{hunt12} have shown that low-mass, low-metallicity galaxies follow a 
``fundamental plane'', whose extension to high masses falls short to account 
for star-forming disks observed in the SDSS,
but does reproduce the properties of the LBG at high-z.
It is not simple to directly compare our results with the result of \cite{hunt12}, 
since they used different metallicity tracers and so the comparison may be subject to
significant offsets introduced by the different calibrations adopted 
for the different samples \citep[see][]{kewley08}.
Moreover, the sample adopted by \cite{hunt12} tends to preferentially select low-metallicity
galaxies, and the stellar mass estimation is also different from ours.
Finally, \cite{hunt12} used the AMAZE and LSD galaxies (the previous subsample of 17 galaxies)
to construct the relationship, which makes claiming a consistency of the AMAZE+LSD sample 
with the work of \cite{hunt12} is somewhat circular.
However, the scenario of a nonevolving population of strongly star-forming galaxies
that deviates from the FMR is a scenario that should be investigated in more detail 
by including additional independent samples.

\begin{figure}[!h]
   \centering
   \includegraphics[width=1.\linewidth]{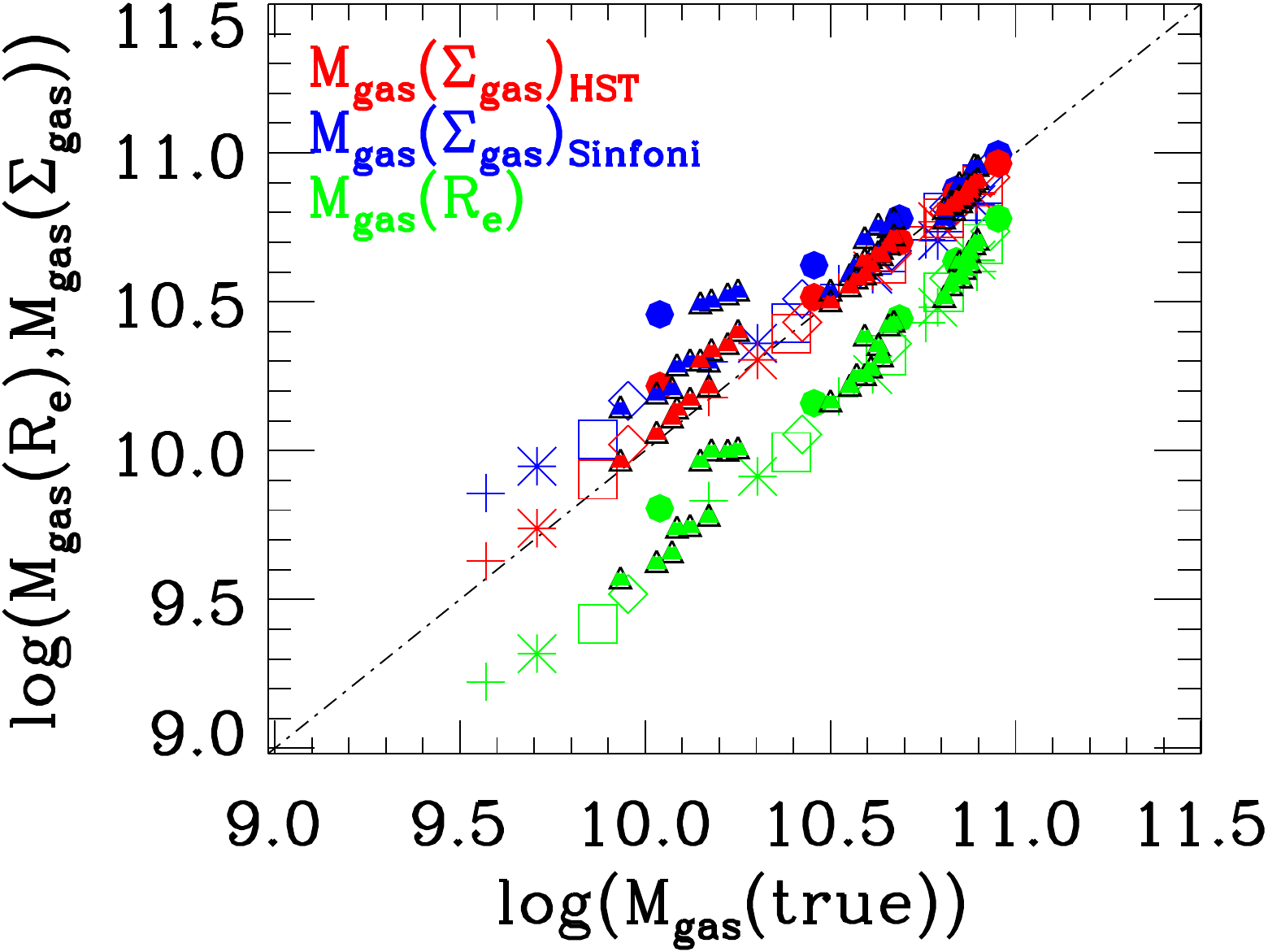}
\caption{Mass of gas of the simulated galaxies with different Sersic profiles ($0.5<n<4$) and 
different effective radii ($5<Re<2.5$~kpc) 
inferred with the two different methods discussed in the text.
The x-axis shows the true gas mass (input in the simulated galaxy),
while the y-axis show either the gas mass calculated with the old method of using the
{\bf total} SFR and effective radius to infer an average gas surface
density ($\rm M_{gas}(R_e)$, green symbols), 
or the gas mass calculated with our new method,
measuring the $\Sigma _{gas}$ pixel by pixel in the data
($\rm M_{gas}(\Sigma _{gas})$, red and blue symbols).
Red symbols show the simulations of HST data and blue symbols show the simulation of SINFONI data.
The green symbols are independent of the method since they use the integrated properties of
galaxies.
Galaxies modeled with Sersic indices $n=0.5,1,2,3,4$ are shown by crosses, asterisks, 
squares, diamonds, and circles, respectively. Interacting systems are marked with black triangles. 
}
  \label{fig_mgas_sim}
\end{figure}

  \begin{figure*}[!ht]
   \centering
 \includegraphics[width=.8\linewidth]{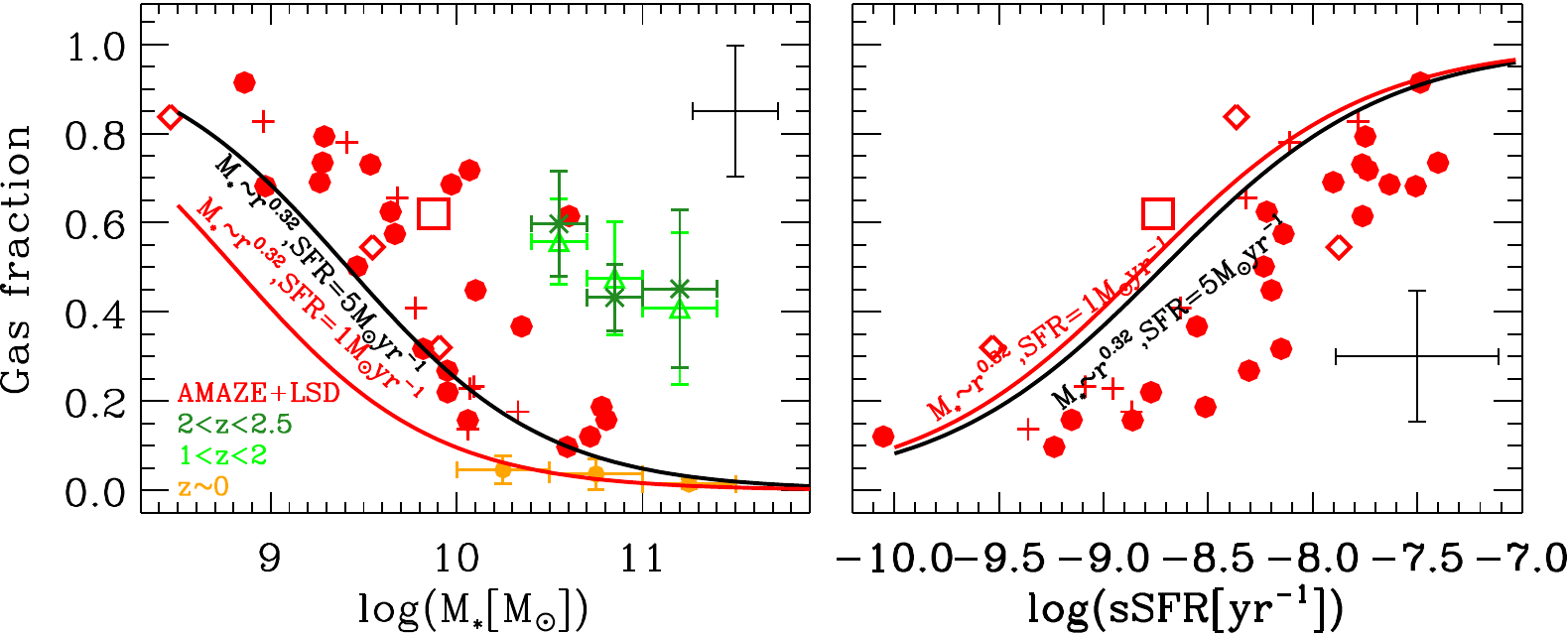}
 \caption{Gas fraction as function of stellar mass and sSFR.
 The red symbols show the AMAZE+LSD galaxies at z$\sim$3--5 
 (same symbol-coding as Fig.\ref{fig_sfrsedhb}).
 The solid lines show the incompleteness limit in the AMAZE unlensed (black) and
 lensed (red) galaxies.
 In the {\it left panel} the orange dots show the average gas fractions of local massive 
 galaxies in bins of stellar mass, while green asterisks and triangles show the averages 
 of the galaxies at 1$<z<$2 and 2$<z<$2.73, respectively, 
 whose gas content has been measured through 
 CO observations \citep{daddi10,saintonge11,tacconi13,saintonge13}. 
 The upper-right error bars show the median error on the gas fraction and on
 the stellar mass in our sample.
 {\it Right panel:} gas fraction as a function of the sSFR, 
 with the same symbol code as in the left panel.
  }
  \label{fig_gasf_mass}
\end{figure*}

\begin{figure*}[!ht]
  \centering
\includegraphics[width=.8\linewidth]{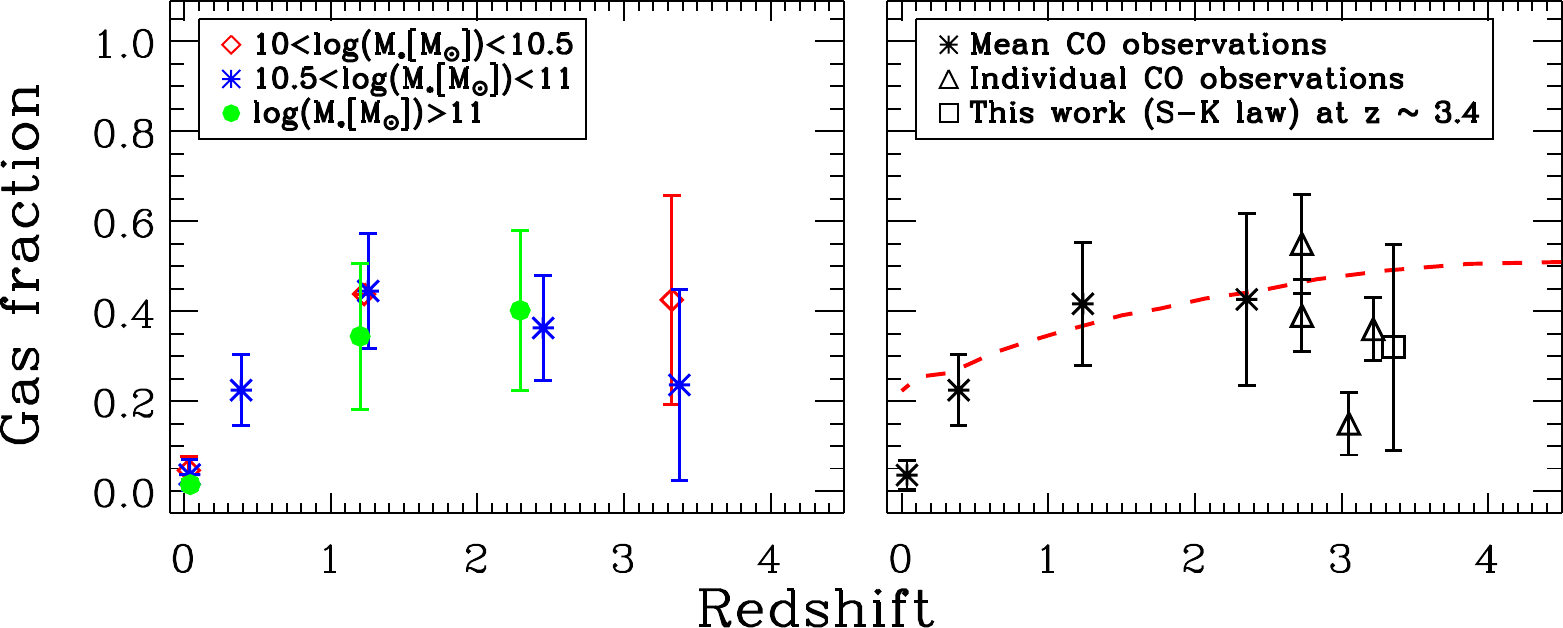}
\caption{Gas fraction as a function of redshift from a compilation 
of measurements in the redshift range 0$\leq$z$\leq$3.7 (same data of Fig.~ \ref{fig_gasf_mass})
and including our new results.
{\it Left panel:} results averaged in bins of stellar mass. 
{\it Right panel:} black asterisks show the average (over the whole mass range at
$\rm M_* > 10^{10}~M_{\odot}$) evolution of the gas fraction up to z$\sim$3.4. 
At z$\ge$2.7, the triangles show the individual values inferred from direct CO 
observations \citep{riechers10,magdis12a,saintonge13}.
The hollow square is the average measurement at z$\sim$3.4 determined 
in this work by inverting the S-K law.
The red dashed line shows the evolution of the gas fraction from the semianalytic models of 
\cite{lagos11a}.}
  \label{fig_gasf_z} 
  \end{figure*}

\section{Gas content}\label{s_gas}
 
As we discuss in the next sections, to properly interpret the metallicity evolution in galaxies,
it is extremely useful to have information on the gas content. 
This additional information allows us to partly remove
degeneracies between various scenarios (e.g., closed-box, inflows, outflows). 
However, with the exclusion of a few powerful SMGs and QSO host galaxies, 
the direct measurement of the gas content in galaxies at z$>$3 is still extremely challenging and,
so far, only a few detections have been obtained \citep{riechers10,magdis12a}, 
only one of which (the Cosmic Eye) overlaps with our sample.
An alternative method used by various authors \citep[e.g.][]{conselice13,vergani12,tan13,erb06b,mannucci09}
to infer the gas content, is to invert the S-K law
although this is indirect and subject to larger uncertainties.

In the local Universe, the S-K law relates the gas surface density ($\Sigma_{gas}$) 
with the star-formation surface density \citep[$\Sigma _{SFR}$,][]{schmidt59,kennicutt98,bigiel08,schruba11}.
Recently, \cite{genzel10}, \cite{daddi10}, \cite{tacconi10}, \cite{tacconi13}, and \cite{saintonge13}
have measured large amounts of molecular gas in star-forming disks at $z\sim 1-2.7$ 
and have found that the S-K law was already in place even at those early epochs of the Universe.
Similar results have been obtained by using the dust mass as a proxy
of the gas content in high-z galaxies \citep{magdis12b,santini13}.

Therefore, the currently favored scenario is that
the high SFRs measured at z$\sim 1-2.7$ are caused by the large
amount of molecular gas and not by a more efficient mechanism that converts gas into stars.

If the AMAZE galaxies are considered as typical star-forming disks at $z\sim$3 (i.e., assuming
the validity of the S-K law), and assuming that the S-K relation does not evolve out to $z\sim$3
\footnote{ {\bf Some indications of evolution of the S-K at high-z were found by \cite{tacconi13}
and \cite{santini13}; however, as pointed out by \cite{santini13}, the evolution, if any, 
is mild and galaxies at all redshifts can be fitted reasonably well with a single (integrated) 
S-K law with a slope of 1.5.} },
we can infer the $\Sigma_{gas}$ from 
the $\Sigma _{SFR}$, and finally derive the gas mass.
A similar approach was adopted by \cite{erb06b} and \cite{vergani12} for galaxies at
z$\sim$1.5, by \cite{conselice13} at z$\sim$1.5--3, and by \cite{mannucci09} for galaxies at z$\sim$3.
Most of these previous studies do not really map the $\Sigma _{SFR}$;
instead, they measure a ``size'' of the galaxies and infer an average $\Sigma _{SFR}$
by dividing the total SFR by the average ``size'' of the galaxy,
and from this they derive an average $\Sigma_{gas}$.
By following this method, assuming an idealized galaxy with a constant 
$\Sigma _{SFR}$ within a radius $r$
(by using the S-K law and adopting a Chabrier IMF), we obtain the relation

\begin{equation}
\Sigma_{gas}(M_\odot/{\rm pc^2})= 241\ \left(\frac{\rm SFR}{\rm M_\odot/yr}\right)^{0.71}\ 
                                   \left(\frac{r}{\rm kpc}\right)^{-1.42}
\label{eqsgas}
\end{equation}

and

\begin{equation}
M_{gas}(M_\odot) = 757 \times 10^6\ \left(\frac{\rm SFR}{\rm M_\odot/yr}\right)^{0.71}\left(\frac{r}{\rm kpc}\right)^{0.58},
\label{eqmgas}
\end{equation}

where $\rm SFR$ is the {\bf total}
galaxy SFR. A factor of 1.7 between the Salpeter \citep[adopted in][]{kennicutt98} 
and Chabrier IMF was applied \citep{pozzetti07}.

However, this method is subject to two main caveats.
First, the previous equations are formally correct for a disk with a flat,
constant distribution of $\Sigma _{SFR}$ within a radius $r$.
It is not obvious that for a distribution of $\Sigma _{SFR}$ typical of star-forming
galaxies (especially those at z$\sim$3) these equations are appropriate, especially given that the
S-K relation is not linear. Secondly, it is not clear which galaxy radius should be used
to infer the average $\Sigma _{SFR}$: as we have discussed in the previous section,
the galaxy radius may change by even a factor of several,
depending on the method adopted to determine the radius and on the adopted definition
of ``galaxy radius''. In galaxies with irregular morphologies (e.g., multiple knots)
or interacting/merging systems it is even more unclear which radius should be taken.

Here we adopted a different method that is much less affected by these problems.
We measured the $\Sigma _{SFR}$ within each individual pixel and from this derived the
$\Sigma_{gas}$ locally by applying the S-K relation, and the gas mass sampled by each pixel
by simply multiplying the local $\Sigma_{gas}$ by the physical area sampled by the pixel.
We then inferred the total gas mass by simply combining the gas masses inferred in each pixel.
This method is rigorous when the galaxy is fully resolved.
It totally bypasses all problems associated with the radius definition,
irregular morphologies, and interacting systems that affect the previous method.
Obviously, when the galaxy is not resolved or is only marginally resolved,
this method incurs problems similar to the previous one.
However, as discussed below, for the class of galaxies observed by us,
which are resolved, we have tested that 
our new method provides a much more accurate determination of the gas content.

We simulated galaxies with different Sersic profiles, different effective radii 
($R_e=0.5,1.,1.5,2.,2.5$ kpc), and different Sersic indices ($n=0.5,1,2,3,4$).
We also simulated interacting/merging systems by combining two of the previously
simulated galaxies, separated by 1 to 2 kpc.
These simulated images describe the true (intrinsic) distribution of $\Sigma _{SFR}$ and,
by inverting the S-K law, we can derive the true distribution of $\Sigma _{gas}$ and
therefore the true gas mass.
Then, these images were smoothed to the PSF of SINFONI or HST 
and resampled to the pixel size of SINFONI or HST. 
We also added noise at a level typical of our poorest data, to be conservative.
In these images that match the SINFONI-HST resolution, the gas mass is
estimated by using the pixel surface brightness method suggested by us
(i.e., by adding the $\Sigma_{gas}$ inferred by the $\Sigma _{SFR}$
measured in each pixel).
The resulting gas mass inferred for the AMAZE galaxies 
by using our method are reported in Table \ref{table_mgas_met}, Col. $2$.
We also applied the method adopted previously of inferring 
the average $\Sigma _{gas}$ from the effective radius (Eq. \ref{eqsgas}).
In the latter case we have been optimistic,
in the sense that we have bypassed the measurement of radius 
in the simulated images (which introduces additional uncertainties)
by directly taking the effective radius given in the simulation.

In Fig.~\ref{fig_mgas_sim}, a comparison of these two methods is shown.
The dash-dotted line indicate the 1:1 relation, 
where the measurements would recover exactly the true gas mass given in the simulation.
Clearly, the old ($R_e$) method significantly underestimates the true gas mass by
as much as a factor of three. Instead,
our new method recovers the true gas mass with an accuracy better than 20\% in most cases.

The resulting gas mass $M_{gas}$ can be compared with the total baryonic mass 
$M_*+M_{gas}$ to estimate the gas mass fraction.
The gas fraction $f_{gas} = M_{gas}/(M_{gas}+M_*)$ is an indicator of the
galaxy evolutionary stage.
Galaxies with a low gas fraction are systems that already used their gas to form stars
and are becoming old \citep[such as most local galaxies in ][]{saintonge11}, 
while galaxies with a low gas fraction still have to convert the bulk
of the available gas into stars.
In figure \ref{fig_gasf_mass}, left panel, the resulting gas fractions are plotted
as a function of the galaxy mass.
The red symbols denote the AMAZE and LSD galaxies at z$\sim$3.4 and at z$\sim$5. 
For comparison, the yellow solid circles show the average gas fractions of 
local massive galaxies (with $M>10^{10}~M_{\odot}$) in bins of stellar mass, 
whose gas content has been measured by \cite{saintonge11} through direct CO observations.
The green triangles and asterisks show the average gas fractions, also measured through direct CO observations,
of galaxies at z$\sim$1-2.73 in bins of stellar mass \citep{daddi10,tacconi13,saintonge13}.

There is an anticorrelation between gas content and galaxy stellar mass, 
which was previously noticed in samples observed directly in CO \citep{tacconi13} at lower
redshifts (and high stellar masses) and also observed by using the dust mass as a proxy
of the gas mass at 0$<$z$<$2.5 \citep{santini13}.
However, part of this anticorrelation may be associated with bias effects at low masses.
Indeed, one should take into account that at low stellar masses our (indirect) method
of deriving the gas mass is subject to incompleteness. Indeed, our sample is limited in SFR;
since we estimate the gas mass through the $\Sigma_{SFR}$ (by inverting the S-K law), 
our SFR limit indirectly translates into a lower limit on the gas mass that we can probe.
Moreover, since $\rm \Sigma_{SFR}\propto SFR/r^2$ and since the galaxy mass is known to correlate 
with galaxy radius ($M_{*}\propto r^{0.4}$), 
our lower limit on the gas content also indirectly depends on the galaxy mass.
The solid lines in Fig.\ref{fig_gasf_mass} show the inferred incompleteness limit for the AMAZE
sample of unlensed galaxies
(black line), which has a minimum SFR of 5~$M_{\odot}~yr^{-1}$, and for the lensed galaxies in the 
AMAZE sample (red line), which have a minimum SFR of 1~$M_{\odot}~yr^{-1}$.
The lack of AMAZE and LSD galaxies below these lines is probably not intrinsic to the distribution 
of galaxies at z$\sim$3, but is possibly due to incompleteness.
However, we note that the location of the four lensed galaxies in AMAZE is significantly above 
the incompleteness line, and they still show a clear drop of gas fraction as a function of stellar mass.
Therefore, the anticorrelation between gas fraction and stellar mass may indeed be real,
and would be in line with the findings of direct CO observations at lower redshifts \citep{tacconi13},
although at higher stellar masses and lower redshifts.
This result would support the downsizing scenarios,
where massive galaxies at z$\sim$3 have already consumed or ejected a significant fraction of gas,
while low-mass galaxies are evolving more slowly.

The right panel of Fig. \ref{fig_gasf_mass} shows the gas fraction as a function of
$\rm sSFR = SFR/M_{star}$, which is a tracer of the evolutionary stage of galaxies. As recently
found by \cite{tacconi13} at lower redshifts, there is a correlation between gas fraction
and sSFR, suggesting that more evolved galaxies (lower sSFR) have lower gas fraction, because gas
has been consumed by star formation (and expelled by the supernovae activity).
However, in this case as well, our results are affected by incompleteness at low sSFR
as a consequence of the SFR limit of our sample.

Either samples extending to lower SFR or direct measurements of the gas content 
(e.g., through CO mm observations) at z$>$3 are required to validate these trends and 
verify that they are not a consequence of selection effects.

At masses higher than $10^{10}~M_{\odot}$ the incompleteness of our sample,
in terms of gas fraction, is very low. Hence our data at $M_{*}>10^{10}~M_{\odot}$ can be used, 
along with results at lower redshift, to constrain the evolution of the gas fraction in galaxies out to z$\sim$3.4.

The left panel of Fig.~\ref{fig_gasf_z} shows the evolution of the gas fraction
in galaxies divided into different bins of stellar mass \citep[to minimize the variation effect
of the gas fraction with stellar mass,
which is found to occur also at these high masses][]{santini13}. 
Although with high dispersion, 
the average gas fraction in massive galaxies shows a rapid increase
by over an order of magnitude
from z$=$0 to z$\sim$2, which was previously reported
\citep{tacconi10,daddi10,tacconi13}.
At higher redshift our data show that the evolution of the gas fraction flattens,
and there is a mild indication that it may even begin to decrease.
This finding is in line with the result by \cite{magdis12a},
who found indications for a flattening of the evolution of the gas fraction 
in galaxies at z$\sim$3 based on CO observations of two galaxies (one of which is an upper limit).
The evolution of the gas fraction in massive galaxies strongly resembles the evolution of the
cosmic density of the star formation rate \citep{hopkinsbeacom06} and provides additional support to the 
scenario in which the cosmic evolution of the SFR in galaxies is driven by the evolution of 
their molecular content, as suggested by \cite{or09}.

The right panel of Fig.~\ref{fig_gasf_z} shows the inferred evolution of the
gas fraction in galaxies, where we took the average in different redshift bins from the left panel.
While the asterisks show the direct CO measurements (averages at z$<$3 and individual
detections at z$>$3),
the hollow square indicates the average of our measurements at z$\sim$3.4, 
obtained by inverting the S-K law.
The prediction of the cosmological models obtained by \cite{lagos11a} is indicated with a red dashed line, 
which shows the expected average for the total population of star-forming galaxies
with vigorous SFR ($> 10~M_{\odot}~yr^{-1}$), that is, a range similar to our sample.
The models described by \cite{lagos11a} do indeed associate the evolution of the SFR
with the evolution of the content of molecular gas in galaxies.
The model expectations for strongly star-forming galaxies is consistent,
within the dispersion and error bars (although on the high side),
with the observed evolution of the gas fraction in massive galaxies.

\section{Comparison with simple models} 
\label{sec_models}

Several models have been proposed to explain the metallicity evolution
in galaxies. Here we attempt a very simple modeling to constrain the
inflow and outflow in galaxies at z$>$3.
Key quantities for investigating these questions are the metallicity, 
the gas fraction, and the effective yield,
that is, the amount of metals produced and retained in the ISM per
unit mass of formed stars.
The latter quantity deserves a few introductory words.

In a closed-box model the metallicity is directly related to the gas fraction 
by the equation
\begin{equation}
Z=y \, ln(1/f_{gas}),
\label{eq:yields}
\end{equation}
where $y$ is the true stellar yield, that is, the ratio between the amount of
metals produced and returned into the ISM and the mass of stars.
Nevertheless, during their lifetime, 
galaxies experience outflows (supernovae explosions, winds),
inflows (pristine or enriched gas) and merging events.
Therefore, the resulting metallicity generally differs from Eq. \ref{eq:yields}.
One can invert Eq. \ref{eq:yields} by defining the effective yield as

\begin{equation}
y_{eff}=Z / ln(1/f_{gas})~.
\label{eq:yields_eff}
\end{equation}

The effective yield is equal to the true yield for a galaxy 
that evolves like a close box, 
while it may differ substantially in the case of metal enriched outflows and 
in the case of inflows of metal-poor gas.
The effective yields computed for the AMAZE and LSD galaxies
are given in Table~\ref{table_mgas_met}.

To constrain inflow and outflow properties of z$>$3 galaxies,
we compared our results with the simple model described in \cite{erb08}.
This model considers pristine gas accretion at a rate that is 
a constant fraction $f_i$ of the SFR
(i.e., as a continuous process or average of many minor events)
and also an outflow rate $f_o$ (of gas with the same metallicity as the current
galaxy average) at a constant fraction of the SFR 
(i.e., the so-called mass-loading factor of the outflow).

 \begin{figure}[!ht]
 \centering
  \includegraphics[width=1.\linewidth]{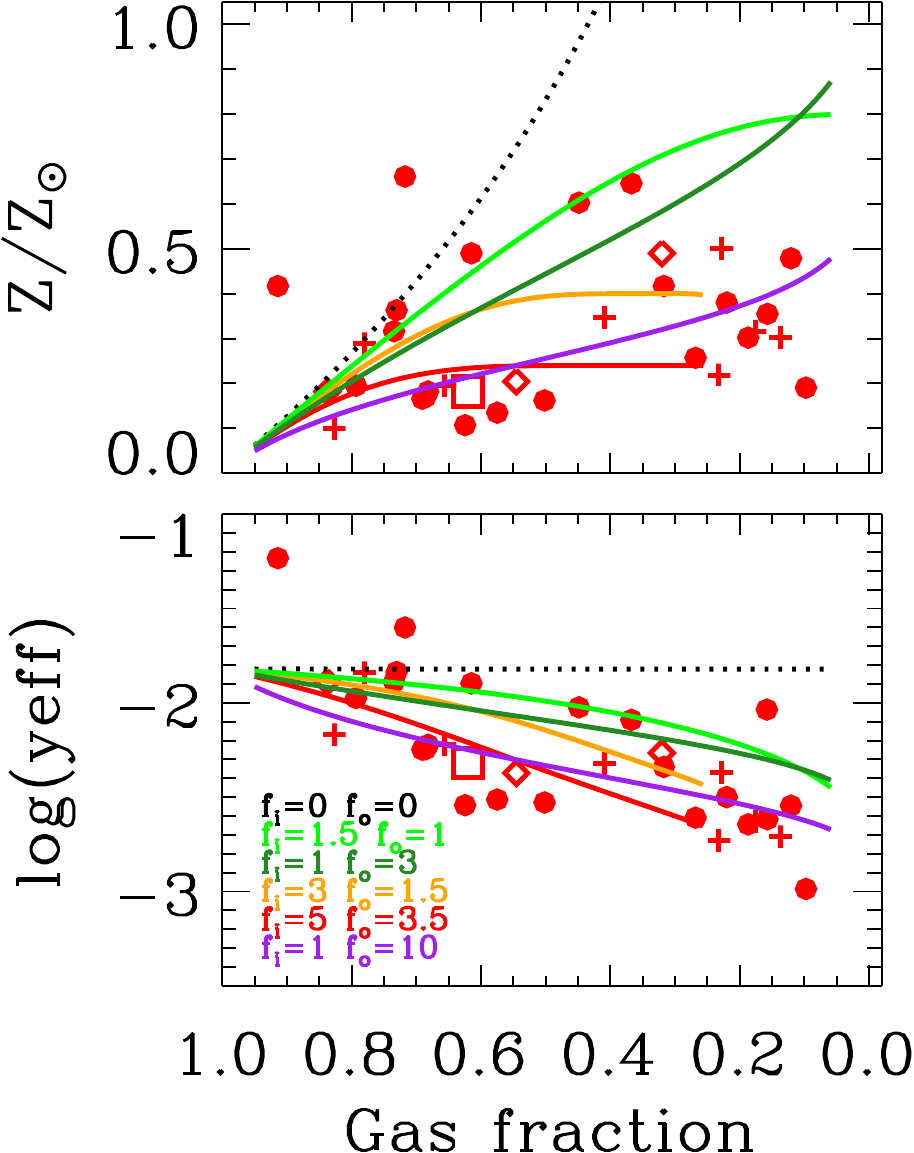}
\caption{Metallicities and effective yields as a function of gas fraction
for the galaxies in our sample at z$\sim$3,
with the same symbols as in the previous figures.
In both panels the \cite{erb08} models for various inflows $f_i$,
outflows $f_o$ rates (relative to the SFR) are shown.
The black dotted line is the closed-box model ($f_i=f_o=0$). 
In the bottom panel (lower-right corner), 
the inflows and outflows rates in units of the SFR are shown for
every model (with different color-coding).
}
  \label{fig_ymgasf}
  \end{figure}

 \begin{figure*}
  \centering
  \includegraphics[width=0.7\linewidth]{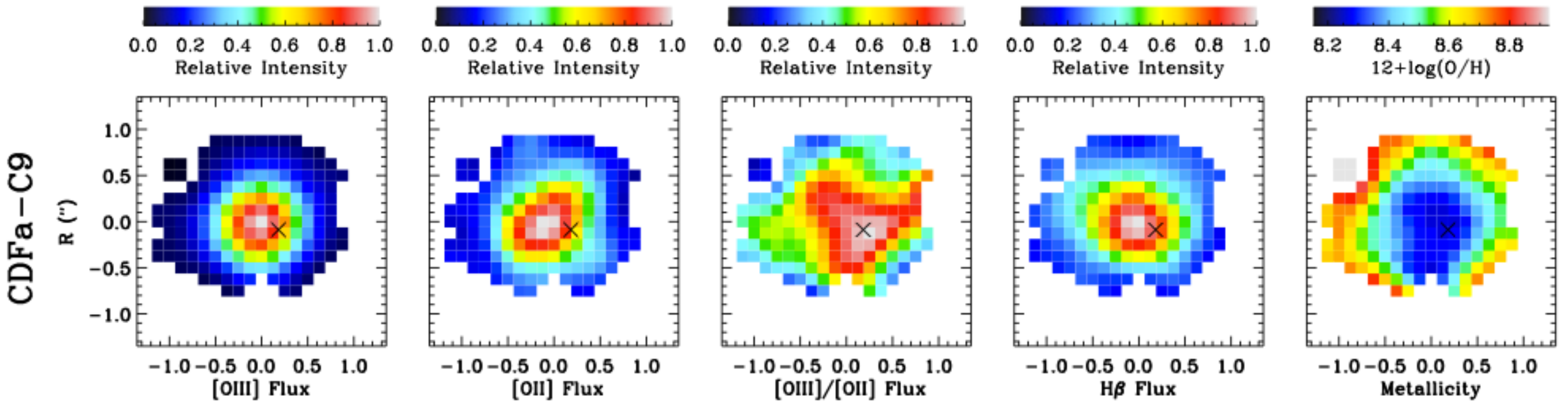}
  \includegraphics[width=0.7\linewidth]{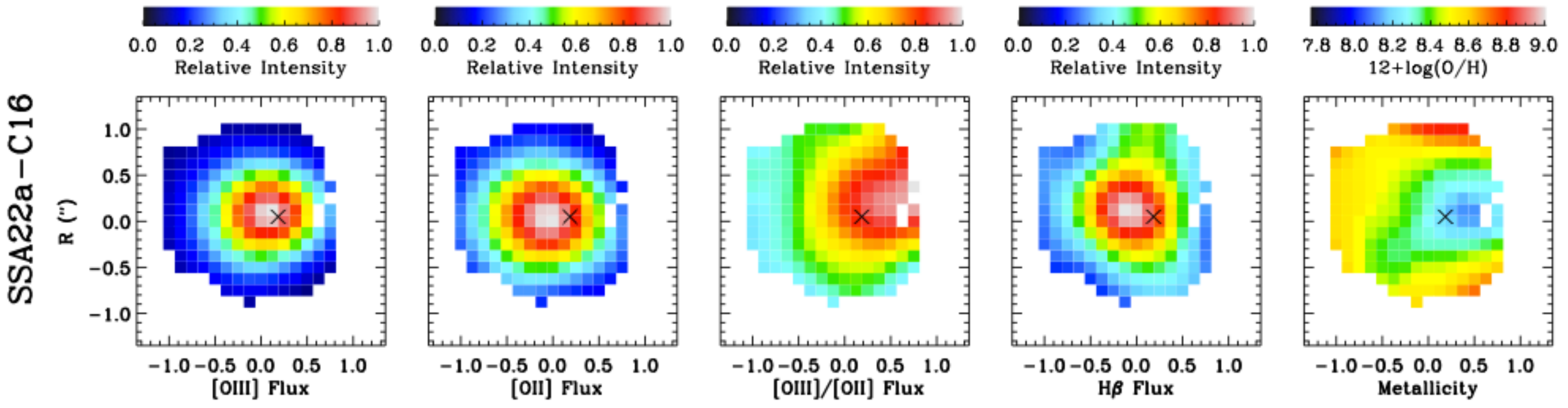}
  \includegraphics[width=0.7\linewidth]{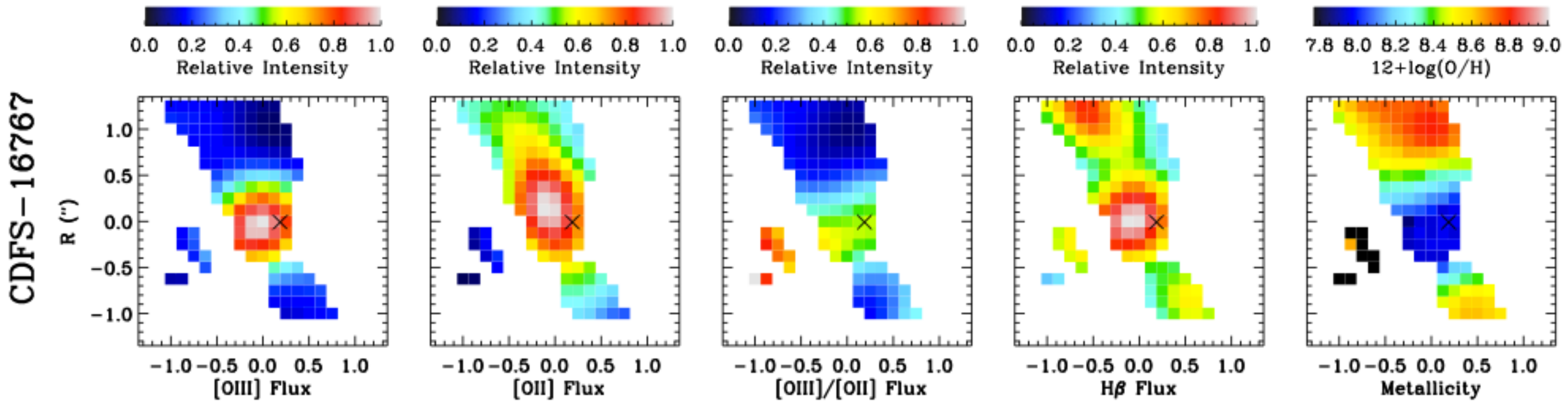}
  \includegraphics[width=0.7\linewidth]{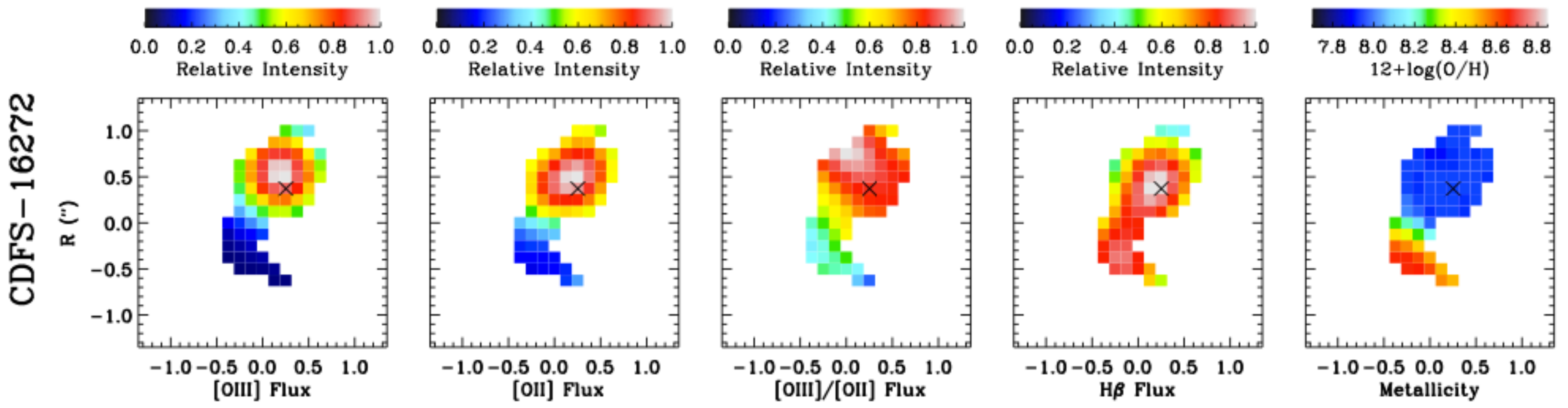}
 \caption{
 [OIII]5007, [OII]3727,  H$\beta$ flux maps, [OIII]5007/[OIII]3727 ratio,
 and metallicity maps for the galaxies
 CDFa-C9, SSA22a-C16, CDFS-11991, and CDFS-16272.
 }
 \label{fig_metmap1}
 \end{figure*}
 
\begin{figure*}
  \centering
  \includegraphics[width=0.7\linewidth]{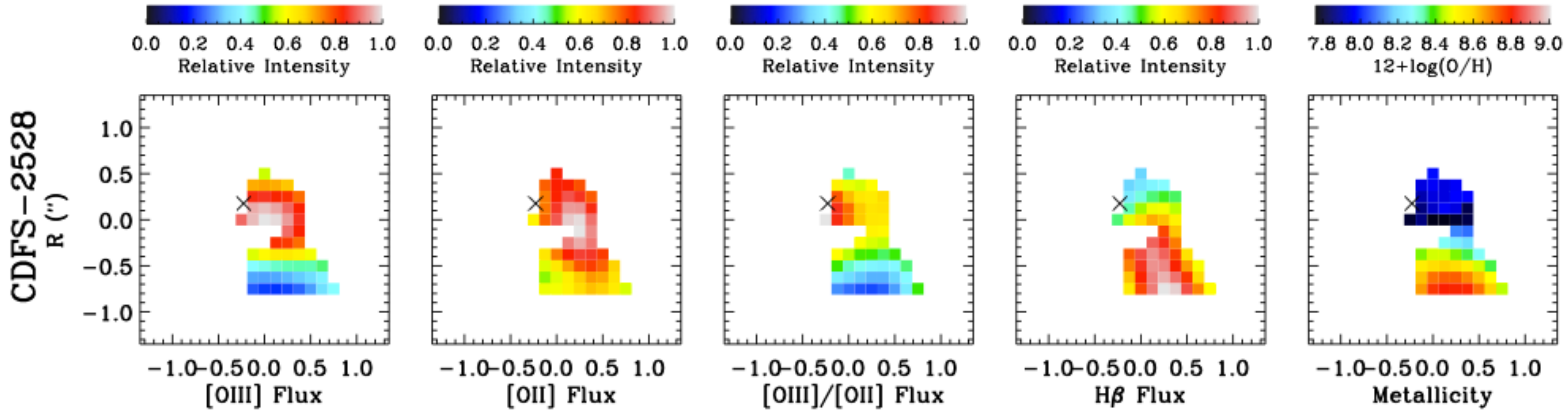}
  \includegraphics[width=0.7\linewidth]{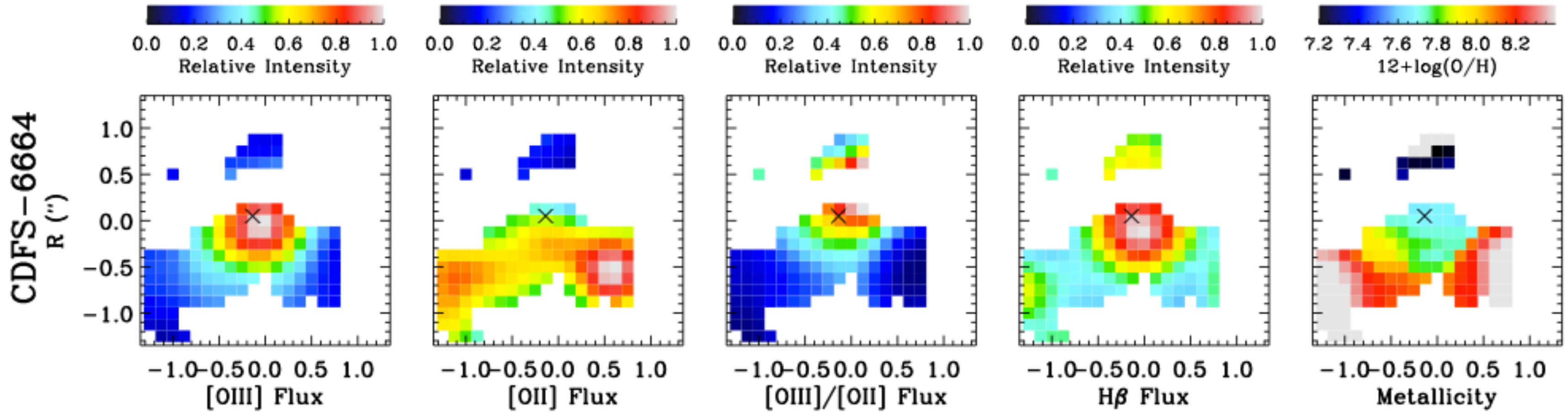} 
  \includegraphics[width=0.7\linewidth]{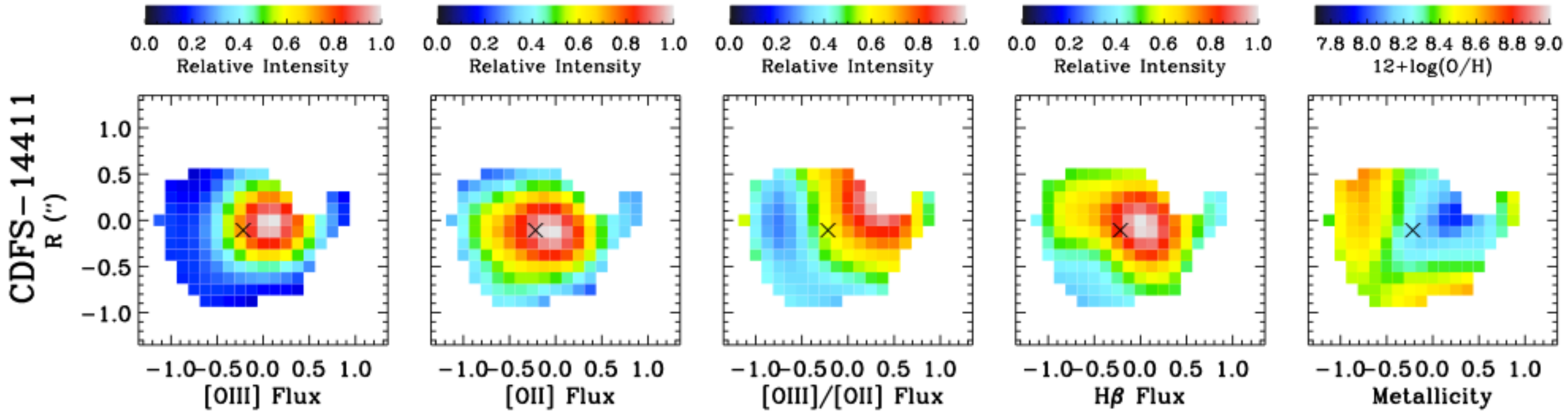}
   \includegraphics[width=0.7\linewidth]{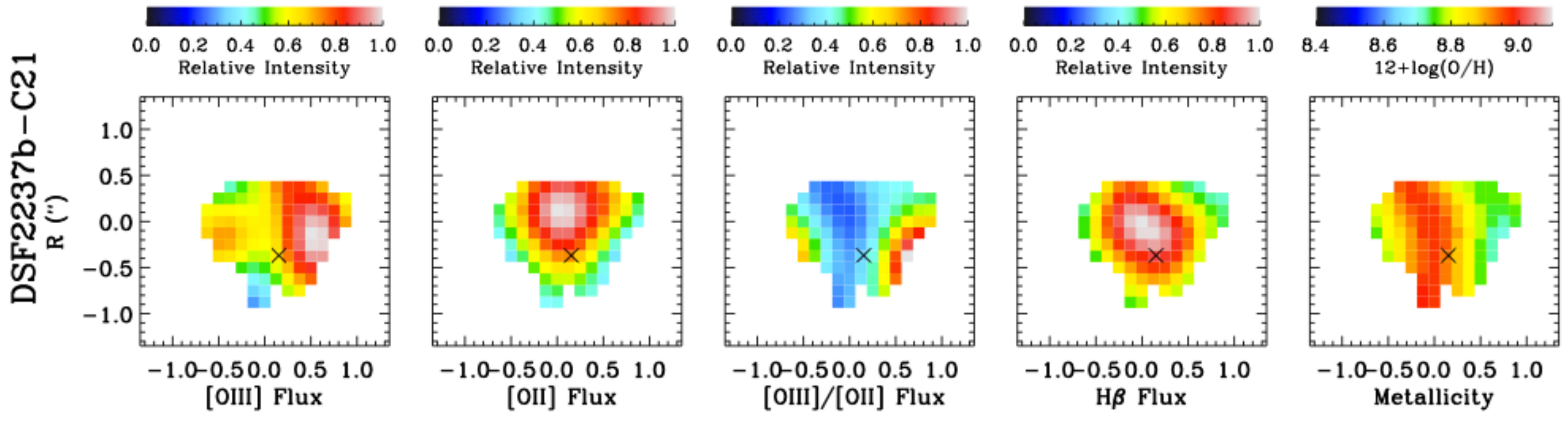}
 \caption{
 [OIII]5007, [OII]3727,  H$\beta$ flux maps, [OIII]5007/[OIII]3727 ratio,
 and metallicity maps for the galaxies
 CDFS-2528, CDFS-6664, CDFS-14411, and DSF2237b-C21.
 }
 \label{fig_metmap2}
  \end{figure*}

\begin{figure*}
 \centering
  \includegraphics[width=0.7\linewidth]{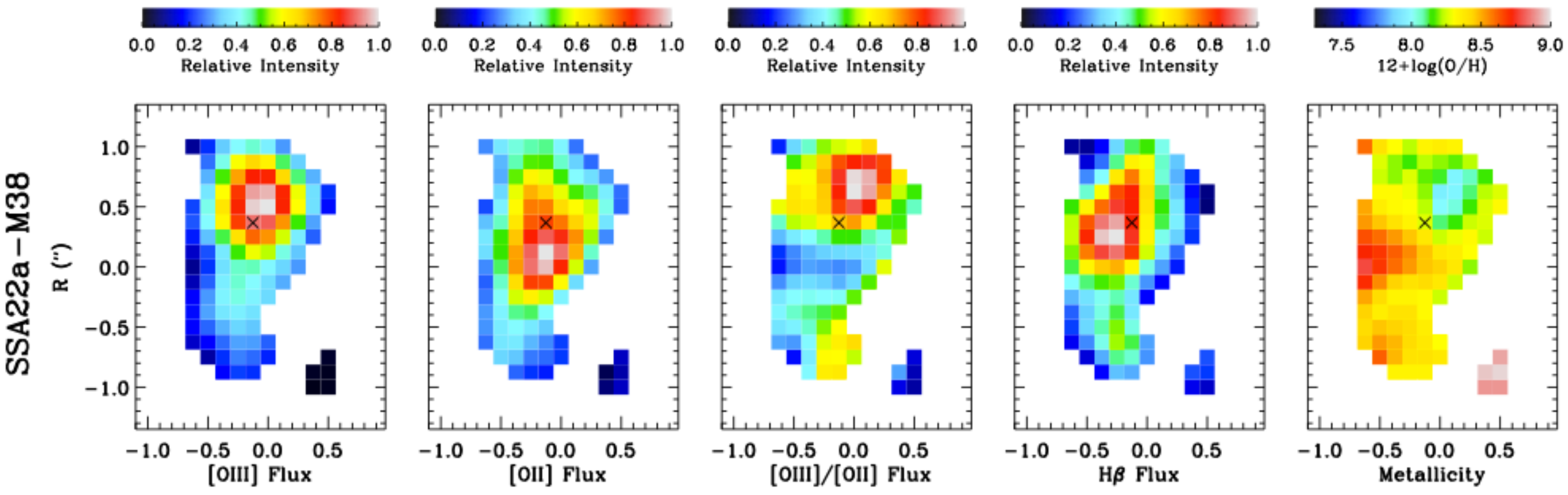}
  \includegraphics[width=0.7\linewidth]{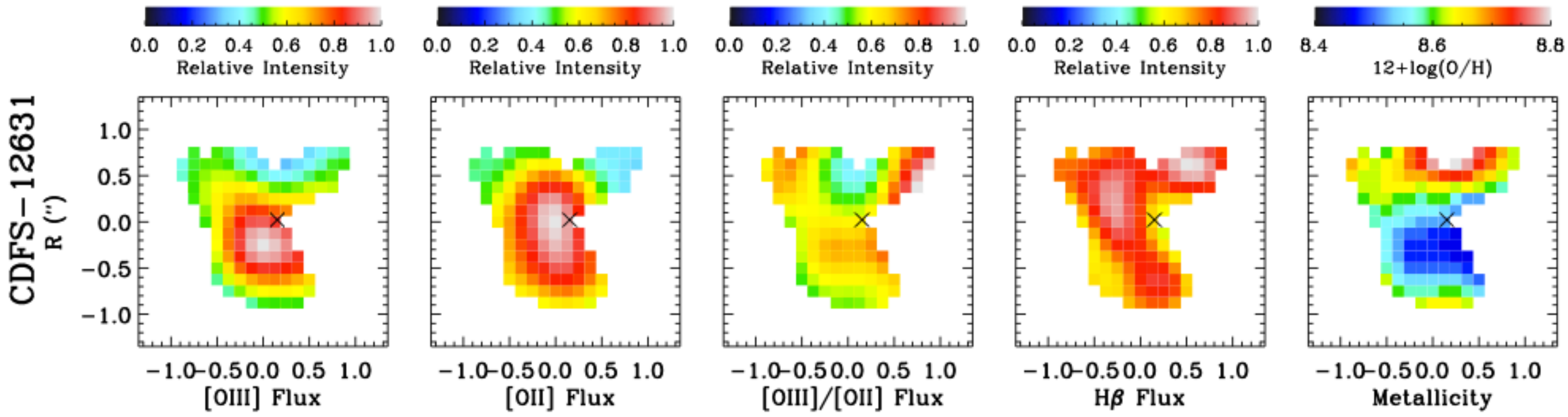}
    \includegraphics[width=0.7\linewidth]{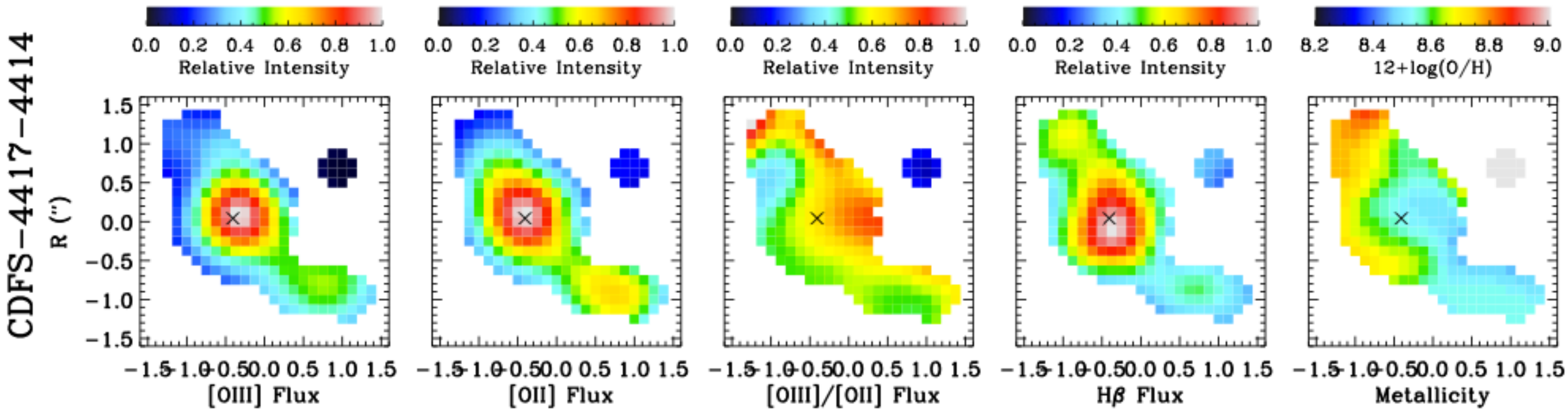}
 \caption{
 [OIII]5007, [OII]3727,  H$\beta$ flux maps, [OIII]5007/[OIII]3727 ratio,
 and metallicity maps for the galaxies
 SSA22a-M38, CDFS-12631, and the interacting galaxies CDFS-4417/CDFS-4414.
 }
  \label{fig_metmap3}
  \end{figure*}

In Figure \ref{fig_ymgasf}, the metallicity and effective yields 
are plotted as a function of the gas fraction.
Red symbols show the metallicity and
effective yields for the AMAZE and LSD galaxies.

For comparison, the bulk of the local disk galaxies (e.g., those in the SDSS sample), 
especially those with $\rm M>10^{10}~M_{\odot}$, 
have a low gas fraction ($<$0.2), metallicities around solar,
and high effective yields,
while massive galaxies at z$>$3 with similar gas fractions have much
lower metallicities and much lower effective yields.

The \cite{erb08} models are overplotted with various inflow and
outflow rates. The black dotted line is the closed-box model (i.e., $f_i=f_o=0$), 
which falls far short of reproducing the bulk behavior of the AMAZE and LSD galaxies.
The most suitable models for the bulk of the sample are those
with both massive inflow rates $f_i=1-5 \times SFR$ {\it and} massive
outflow rates $f_o=1.5-3.5 \times SFR$.

Massive cool inflows are indeed expected in the early Universe
according to recent cosmological models 
\citep[e.g.][and references therein]{dekel09,vandevoort12}, 
although direct observational evidence is still scarce.
In particular, no direct measurement of the inflow rate is available
that could be compared with the high values inferred by us in this paper. 
Outflows are ubiquitously observed, both locally and at high redshift. 
However, in high-z star-forming galaxies the typical load factor ($f_o$)
measured in observations is generally about one or a few \citep{steidel10,genzel11,newman12}. 
Our data require a load factor of at least a few,
and even $f_o \ge 3$ for the most extreme galaxies at z$>$3.

Achieving such a high outflow rate in models of star-forming galaxies is difficult.
However, a possibility is that this high outflow rate might be
associated with AGN/quasar activity.
Very high quasar-driven outflow rates have been observed both locally 
\citep{feruglio10,sturm11,rupkeveilleux11,cicone12} and at high redshift 
\citep{canodiaz12,maiolino12b}.
Although the galaxies in the AMAZE sample show no evidence for AGN activity, 
black-hole feedback may have occurred in cyclic episodes and driven enriched gas out of the galaxy.
Such massive inflows and outflows may be responsible for bringing
the early galaxies out of the equilibrium that characterizes galaxies
at lower redshifts and causes them to deviate from the FMR.

\section{Metallicity gradients at z$\sim$3}
\label{sec_metgrad}

As mentioned in the introduction, metallicity gradients provide important
information on the formation mechanism of galaxies.
Mapping the gas metallicity in high-z galaxies is extremely challenging,
because the surface brightness of the emission lines in galaxies is low and 
the emission is often unresolved in seeing-limited images.
\cite{cresci10} presented the first metallicity maps at z$>$3
by exploiting data of three bright and resolved AMAZE galaxies, out of the
first subsample observed initially \citep{maiolino08}.
Here we present resolved metallicity maps for 
a total of eleven galaxies in the AMAZE sample,
for which the S/N of each spectral-pixel and 
the spatial extension allow us to extract information
on their spatially resolved metal content.

In the same way as for the integrated metallicities, 
the metallicity maps were determined by 
means of a combination of strong line-diagnostics based on H$\beta$ and 
[OIII]$\lambda$5007 shifted into the K band, and [OII]$\lambda$3727 
shifted into the H band for sources at $3<z<3.7$, 
as mentioned at the beginning of section \S \ref{met_z3}
and as detailed in Appendix B.
The derived metallicity maps and flux maps are shown in figures 
\ref{fig_metmap1}, \ref{fig_metmap2}, and \ref{fig_metmap3}.
In the same figures we also show the [OIII]/[OII] line ratio, 
which is a proxy of the ionization parameter,
but which also correlates with the metallicity, although with a high dispersion,
which is useful to remove the degeneracy in the R$_{23}$ parameter.
Regions with an error on the metallicity larger than about 0.15-0.2~dex are
masked out (for strongly asymmetric error bars the points where
retained when the shortest error bar provided evidence of a clear metallicity 
variation with respect to other points in the galaxy).
The black cross in each map indicates the peak of the continuum flux.
As found by \cite{cresci10} for the initial subsample of three galaxies,
the peak of the star formation activity, as traced by H$\beta$ and
[OIII]$\lambda$50007 emission, is spatially correlated with the region
of the galaxy with the lowest metallicity.
A clear exception is CDFS-4414 where the metallicity tends to be flat.

  \begin{figure}[!h]
    \centering
  \includegraphics[width=1.0\linewidth]{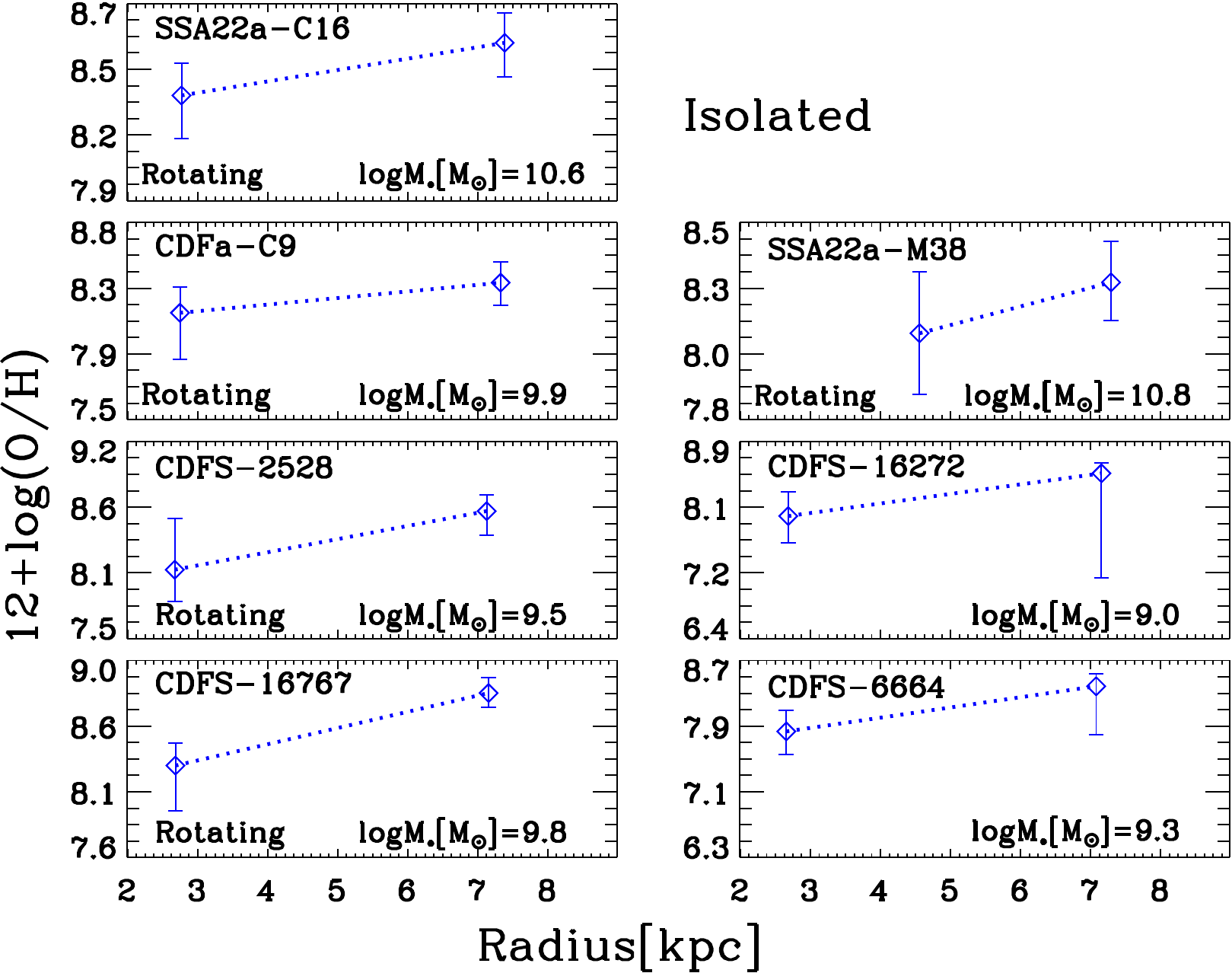} 
  \caption{Azimuthally averaged metallicity
  as a function of the galaxy radius for galaxies classified as isolated  
  by visual inspection of F160W-band (H band) HST images. 
  The galaxies are dynamically classified according to \cite{gnerucci11a}.
  Note, however, that the azimuthal averages tend to dilute the nonaxisymmetric metallicity
  variations that characterize these objects.
  }
  \label{figgrad_iso}
  \end{figure}

  \begin{figure}[!h]
  \includegraphics[width=1.0\linewidth]{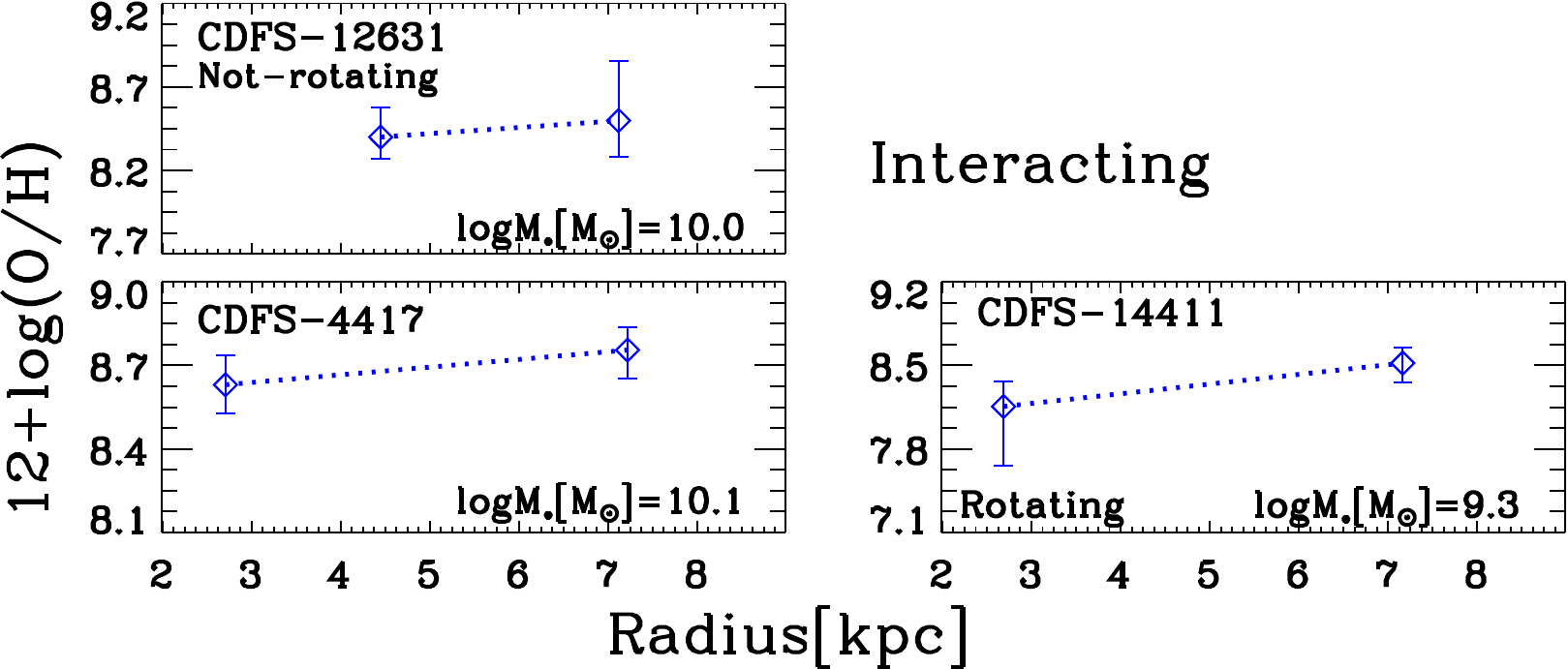} 
  \caption{Azimuthally averaged metallicity
  as a function of the galaxy radius for galaxies classified as interacting 
  by visual inspection of F160W-band (H band) HST images.
  The galaxies are dynamically classified according to \cite{gnerucci11a}.
  Note, however, that the azimuthal averages tend to dilute the nonaxisymmetric metallicity
  variations that characterize these objects.
  }
    \label{figgrad_int}
  \end{figure}

To increase the S/N on the outer region we also obtained
the radial metallicity variation by integrating the line fluxes within
a central aperture (a few kpc in radius) and in an outer ring (with an
outer radius typically extending to about 8-10 kpc).
The resulting metallicity radial profiles are shown
in Figs.~\ref{figgrad_iso} and \ref{figgrad_int}.
For DFS2237b-C21 it was not possible to determine the metallicity
of the outer region because of the low S/N even in the spectrum integrated in
the annulus.
The resulting radial metallicity gradients are reported in Table \ref{table_gradients}.
Most of the gradients are flat or inverted (positive).

For the flat gradients one should keep in mind that as pointed out by \cite{yuan13},
these shallow gradients may be partially due to the low angular resolution.
However, we also note that the two-dimensional metallicity maps show complex, nonradially
symmetric patterns, which were averaged out when we obtained averages in radial annuli.
We found no significant dependence of the metallicity gradients
as a function of stellar mass, SFR, and sSFR.

We also investigated the metallicity gradients as a function
of morphology and the dynamical status.
Fig.~\ref{figgrad_iso} gathers all galaxies that according to the HST images are isolated,
while Fig.~\ref{figgrad_int} include all galaxies that are
in close interaction (a companion within 30 kpc) or with strongly irregular
morphology (in the H band), presumably associated with late mergers.
There is no relation between the morphological properties (isolated versus
interacting) and the trend of the metallicity gradient.
The dynamical status (rotating disk versus nonrotating),
as inferred by \cite{gnerucci11a}, is also reported in each panel, when available. 
It is interesting to note that most of the systems with inverted gradients
are associated with rotating disks, as was suggested by 
\cite{cresci10} for the first subsample of three galaxies.
At least at z$>$3, these results do not fully support
the scenario in which inverted gradients are
preferentially associated with galaxy interactions/mergers that
according to some models \citep[e.g.][]{rupke10,perez11,pilkington12,torrey12}
are supposed to drive pristine gas from the outskirts to the inner region.
Instead, smooth inflows of pristine gas toward the central region,
both diluting the gas metallicity and enhancing star formation,
may be responsible for the gradients observed at z$>$3.
The chemical evolutionary model described by \cite{mott13} 
do indeed fit our oxygen abundance gradients at z$\sim$3 and also reproduce the oxygen and iron gradients of the Milky Way
when an inside-out formation of the disk, 
a constant star formation efficiency, and time-dependent velocity radial flows are assumed.

Nevertheless, it is clear from figures \ref{fig_metmap1} to \ref{fig_metmap3}
that the metallicity gradients are generally not radially symmetric,
hence the radial averages shown obtained
in Figs.~\ref{figgrad_iso}-\ref{figgrad_int} and in Table \ref{table_gradients}
can be deceiving, and moreover,
tend to dilute more pronounced gradients observed in the metallicity maps of galaxies.

\begin{figure}[!h]
  \includegraphics[width=1.\linewidth]{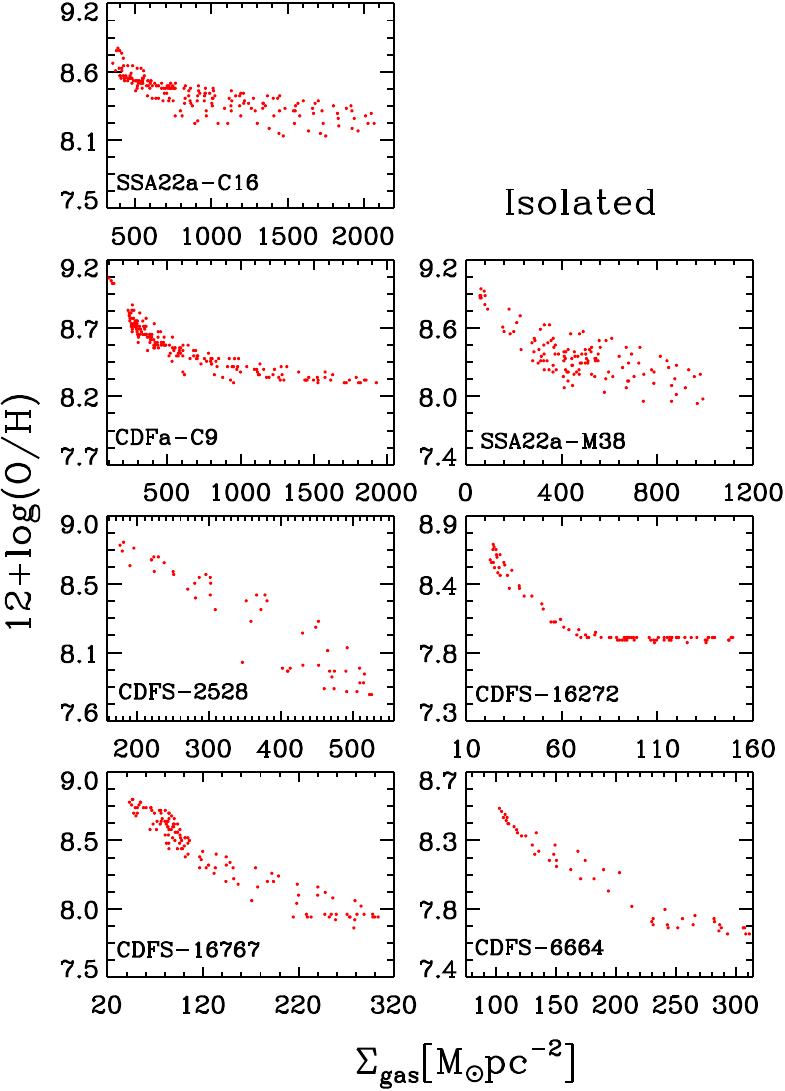} 
  \caption{Spatially resolved metallicities as a function of the gas surface density 
  of isolated galaxies.}
  \label{fig_anticor_sfrmet_iso}
  \end{figure}

\begin{figure}[!h]
  \includegraphics[width=1.\linewidth]{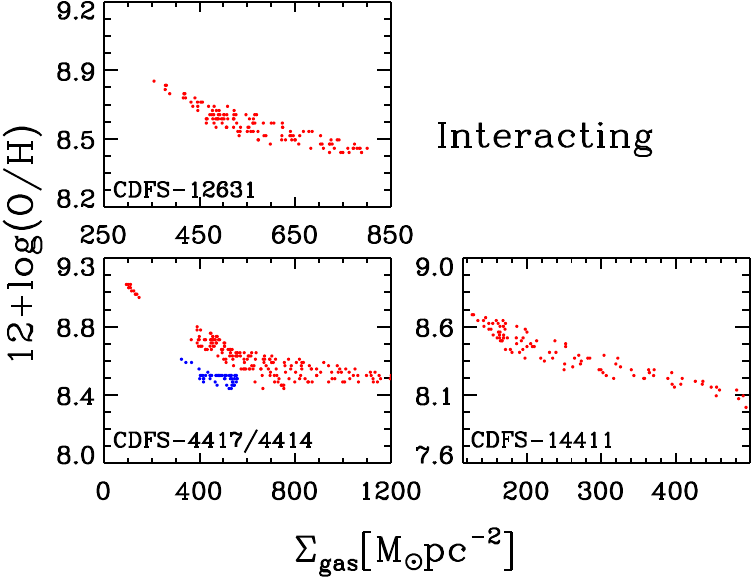} 
  \caption{Spatially resolved metallicities as a function of the gas surface density 
  of interacting galaxies.
  Blue dots show the distribution of the galaxy CDFS-4414 that interacts with CDFS-4417.}
  \label{fig_anticor_sfrmet_int}
  \end{figure}

To avoid the radial-average approach,
and to investigate in more detail the scenario in which lower
metallicity regions are associated with star-forming regions by an
excess of (pristine) gas that both dilutes metallicity and boosts star formation,
we compared the spatially resolved metallicity with the
surface gas density (obtained by inverting the Schmidt-Kennicutt law).
This comparison is shown for each galaxy in Figs. \ref{fig_anticor_sfrmet_iso} 
and \ref{fig_anticor_sfrmet_int},
where each point shows the metallicity and $\Sigma _{gas}$ at each pixel.
We verified that the correlated errors on the two axes 
(because the H$\beta$ flux was used on both axes) 
does not introduce a significant artificial correlation.
Moreover, not all of the plotted points are fully independent 
of each other because of the seeing.
However, in most galaxies there is a clear anticorrelation
between metallicity and $\Sigma _{gas}$,
which additionally supports the inflow/dilution scenario.

  \begin{figure}[!h]
  \centering
  \includegraphics[width=0.9\linewidth]{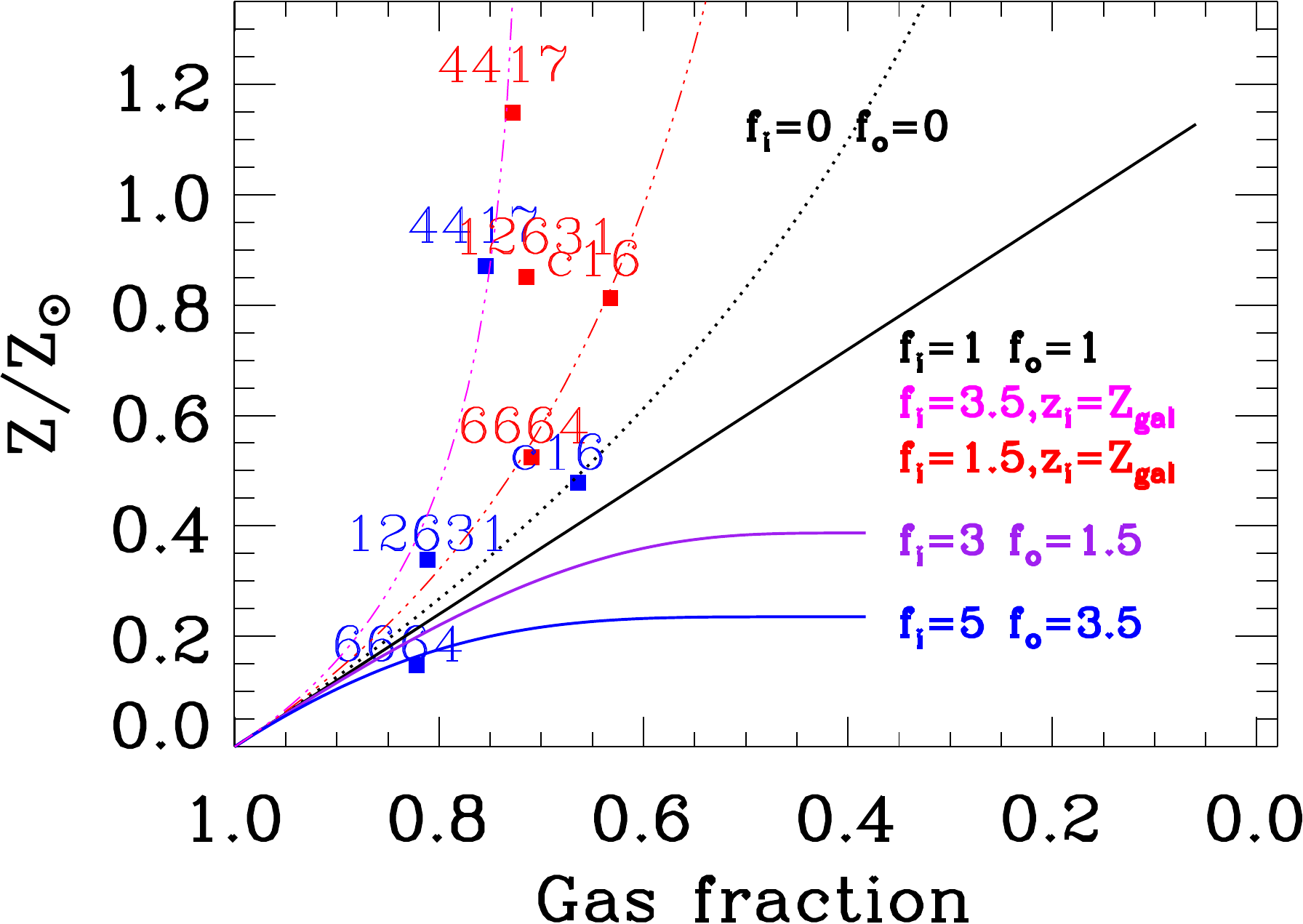}
  \caption{Metallicity as a function of the gas fraction for the inner/outer galaxy regions. 
  Blue squares show the inner regions, while red squares show the outer regions. 
  Blue and purple solid lines show models with outflows and inflows of pristine gas.
  The black dotted line shows the closed-box model. 
  Red and magenta dashed-dotted lines show models with enriched inflows 
  (with the same metallicity as the galaxy) and no outflows.
  }
  \label{gasf_inner_outer}
  \end{figure}

However, $\Sigma _{gas}$ is not necessarily illustrative of the evolutionary
stage of individual regions of each galaxy. The gas fraction is more
useful to constrain the local evolutionary processes. 
But it is difficult to obtain resolved information on the gas fraction,
since it is not easy to determine the stellar mass surface density.
Yet, for a few galaxies for which we resolved the distribution of the nebular
lines (for which we hence have spatially resolved information on $\Sigma _{gas}$
by inverting the S-K relation) we also have deep HST-WFC3 images in the H band,
sampling the V-band rest frame. To a first approximation, we can assume that
the $\Sigma _{star}$ scales proportionally to the rest-frame V-band surface brightness
(the normalization factor is given by the constraint that the total V-band light should
be associated with the total stellar mass inferred by the detailed multiband
SED modeling). We could therefore estimate the gas fraction
in the same inner and outer regions for which the metallicity could be inferred.

Fig.~\ref{gasf_inner_outer} shows the metallicity as a function
of the gas fraction for the four galaxies for which we can resolve both
the metallicity and the gas fraction. Blue symbols indicate
the inner regions, while red symbols indicate the outer regions.
In all these galaxies the inner region has a lower metallicity
and higher gas fraction than the outer region.
This finding is in line with the result of \cite{cresci10}, 
suggesting that the inner regions suffer from metallicity dilution from inflow of
pristine gas, 
which centrally both increases the gas fraction and dilutes the metallicity.

However, we note that several points, especially those in the outer regions,
have higher metallicities than those of the closed-box model. 
This may indicate that our metallicity calibration is biased high
or, alternatively, that in these regions the stellar yields are higher than the average
of a normal stellar population (e.g., through recent injection of metals
by massive stars, which have higher yields than average).
An additional,
alternative explanation for the high metallicity in these regions, especially
in the outer ones, is inflow of metal-enriched gas, which can give a higher metal
content than expected from the closed-box model.
Indeed, in several recent models of galactic fountains,
a significant fraction of the the metals ejected from
the central winds falls back onto the outer parts of the galaxy.

We explored the latter scenario by using the model of \cite{erb08}.
We constructed additional models in which the gas
inflow is not pristine, but has a metallicity equal to the metallicity
of the bulk of the galaxy.
These models are shown with red and magenta lines in
Fig.~\ref{gasf_inner_outer} and nicely reproduce the observed values
in the outer regions of several of these galaxies. 
Therefore, enhanced metallicity transport from the inner to the outer regions
through galactic fountains may be an additional mechanism responsible for 
the inverted gradients in these young systems.

\section{Conclusions}

We have reported the results of a program aimed at measuring
optical nebular emission lines in a sample of 35 star-forming galaxies at $z\sim3.4$ by using SINFONI,
the near-infrared integral field spectrograph at the VLT.
The integrated and spatially resolved spectral information allowed us to constrain 
the metallicity and, {\bf indirectly}, the amount of gas hosted in these galaxies.
The main results are summarized as follows:

\begin{itemize}

\item On average the galaxies in our sample at z$\sim$3.4 are metal poorer,
by $-0.43^{+0.16}_{-0.18}~dex$, than the fundamental relation between metallicity,
stellar mass, and SFR (FMR) that characterizes local and lower-redshift galaxies (0$<$z$<$2.5).
Galaxies at z$\sim$3.4 have a metallicity dispersion of about 0.25~dex in the FMR,
much higher than local galaxies (which have $\sigma \sim $0.05),
which probably reflects a mixture of different stages of unsteady chemical 
evolutionary processes at this epoch, in contrast to what is observed at later epochs.

\item There is no significant correlation between the dynamical state of these galaxies and 
the deviations from the FMR. 
More specifically, galaxies characterized by regular rotation patterns have, on average, 
the same metallicity deviations from the FMR as galaxies with irregular kinematics 
(indicative of recent/ongoing merging).
These results suggest that the enhanced merging rate at z$>$3 is unlikely to be the
main reason for the deviations from the FMR at z$>$3.

\item We found that the deviations from the FMR at z$\sim$3.4 correlate with
$\rm \mu _{32} = \log{(M_*)}- 0.32\, \log{(SFR)}$, which is the parameter giving the
projection of the FMR that minimizes the dispersion, and which is associated with
the star formation efficiency as well as with the presence of outflows and inflows.

\item By mapping the optical nebular emission lines and by inverting the 
Schmidt-Kennicutt relation, we inferred the amount of molecular gas hosted in these galaxies.
At z$\sim$3, the average gas fraction of massive galaxies ($M_* > 10^{10} M_{\sun}$) 
does not follow the steep increasing evolution observed from z=0 to z$\sim$2. 
Between z$\sim$2 and z$\sim$3 the average gas fraction in galaxies
remains constant or possibly even decreases.
Our results support the scenario in which the evolution of cosmic star formation in
galaxies is primarily driven by the evolution of the amount of gas in galaxies,
and not by an evolution in the efficiency of star formation.

\item The observed anticorrelation between gas fraction and stellar mass, 
as well as the correlation between gas fraction and $sSFR$,
may support the scenario in which downsizing is in place at z$\sim$3. 
However, new observations are required to verify this trend and validate 
that it is not a consequence of selections effects.

\item Models with both high inflows and outflows rates ($\rm
\sim 2-5 \times SFR$)  are necessary to reproduce
the measured galaxy properties, and in particular the metallicities,
gas fraction, and effective yields.
These massive flows in the early Universe are most likely responsible for the 
different properties and deviations of galaxies
at z$\sim$3 compared with local and lower redshift galaxies.

\item By mapping the distribution of the star formation and metallicity 
in 10 out of 34 galaxies at z$\sim$3, we found an spatial anticorrelation 
between the peak of the SFR and the lowest metallicity region.
We furthermore found within each galaxy an anticorrelation of metallicity and
gas surface density.
This result supports the models in which smooth gas inflows feed galaxies at high redshift.
In this scenario the pristine infall both boosts star formation (through the
Schmidt-Kennicutt law) and dilutes the metallicity, generating the observed anticorrelation. 

\item For four AMAZE galaxies, it was possible to determine the metallicity and 
gas fraction in the inner (r<3~kpc) and in the out outer (3<r<8~kpc) galaxy regions. 
In all these galaxies, the inner region has a lower metallicity and 
higher gas fraction than the outer region. 
This finding suggests that the inner regions suffer from metallicity dilution
from inflows of pristine gas, 
which centrally increases the gas fraction and dilutes the metallicity.
However, nuclear enriched outflows probably contribute to lower 
the metallicity in the central region.
Additional modeling supported the galactic-fountain scenario in which
outflows of enriched material are expelled from the inner region
and fall back, as inflows of enriched material, onto the galaxy outskirts. 

\end{itemize}

\begin{acknowledgements}
This work was funded by the Marie Curie Initial Training Network ELIXIR 214227 of the European Commission. 
We also acknowledge partial support by INAF. 
Alessandro Marconi acknowledges support from grant PRIN-MIUR 2010-2011 
``The dark Universe and the cosmic evolution of baryons: from current surveys
to Euclid''.
\end{acknowledgements}
  
\begin{table*}
  \caption[!h]{Galaxy sample.  } 
\label{table_sample}
  {\centering
  \begin{tabular}{l c c c c c c}
    \hline
    \hline
    \noalign{\smallskip}
     Object & sample & R.A. & Dec. & z & Texp(min) & R$_{AB}$ \\
    \hline
    \hline
    \noalign{\smallskip}
    CDFa-C9 	      & AMAZE & 00:53:13.7  & +12:32:11.1 &  $3.2119$ & 250 &23.99   \\
    CDFS-4414	      & AMAZE & 03:32:23.2  & -27:51:57.9 &  $3.4714$ & 350 &24.18   \\
    CDFS-4417$^{(1)}$ & AMAZE & 03:32:23.3  & -27:51:56.8 &  $3.4733$ & 350 &23.42   \\
    CDFS-6664 	      & AMAZE & 03:32:33.3  & -27:50:7.4  &  $3.7967$ & 500 &24.80  \\
    CDFS-16767	      & AMAZE & 03:32:35.9  & -27:41:49.9 &  $3.6239$ & 300 &24.13  \\
    CDFS-13497        & AMAZE & 03:32:36.3  & -27:44:34.6 &  $3.4138$ & 150 &24.21  \\
    CDFS-11991	      & AMAZE & 03:32:42.4  & -27:45:51.6 &  $3.6110$ & 450 &24.23  \\
    CDFS-2528 	      & AMAZE & 03:32:45.5  & -27:53:33.3 &  $3.6872$ & 350 &24.64   \\
    CDFS-16272        & AMAZE & 03:32:17.1  & -27:42:17.8 &  $3.6195$ & 350 &25.08  \\
    CDFS-9313         & AMAZE & 03:32:17.2  & -27:47:54.4 &  $3.6545$ & 250 &24.82 \\
    CDFS-9340$^{(1)}$ & AMAZE & 03:32:17.2  & -27:47:53.4 &  $3.6578$ & 250 &25.85  \\
    CDFS-12631        & AMAZE & 03:32:18.1  & -27:45:19.0 &  $3.7090$ & 250 &24.72 \\
    CDFS-14411 	      & AMAZE & 03:32:20.9  & -27:43:46.3 &  $3.5989$ & 200 &24.57  \\
    CDFS-5161         & AMAZE & 03:32:22.6  & -27:51:18.0 &  $3.6610$ & 300 &24.96  \\
    LnA1689-2$^{(2)}$ & AMAZE & 13:11:25.5  & -01:20:51.9 &  $4.8740$ & 400 &23.31  \\
    LnA1689-4$^{(2)}$ & AMAZE & 13:11:26.5  & -01:19:56.8 &  $3.0428$ & 240 &22.40  \\
    LnA1689-1$^{(2)}$ & AMAZE & 13:11:30.0  & -01:19:15.3 &  $3.7760$ & 300 &24.20  \\
    LnA1689-3$^{(2,4)}$ & AMAZE & 13:11:35.0  & -01:19:51.6 &  $5.120$ & 225 &25.0   \\
    Q1422-D88  	     & AMAZE & 14:24:37.9  & +23:00:22.3 & $3.7520$ & 250 &24.44  \\
    3C324-C3    	     & AMAZE & 15:49:47.1  & +21:27:05.0 & $3.2890$ & 150 &24.14  \\
    Cosmic Eye$^{(2,3)}$ & AMAZE & 21:35:12.7  & -01:01:42.9 & $3.0755$ & 200 &20.54  \\
    SSA22a-M38 	     & AMAZE & 22:17:17.7  & +00:19:00.7 & $3.2928$ & 400 &24.11  \\
    SSA22a-C48$^{(4)}$ & AMAZE & 22:17:18.6  & +00:18:16.2 & $3.079$ & 250 &24.71  \\
    SSA22a-D17$^{(1)}$ & AMAZE & 22:17:18.9  & +00:18:16.8 & $3.0870$ & 250 &24.27 \\
    SSA22a-aug96M16   & AMAZE & 22:17:30.9  & +00:13:10.7 &  $3.2920$ & 250 &23.83  \\
    SSA22a-G03$^{(4)}$ & AMAZE & 22:17:30.8  & +00:12:51.0 & $4.527$ & 225 &25.03  \\
    SSA22a-C16        & AMAZE & 22:17:32.0  & +00:13:16.1 &  $3.0675$ & 350 &23.64 \\
    SSA22a-C36 	        & AMAZE & 22:17:46.1  & +00:16:43.0 &  $3.0630$ & 100 & 24.06  \\
    DFS2237b-C21	& AMAZE & 22:39:29.0  & +11:50:58.0 &  $3.4029$ & 200 & 23.50  \\
    DFS2237b-D29 	& AMAZE & 22:39:32.7  & +11:55:51.7 &  $3.3709$ & 250 & 23.70  \\
    Q0302-C131  	& LSD   & 03:04:35.0  & -00:11:18.3 &  $3.2350$ & 240 & 24.5  \\
    Q0302-C171     	& LSD   & 03:04:44.3  & -00:08:23.2 &  $3.3280$ & 240 & 24.6  \\
    Q0302-M80  	    & LSD   & 03:04:45.7  & -00:13:40.6 &  $3.4160$ & 240 & 24.1  \\
    Q0302-MD287$^{(4)}$ & LSD & 03:04:52.8  & -00:09:54.6 &  $2.395$  & 160 & 24.8  \\
    SSA22a-C30 	    & LSD   & 22:17:19.3  & +00:15:44.7 &  $3.1025$ & 240 & 24.2  \\
    SSA22a-C6  	    & LSD   & 22:17:40.9  & +00:11:26.0 &  $3.0970$ & 280 & 23.4  \\
    SSA22a-M4$^{(1)}$ & LSD   & 22:17:40.9  & +00:11:27.9 &  $3.0972$ & 280 & 24.8  \\
    SSA22b-C5 	    & LSD   & 22:17:47.1  & +00:04:25.7 &  $3.1120$ & 240 & 22.0  \\
    DSF2237b-D28 	    & LSD   & 22:39:20.2  & +11:55:11.3 &  $2.9323$ & 240 & 24.5  \\
    DSF2237b-MD19$^{(4)}$& LSD & 22:39:21.1 & +11:48:27.7 &  $2.616$ & 200 & 24.5  \\
    \hline    \hline
  \end{tabular} 
  } \\
 Col. 1, object name; Col.2, sample name, Cols. 3, 4, coordinates (J2000); 
 Col. 5, redshift spectroscopically determined through the [OIII]5007 line in our spectra; 
 Col. 6, on-source integration time (in unit of minutes);
 Col. 7, R band AB magnitude.
 Notes:$^{(1)}$ The object is in the same field of view as the object on the previous line.
$^{(2)}$ Lensed objects. $^{(3)}$ Data were taken from the archive.
$^{(4)}$ Undetected sources. Redshift was taken from the literature.
\end{table*}

\begin{table*}[!ht]
\caption{Line fluxes inferred from the near-IR spectra.}
\label{table_flux}
{\centering
\begin{tabular}{lcccc}
\hline\hline                 
Name & F([OIII]5007) & F(H$\beta$) & F([OII]3727) & F([NeIII]3870)  \\
     & \multicolumn{4}{c}{$\rm 10^{-17}~erg~s^{-1}~cm^{-2}$}  \\
\hline
    CDFa-C9 &7.23$\rm {^+_-0.09} $ &1.49$\rm {^+_-0.09} $ &2.36$\rm {^+_-0.11} $ &0.60$\rm {^+_-0.09} $   \\
  CDFS-4414 &1.06$\rm {^+_-0.08} $ &0.42$\rm {^+_-0.06} $ &0.89$\rm {^+_-0.08} $ &0.12$\rm {^+_-0.04} $   \\
  CDFS-4417 &1.77$\rm {^+_-0.08} $ &0.88$\rm {^+_-0.09} $ &1.39$\rm {^+_-0.09} $ &0.31$\rm {^+_-0.07} $   \\
  CDFS-6664 &3.38$\rm {^+_-0.15} $ &0.34$\rm {^+_-0.09} $ &0.29$\rm {^+_-0.04} $ &                   --   \\
 CDFS-16767 &1.99$\rm {^+_-0.08} $ &0.46$\rm {^+_-0.09} $ &0.60$\rm {^+_-0.05} $ &                   --   \\
 CDFS-11991 &3.02$\rm {^+_-0.10} $ &0.28$\rm {^+_-0.09} $ &0.52$\rm {^+_-0.05} $ &0.22$\rm {^+_-0.07} $   \\
  CDFS-2528 &1.61$\rm {^+_-0.11} $ &0.35$\rm {^+_-0.10} $ &0.61$\rm {^+_-0.07} $ &                   --   \\
 CDFS-16272 &3.62$\rm {^+_-0.09} $ &0.48$\rm {^+_-0.13} $ &0.66$\rm {^+_-0.05} $ &0.23$\rm {^+_-0.09} $   \\
  CDFS-9313 &4.45$\rm {^+_-0.13} $ &0.54$\rm {^+_-0.12} $ &0.72$\rm {^+_-0.06} $ &                   --   \\
  CDFS-9340 &1.64$\rm {^+_-0.10} $ &0.37$\rm {^+_-0.09} $ &0.40$\rm {^+_-0.06} $ &                   --   \\
 CDFS-12631 &2.98$\rm {^+_-0.25} $ &0.49$\rm {^+_-0.12} $ &2.09$\rm {^+_-0.19} $ &1.13$\rm {^+_-0.13} $   \\
 CDFS-14411 &3.88$\rm {^+_-0.13} $ &0.75$\rm {^+_-0.11} $ &1.02$\rm {^+_-0.07} $ &0.20$\rm {^+_-0.06} $   \\
  CDFS-5161 &1.23$\rm {^+_-0.17} $ &0.17$\rm {^+_-0.09} $ &0.17$\rm {^+_-0.04} $ &0.11$\rm {^+_-0.05} $   \\
 LnA1689-2$^{a}$ &                    - &                    - &0.16$\rm {^+_-0.03} $ &0.05$\rm {^+_-0.03} $   \\
 LnA1689-4$^{a}$ &0.45$\rm {^+_-0.01} $ &0.05$\rm {^+_-0.00} $ &0.06$\rm {^+_-0.01} $ &0.02$\rm {^+_-0.00} $   \\
 LnA1689-1$^{a}$ &0.15$\rm {^+_-0.02} $ &0.04$\rm {^+_-0.01} $ &0.06$\rm {^+_-0.01} $ &                   --   \\
  Q1422-D88 &4.31$\rm {^+_-0.21} $ &0.55$\rm {^+_-0.17} $ &0.55$\rm {^+_-0.08} $ &                   --   \\
   3C324-C3 &1.15$\rm {^+_-0.08} $ &0.26$\rm {^+_-0.06} $ &0.46$\rm {^+_-0.12} $ &0.31$\rm {^+_-0.11} $   \\
 Cosmic Eye$^{a}$ &0.78$\rm {^+_-0.02} $ &0.09$\rm {^+_-0.03} $ &0.24$\rm {^+_-0.03} $ &                   --   \\
 SSA22a-M38 &5.56$\rm {^+_-0.14} $ &1.40$\rm {^+_-0.18} $ &3.12$\rm {^+_-0.18} $ &1.03$\rm {^+_-0.16} $   \\
 SSA22a-D17 &1.73$\rm {^+_-0.11} $ &0.36$\rm {^+_-0.09} $ &1.04$\rm {^+_-0.10} $ &                   --   \\
 SSA22a-aug96M16 &1.82$\rm {^+_-0.08} $ &0.38$\rm {^+_-0.06} $ &0.47$\rm {^+_-0.08} $ &                   --   \\
 SSA22a-C16 &5.45$\rm {^+_-0.09} $ &1.47$\rm {^+_-0.12} $ &2.59$\rm {^+_-0.09} $ &                   --   \\
 SSA22a-C36 &1.81$\rm {^+_-0.24} $ &0.48$\rm {^+_-0.21} $ &2.12$\rm {^+_-0.32} $ &0.34$\rm {^+_-0.17} $   \\
 DSF2237b-C21 &0.70$\rm {^+_-0.12} $ &1.00$\rm {^+_-0.17} $ &1.85$\rm {^+_-0.14} $ &0.35$\rm {^+_-0.13} $   \\
 DSF2237b-D29 &0.93$\rm {^+_-0.09} $ &0.14$\rm {^+_-0.07} $ &0.76$\rm {^+_-0.09} $ &                   --   \\
 Q0302-C131 &2.62$\rm {^+_-0.52} $ &0.46$\rm {^+_-0.16} $ &0.80$\rm {^+_-0.32} $ &                   --   \\
 Q0302-C171 &1.10$\rm {^+_-0.12} $ &0.18$\rm {^+_-0.08} $ &0.56$\rm {^+_-0.24} $ &                   --   \\
  Q0302-M80 &1.32$\rm {^+_-0.26} $ &0.40$\rm {^+_-0.12} $ &0.56$\rm {^+_-0.18} $ &                   --   \\
 SSA22a-C30 &1.28$\rm {^+_-0.22} $ &0.26$\rm {^+_-0.10} $ &0.30$\rm {^+_-0.12} $ &                   --   \\
  SSA22a-C6 &5.50$\rm {^+_-0.62} $ &0.76$\rm {^+_-0.45} $ &1.21$\rm {^+_-0.38} $ &                   --   \\
  SSA22a-M4 &3.66$\rm {^+_-0.49} $ &0.69$\rm {^+_-0.25} $ &1.46$\rm {^+_-0.23} $ &                   --   \\
  SSA22b-C5 &3.28$\rm {^+_-0.24} $ &0.50$\rm {^+_-0.14} $ &0.27$\rm {^+_-0.09} $ &0.38$\rm {^+_-0.14} $   \\
 DSF2237b-D28 &1.86$\rm {^+_-0.36} $ &0.16$\rm {^+_-0.08} $ &0.74$\rm {^+_-0.16} $ &                   --   \\
{\it Composite}$^b$ & 6.13$\pm$0.14 & 1.00$\pm$0.10 & 1.70$\pm$0.12 & 0.39$\pm$0.08 \\
\hline\hline                
\end{tabular} 
} \\
Col. 1, object name;
Cols. 2, 3, 4, and 5, emission line fluxes.
Notes:$^{a}$ values corrected for a magnification factor of $ 5.9 \pm 1.9, 43.5\pm 4.1, 6.7
\pm 2.1 $ (see text), and $28 \pm 3$ \citep{smail07}, respectively.
$^b$ In the case of the composite spectrum fluxes were normalized to the H$\beta$ flux 
(which is also subject to an error, as listed in the corresponding column).
\end{table*}

\begin{table*}[!ht]
\caption{Physical properties of the sample inferred from their SED and metallicity.}
\label{tab_sed}
{\centering
\begin{tabular}{lccccc}
\hline\hline
\noalign{\smallskip}
Name & $\rm log M_*$ (BC03) & $\rm log M_*$ (M05)  &E(B-V)$_*$ 	& log SFR (SED)  		& log SFR ($H\beta$)			\\
     & $\rm [M_{\odot}]$    & $\rm [M_{\odot}]$    & [mag]	& $\rm [M_{\odot} yr^{-1}]$ 	&$\rm [M_{\odot} yr^{-1}]$      \\
\noalign{\smallskip}

    CDFa-C9 &   9.95$\rm ^{+ 0.40 }_{- 0.08 } $ &   9.90$\rm ^{+ 0.30 }_{- 0.05 } $ & 0.00 & 2.19$\rm ^{+ 0.18 }_{- 0.30 } $ & 1.65$\rm {^+_- 0.14 } $   \\  \noalign{\smallskip}
  CDFS-4414 &  10.35$\rm ^{+ 0.19 }_{- 0.23 } $ &  10.25$\rm ^{+ 0.07 }_{- 1.24 } $ & 0.20 & 1.82$\rm ^{+ 0.25 }_{- 0.62 } $ & 1.80$\rm {^+_- 0.31 } $   \\  \noalign{\smallskip}
  CDFS-4417 &  10.07$\rm ^{+ 0.38 }_{- 0.11 } $ &  10.09$\rm ^{+ 0.12 }_{- 0.05 } $ & 0.25 & 2.41$\rm ^{+ 0.01 }_{- 0.63 } $ & 2.33$\rm {^+_- 0.31 } $   \\  \noalign{\smallskip}
  CDFS-6664 &   9.26$\rm ^{+ 0.11 }_{- 0.23 } $ &   9.07$\rm ^{+ 0.22 }_{- 0.08 } $ & 0.10 & 1.31$\rm ^{+ 0.27 }_{- 0.31 } $ & 1.36$\rm {^+_- 0.33 } $   \\  \noalign{\smallskip}
 CDFS-16767 &   9.82$\rm ^{+ 0.10 }_{- 0.16 } $ &   9.67$\rm ^{+ 0.13 }_{- 0.13 } $ & 0.15 & 1.69$\rm ^{+ 0.31 }_{- 0.05 } $ & 1.67$\rm {^+_- 0.31 } $   \\  \noalign{\smallskip}
 CDFS-11991 &   9.47$\rm ^{+ 0.13 }_{- 0.16 } $ &   9.42$\rm ^{+ 0.18 }_{- 0.08 } $ & 0.10 & 1.51$\rm ^{+ 0.26 }_{- 0.24 } $ & 1.23$\rm {^+_- 0.34 } $   \\  \noalign{\smallskip}
  CDFS-2528 &   9.54$\rm ^{+ 0.09 }_{- 0.08 } $ &   9.48$\rm ^{+ 0.16 }_{- 0.00 } $ & 0.20 & 1.77$\rm ^{+ 0.02 }_{- 0.33 } $ & 1.77$\rm {^+_- 0.33 } $   \\  \noalign{\smallskip}
 CDFS-16272 &   8.97$\rm ^{+ 0.18 }_{- 0.08 } $ &   9.00$\rm ^{+ 0.19 }_{- 0.09 } $ & 0.10 & 1.22$\rm ^{+ 0.02 }_{- 0.30 } $ & 1.46$\rm {^+_- 0.33 } $   \\  \noalign{\smallskip}
  CDFS-9313 &   9.29$\rm ^{+ 0.29 }_{- 0.28 } $ &   9.29$\rm ^{+ 0.25 }_{- 0.27 } $ & 0.10 & 1.27$\rm ^{+ 0.29 }_{- 0.63 } $ & 1.54$\rm {^+_- 0.32 } $   \\  \noalign{\smallskip}
  CDFS-9340 &   8.86$\rm ^{+ 0.39 }_{- 0.28 } $ &   8.85$\rm ^{+ 0.32 }_{- 0.26 } $ & 0.10 & 0.84$\rm ^{+ 0.29 }_{- 0.65 } $ & 1.37$\rm {^+_- 0.32 } $   \\  \noalign{\smallskip}
 CDFS-12631 &   9.97$\rm ^{+ 0.45 }_{- 0.10 } $ &  10.18$\rm ^{+ 0.10 }_{- 0.29 } $ & 0.30 & 2.17$\rm ^{+ 0.05 }_{- 0.86 } $ & 2.34$\rm {^+_- 0.32 } $   \\  \noalign{\smallskip}
 CDFS-14411 &   9.28$\rm ^{+ 0.08 }_{- 0.03 } $ &   9.30$\rm ^{+ 0.05 }_{- 0.05 } $ & 0.15 & 1.62$\rm ^{+ 0.05 }_{- 0.02 } $ & 1.88$\rm {^+_- 0.31 } $   \\  \noalign{\smallskip}
  CDFS-5161 &   9.65$\rm ^{+-0.58 }_{- 0.24 } $ &   9.67$\rm ^{+ 0.29 }_{- 0.24 } $ & 0.20 & 1.59$\rm ^{+ 0.27 }_{- 1.21 } $ & 1.43$\rm {^+_- 0.40 } $   \\  \noalign{\smallskip}
 LnA1689-2 $^a$ &   9.88$\rm ^{+ 0.21 }_{- 0.36 } $ &   9.79$\rm ^{+ 0.10 }_{- 0.32 } $ & 0.06 & 1.14$\rm ^{+ 0.18 }_{- 0.30 } $ & 1.31$\rm {^+_- 0.34 } $   \\  \noalign{\smallskip}
 LnA1689-4 $^a$ &   8.46$\rm ^{+ 0.26 }_{- 0.39 } $ &   8.12$\rm ^{+ 0.53 }_{- 0.03 } $ & 0.06 & 0.10$\rm ^{+ 0.18 }_{- 0.30 } $ & 0.04$\rm {^+_- 0.24 } $   \\  \noalign{\smallskip}
 LnA1689-1 $^a$ &   9.91$\rm ^{+ 0.06 }_{- 0.10 } $ &   9.63$\rm ^{+ 0.21 }_{- 0.09 } $ & 0.00 & 0.38$\rm ^{+ 0.18 }_{- 0.30 } $ & 0.09$\rm {^+_- 0.19 } $   \\  \noalign{\smallskip}
 Q1422-D88 &  10.60$\rm ^{+ 0.02 }_{- 0.23 } $ &  10.43$\rm ^{+ 0.11 }_{- 0.10 } $ & 0.00 & 0.94$\rm ^{+ 0.30 }_{- 0.05 } $ & 1.36$\rm {^+_- 0.19 } $   \\  \noalign{\smallskip}
 3C324-C3 &   9.67$\rm ^{+ 0.33 }_{- 0.23 } $ &   9.62$\rm ^{+ 0.28 }_{- 0.16 } $ & 0.20 & 1.72$\rm ^{+ 0.25 }_{- 0.95 } $  & 1.53$\rm {^+_- 0.32 } $   \\  \noalign{\smallskip}
 Cosmic Eye $^a$ &   9.55$\rm ^{+-0.48 }_{- 0.48 } $ &   9.55$\rm ^{+-0.48 }_{- 0.48 } $ & 0.40 & 1.67$\rm ^{+-0.40 }_{- 0.40 } $ & 1.63$\rm {^+_- 0.36 } $   \\  \noalign{\smallskip}
 SSA22a-M38 &  10.78$\rm ^{+ 0.18 }_{- 0.41 } $ &  10.48$\rm ^{+ 0.19 }_{- 0.40 } $ & 0.20 & 1.83$\rm ^{+ 0.18 }_{- 0.30 } $ & 2.27$\rm {^+_- 0.31 } $   \\  \noalign{\smallskip}
 SSA22a-D17 &   9.95$\rm ^{+ 0.50 }_{- 0.61 } $ &   9.67$\rm ^{+ 0.50 }_{- 0.27 } $ & 0.10 & 1.42$\rm ^{+ 0.18 }_{- 0.30 } $ & 1.18$\rm {^+_- 0.32 } $   \\  \noalign{\smallskip}
 SSA22a-aug96M16 &  10.06$\rm ^{+ 0.20 }_{- 0.21 } $ &   9.92$\rm ^{+ 0.11 }_{- 0.28 } $ & 0.06 & 1.39$\rm ^{+ 0.18 }_{- 0.30 } $ & 1.20$\rm {^+_- 0.24 } $   \\  \noalign{\smallskip}
 SSA22a-C16 &  10.61$\rm ^{+ 0.13 }_{- 0.55 } $ &  10.46$\rm ^{+ 0.11 }_{- 0.37 } $ & 0.35 & 2.58$\rm ^{+ 0.18 }_{- 0.30 } $ & 2.84$\rm {^+_- 0.31 } $   \\  \noalign{\smallskip}
 SSA22a-C36 &  10.10$\rm ^{+ 0.30 }_{- 0.20 } $ &  10.08$\rm ^{+ 0.21 }_{- 0.17 } $ & 0.25 & 2.06$\rm ^{+ 0.27 }_{- 1.46 } $ & 1.91$\rm {^+_- 0.38 } $   \\  \noalign{\smallskip}
 DSF2237b-C21 &  10.80$\rm ^{+ 0.14 }_{- 0.31 } $ &  10.55$\rm ^{+ 0.06 }_{- 0.11 } $ & 0.06 & 1.44$\rm ^{+ 0.74 }_{- 0.56 } $ & 1.65$\rm {^+_- 0.24 } $   \\  \noalign{\smallskip}
 DSF2237b-D29 &  10.72$\rm ^{+ 0.15 }_{- 0.22 } $ &  10.55$\rm ^{+ 0.14 }_{- 0.11 } $ & 0.00 & 1.35$\rm ^{+ 0.50 }_{- 0.56 } $ & 0.66$\rm {^+_- 0.26 } $   \\  \noalign{\smallskip}
 Q0302-C131  $^{(*)}$ &  10.09$\rm ^{+ 0.10 }_{- 0.33 } $ &-- & 0.15 &   1.11$\rm ^{+ 0.16 }_{- 0.26 } $ & 1.00$\rm ^{+ 0.22 }_{- 0.40 } $ \\  \noalign{\smallskip}
 Q0302-C171  $^{(*)}$ &  10.06$\rm ^{+ 0.10 }_{- 0.28 } $ &-- & 0.10 &   1.00$\rm ^{+ 0.13 }_{- 0.18 } $ & 0.70$\rm ^{+ 0.40 }_{- 0.40 } $ \\  \noalign{\smallskip}
  Q0302-M80  $^{(*)}$ &  10.07$\rm ^{+ 0.23 }_{- 0.19 } $ &-- & 0.30 &   1.54$\rm ^{+ 0.26 }_{- 0.70 } $ & 1.11$\rm ^{+ 0.12 }_{- 0.21 } $ \\ \noalign{\smallskip}
 SSA22a-C30  $^{(*)}$ &  10.33$\rm ^{+ 0.31 }_{- 0.38 } $ &-- & 0.40 &   2.00$\rm ^{+ 0.29 }_{- 1.35 } $ & 1.46$\rm ^{+ 0.14 }_{- 0.45 } $ \\ \noalign{\smallskip}
  SSA22a-C6  $^{(*)}$ &   9.68$\rm ^{+ 0.15 }_{- 0.06 } $ &-- & 0.40 &   1.73$\rm ^{+ 0.16 }_{- 0.24 } $ & 1.36$\rm ^{+ 0.32 }_{- 0.46 } $ \\ \noalign{\smallskip}
  SSA22a-M4  $^{(*)}$ &   9.41$\rm ^{+ 0.34 }_{- 0.13 } $ &-- & 0.40 &   1.33$\rm ^{+ 0.27 }_{- 0.89 } $ & 1.30$\rm ^{+ 0.30 }_{- 0.19 } $ \\ \noalign{\smallskip}
  SSA22b-C5  $^{(*)}$ &   8.96$\rm ^{+ 0.38 }_{- 0.22 } $ &-- & 0.25 &   0.93$\rm ^{+ 0.23 }_{- 0.52 } $ & 1.18$\rm ^{+ 0.00 }_{- 0.27 } $ \\ \noalign{\smallskip}
DSF2237b-D28  $^{(*)}$ &   9.78$\rm ^{+ 0.28 }_{- 0.29 } $ &-- & 0.30 &   1.33$\rm ^{+ 0.25 }_{- 0.66 } $ & 1.15$\rm ^{+ 0.11 }_{- 0.24 } $ \\ \noalign{\smallskip}

\hline\hline                 
\noalign{\smallskip}
\end{tabular}
}
\\
Col. 1, object name; 
Col. 2, stellar mass inferred from the galaxy templates of BC03;
Col. 3, stellar mass inferred from the galaxy templates of M05;
Col. 4, dust reddening that affects the stellar light, from the attenuation curve of \cite{calzetti00};
Col. 5, star formation rate (from BC03);
Col. 6, star formation rate (from $H\beta$ flux).
To compare our results with other studies using the Chabrier IMF,
the masses and the star formation rates were calculated from the 
Salpeter IMF corrected for a factor of 1.7.
 The stellar mass, reddening, and star formation rate of the Cosmic Eye 
 correspond to the best-fitting SED values reported in \cite{coppin07}, 
 the $SFR(H\beta)$ was derived with the measured $H\beta$ flux,
 and the reported extinction considering an error of $\pm 0.15~mag$. 
Notes:$^{a}$ values corrected for a magnification factor of $43.5\pm 4.1, 6.7 \pm
 2.1, 5.9 \pm 1.9$ (see text), and
 $28 \pm 3$ \citep{smail07}, respectively. 
 $^*$ LSD galaxies, all values have been taken from \cite{mannucci09}.
\end{table*}

\begin{table*}

\caption[Galaxy mass estimated using OIII images.]
{Galaxy mass estimated using OIII images. Derived quantity yields and metallicities.}

{\centering
\begin{tabular}{lcccc}
\hline\hline                
Name     &log Mgas($\Sigma SFR$) $\, M_{\sun}$ 	   &Gas fraction   &12+log(O/H)	 &log Yields		\\ 		\\
\hline

\noalign{\smallskip}
   CDFa-C9     & 9.51$\pm$0.29     & 0.27$\pm$0.17     &8.07 $^{+0.17 }_{-0.17 }$     &-2.61$\pm$0.54   \\
\noalign{\smallskip}
  CDFS-4414     &10.11$\pm$0.30     & 0.37$\pm$0.20     &8.47 $^{+0.11 }_{-0.10 }$     &-2.09$\pm$0.50   \\
\noalign{\smallskip}
  CDFS-4417     &10.47$\pm$0.30     & 0.72$\pm$0.18     &8.48 $^{+0.09 }_{-0.09 }$     &-1.60$\pm$0.62   \\
\noalign{\smallskip}
  CDFS-6664     & 9.61$\pm$0.30     & 0.69$\pm$0.17     &7.88 $^{+0.17 }_{-0.20 }$     &-2.25$\pm$0.72   \\
\noalign{\smallskip}
 CDFS-16767     & 9.49$\pm$0.30     & 0.32$\pm$0.16     &8.28 $^{+0.16 }_{-0.32 }$     &-2.34$\pm$0.69   \\
\noalign{\smallskip}
 CDFS-11991     & 9.47$\pm$0.30     & 0.50$\pm$0.19     &7.87 $^{+0.20 }_{-0.20 }$     &-2.53$\pm$0.68   \\
\noalign{\smallskip}
  CDFS-2528     & 9.97$\pm$0.30     & 0.73$\pm$0.14     &8.22 $^{+0.24 }_{-0.31 }$     &-1.84$\pm$0.87   \\
\noalign{\smallskip}
 CDFS-16272     & 9.30$\pm$0.30     & 0.68$\pm$0.16     &7.92 $^{+0.23 }_{-0.22 }$     &-2.22$\pm$0.77   \\
\noalign{\smallskip}
  CDFS-9313     & 9.87$\pm$0.30     & 0.79$\pm$0.16     &7.95 $^{+0.23 }_{-0.25 }$     &-1.97$\pm$0.83   \\
\noalign{\smallskip}
  CDFS-9340     & 9.89$\pm$0.30     & 0.91$\pm$0.08     &8.28 $^{+0.14 }_{-0.71 }$     &-1.23$\pm$1.18   \\
\noalign{\smallskip}
 CDFS-12631     &10.31$\pm$0.30     & 0.69$\pm$0.20     &7.89 $^{+0.16 }_{-0.17 }$     &-2.25$\pm$0.69   \\
\noalign{\smallskip}
 CDFS-14411     & 9.72$\pm$0.30     & 0.73$\pm$0.14     &8.16 $^{+0.16 }_{-0.23 }$     &-1.89$\pm$0.74   \\
\noalign{\smallskip}
  CDFS-5161     & 9.87$\pm$0.30     & 0.62$\pm$0.27     &7.69 $^{+0.37 }_{-0.28 }$     &-2.54$\pm$0.93   \\
\noalign{\smallskip}
 LnA1689-2     &10.09$\pm$0.31     & 0.62$\pm$0.23      &7.92 $^{+0.74 }_{-0.39}$     &-2.32$\pm$1.42   \\
\noalign{\smallskip}
 LnA1689-4     & 9.17$\pm$0.29     & 0.84$\pm$0.14     &7.92 $^{+0.11 }_{-0.11 }$     &-1.89$\pm$0.66   \\
\noalign{\smallskip}
 LnA1689-1     & 9.58$\pm$0.30     & 0.32$\pm$0.16     &8.35 $^{+0.19 }_{-0.26 }$     &-2.27$\pm$0.66   \\
\noalign{\smallskip}
  Q1422-D88     & 9.63$\pm$0.30     & 0.10$\pm$0.07     &7.94 $^{+0.24 }_{-0.33 }$     &-2.99$\pm$0.71   \\
\noalign{\smallskip}
  3C324-C3     & 9.80$\pm$0.30     & 0.57$\pm$0.23     &7.79 $^{+0.35 }_{-0.29 }$     &-2.51$\pm$0.91   \\
\noalign{\smallskip}
 Cosmic Eye     & 9.63$\pm$0.29     & 0.55$\pm$0.32     &7.97 $^{+0.32 }_{-0.23 }$     &-2.37$\pm$0.81   \\
\noalign{\smallskip}
 SSA22a-M38     &10.14$\pm$0.30     & 0.19$\pm$0.15     &8.14 $^{+0.11 }_{-0.15 }$     &-2.64$\pm$0.45   \\
\noalign{\smallskip}
 SSA22a-D17     & 9.40$\pm$0.30     & 0.22$\pm$0.25     &8.24 $^{+0.18 }_{-0.17 }$     &-2.50$\pm$0.54   \\
\noalign{\smallskip}
 SSA22a-aug96M16     & 9.33$\pm$0.30     & 0.16$\pm$0.11     &8.21 $^{+0.18 }_{-0.60 }$     &-2.62$\pm$0.95   \\
\noalign{\smallskip}
 SSA22a-C16     &10.81$\pm$0.29     & 0.61$\pm$0.24     &8.35 $^{+0.13 }_{-0.15 }$     &-1.90$\pm$0.62   \\
\noalign{\smallskip}
 SSA22a-C36     &10.01$\pm$0.30     & 0.45$\pm$0.22     &8.44 $^{+0.13 }_{-0.15 }$     &-2.02$\pm$0.57   \\
\noalign{\smallskip}
 DSF2237b-C21     &10.08$\pm$0.30     & 0.16$\pm$0.11     &8.79 $^{+0.08 }_{-0.07 }$     &-2.04$\pm$0.36   \\
\noalign{\smallskip}
 DSF2237b-D29     & 9.85$\pm$0.30     & 0.12$\pm$0.09     &8.34 $^{+0.19 }_{-0.19 }$     &-2.55$\pm$0.52   \\
\noalign{\smallskip}

\hline\hline                 
\end{tabular}
} \\	
Col. 1, object name;
Col. 2, mass of gas in [$M_{\sun}$] obtained by inverting the Schmidt-Kennicutt law and
 adding the surface star formation rate at every galaxy pixel throughout the galaxy extention;
Col. 3, gas fraction;
Col. 4, metallicity at 0.375arcsec using the \cite{maiolino08} metallicity calibration;
Col. 5, effective yields.\\
\label{table_mgas_met}
\end{table*}

\begin{table*}
\label{table_gradients}
\caption{Metallicity differences between inner and outer galaxy regions inferred from the near-IR spectra.}
\begin{tabular}{lccc}
\hline\hline                 
\tiny Name &\tiny{12+log(O/H)$_{inner}$} &\tiny{12+log(O/H)$_{outer}$} &\tiny{$\Delta$ 12+log(O/H)} \\ 
\hline 
\hline
\noalign{\smallskip}
      CDFa-C9  & $  8.18 ^{+  0.17}_{-  0.31}$   &$  8.38 ^{+  0.14}_{-  0.15}$ &$  0.20 ^{+  0.22}_{-  0.34}$ \\
\noalign{\smallskip}
      CDFS-4417  & $  8.60 ^{+  0.10}_{-  0.10}$ &$  8.72 ^{+  0.08}_{-  0.10}$ &$  0.12 ^{+  0.13}_{-  0.14}$ \\
\noalign{\smallskip}
      CDFS-6664  & $  7.83 ^{+  0.26}_{-  0.28}$ &$  8.38 ^{+  0.15}_{-  0.59}$ &$  0.55 ^{+  0.30}_{-  0.65}$ \\
\noalign{\smallskip}
      CDFS-16767  & $  8.28 ^{+  0.16}_{-  0.32}$ &$  8.79 ^{+  0.11}_{-  0.10}$ &$  0.51 ^{+  0.19}_{-  0.34}$ \\
\noalign{\smallskip}
      CDFS-2528  & $  8.10 ^{+  0.42}_{-  0.26}$ &$  8.58 ^{+  0.13}_{-  0.20}$ &$  0.48 ^{+  0.44}_{-  0.33}$ \\
\noalign{\smallskip}
      CDFS-16272  & $  7.95 ^{+  0.31}_{-  0.34}$ &$  8.49 ^{+  0.13}_{-  1.32}$ &$  0.54 ^{+  0.34}_{-  1.36}$ \\
\noalign{\smallskip}
      CDFS-12631 &$ 8.41 ^{+  0.18}_{-  0.13}$   &$  8.51 ^{+0.36}_{-0.22 }$ 	&$0.10^{+0.40}_{-0.26}$ \\
\noalign{\smallskip}
      CDFS-14411  & $  8.13 ^{+  0.21}_{-  0.49}$ &$  8.49 ^{+  0.13}_{-  0.16}$ &$  0.36 ^{+  0.25}_{-  0.52}$ \\
\noalign{\smallskip}
      SSA22a-M38  & $  8.10 ^{+  0.24}_{-  0.24}$ &$  8.30 ^{+  0.16}_{-  0.15}$ &$  0.20 ^{+  0.29}_{-  0.28}$ \\
\noalign{\smallskip}
      SSA22a-C16  & $  8.34 ^{+  0.14}_{-  0.19}$ &$  8.57 ^{+  0.13}_{-  0.15}$ &$  0.23 ^{+  0.19}_{-  0.24}$ \\
\noalign{\smallskip}
{\it Composite}$^a$ &$8.13 ^{+  0.23}_{-  0.18}$ & $8.56 ^{+  0.44}_{-  0.29}$ &$0.43^{+  0.50}_{- 0.34}$	\\ 
\hline\hline                 
\end{tabular}
\\
Col. 1, object name;
Col. 2, gas metallicity of the inner aperture within a radius of 0.375";
Col. 3, gas metallicity of the outer region within a radius of 0.375"$< r < $1";
Col. 4, metallicity difference between inner and outer regions. 
Notes: $^a$ The same values as described above are reported for the composite spectra of the ten galaxies presented in this table. 
 \\
\end{table*}

\vfill
\newpage
\vfill
\newpage

\appendix
\section{Integrated near-IR spectra of the AMAZE and LSD galaxies}

\begin{figure*}
  \centering
  \includegraphics[width=0.49\linewidth]{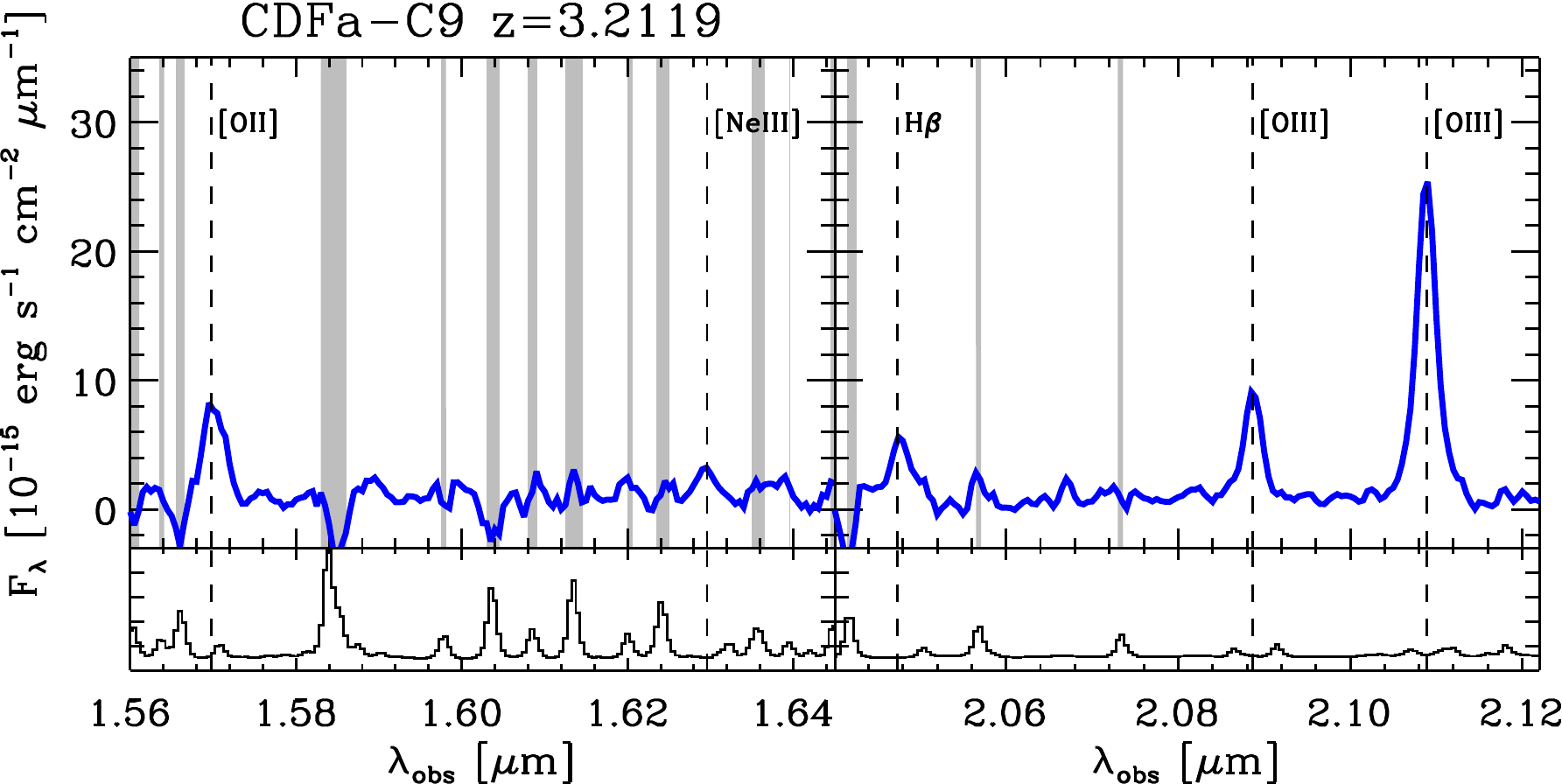} 
  \includegraphics[width=0.49\linewidth]{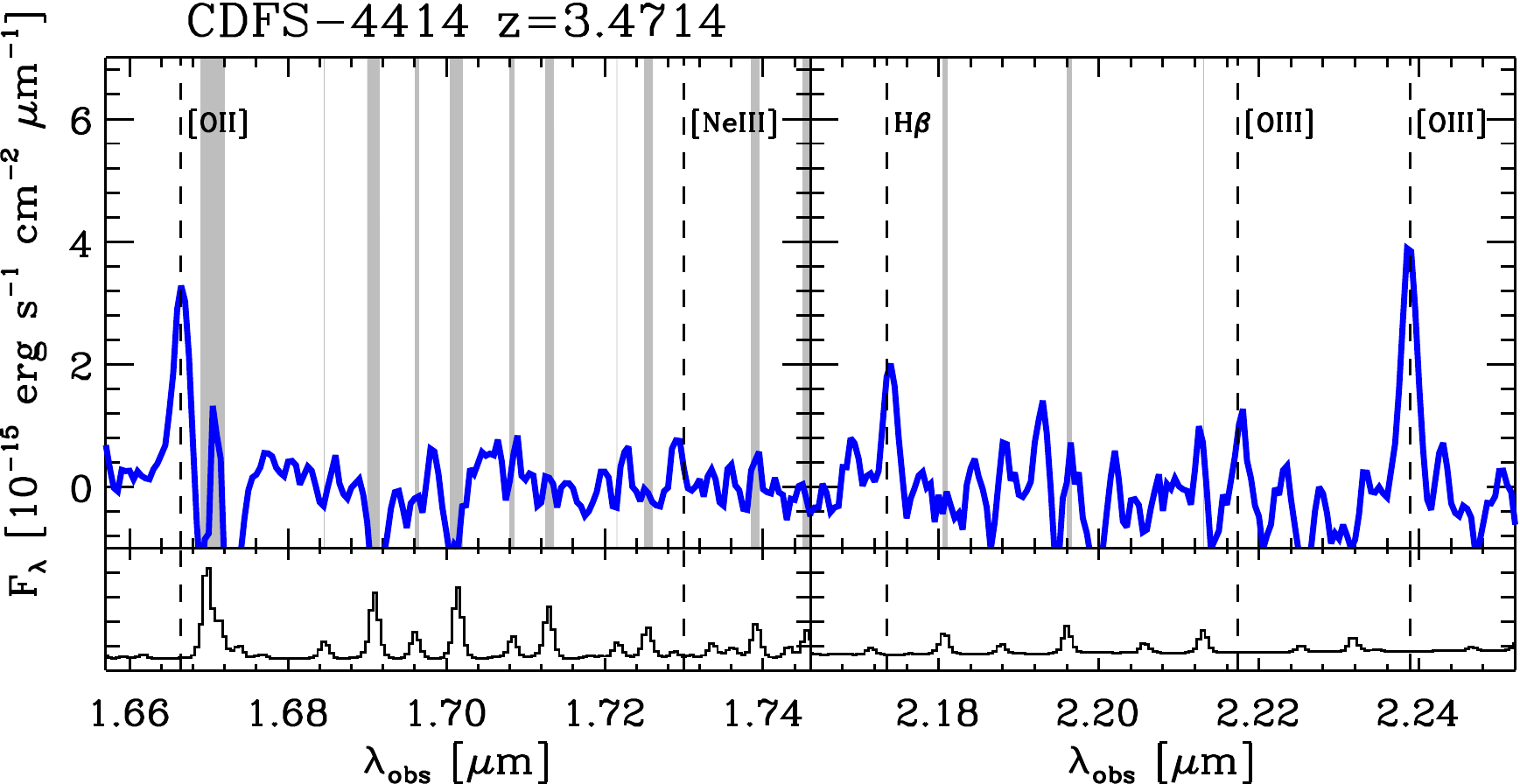}
  \includegraphics[width=0.49\linewidth]{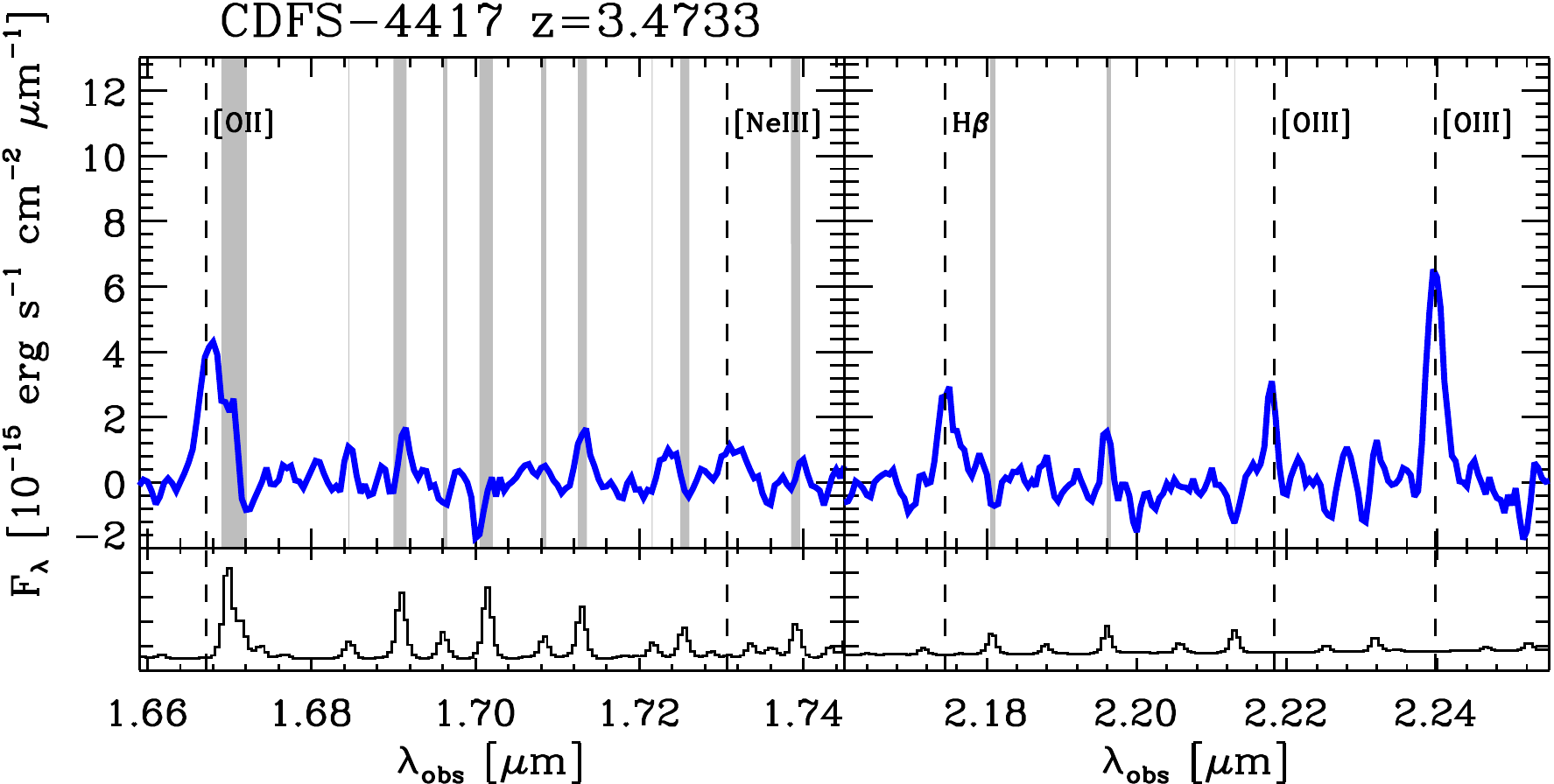}
  \includegraphics[width=0.49\linewidth]{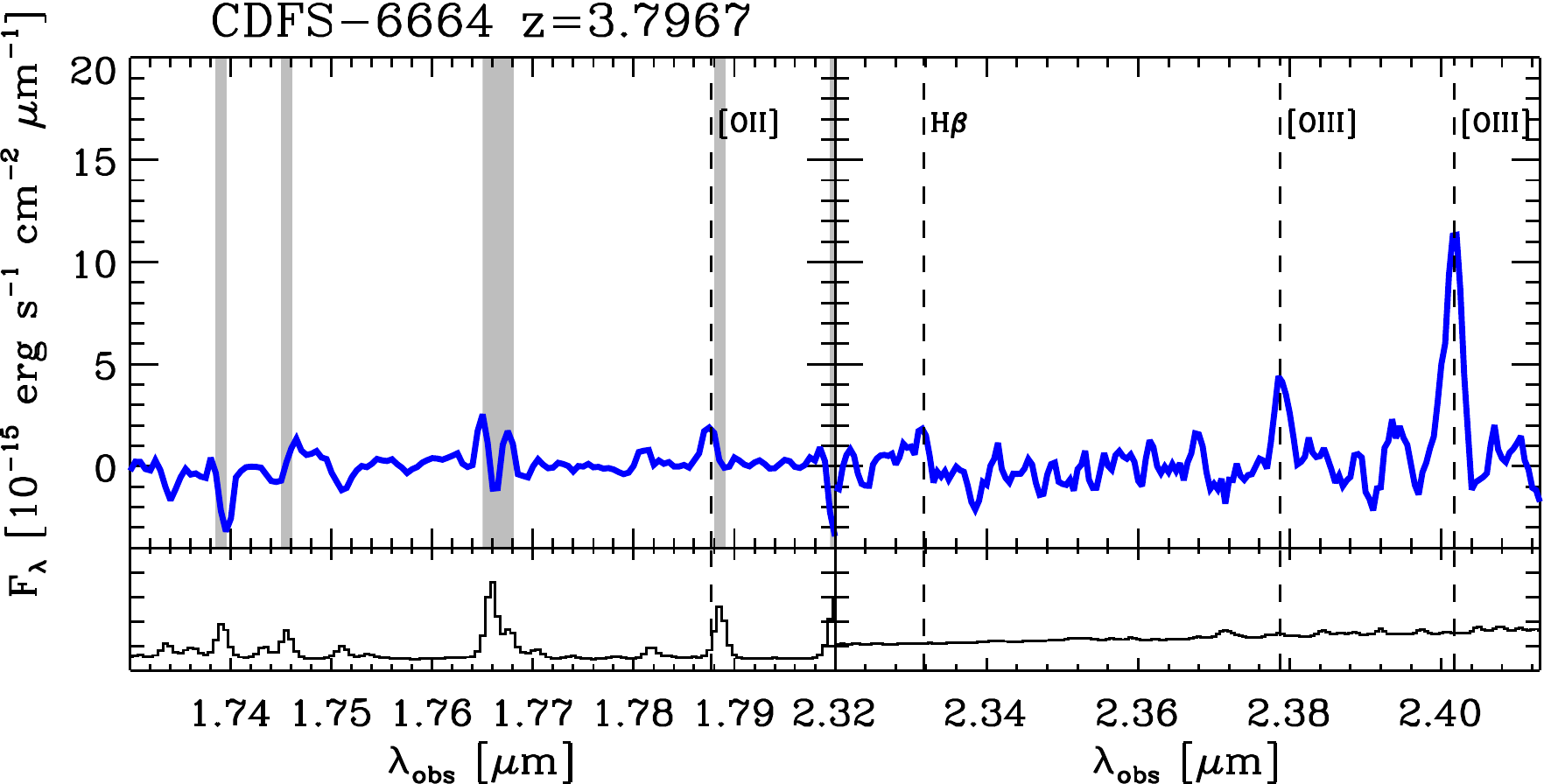}
  \includegraphics[width=0.49\linewidth]{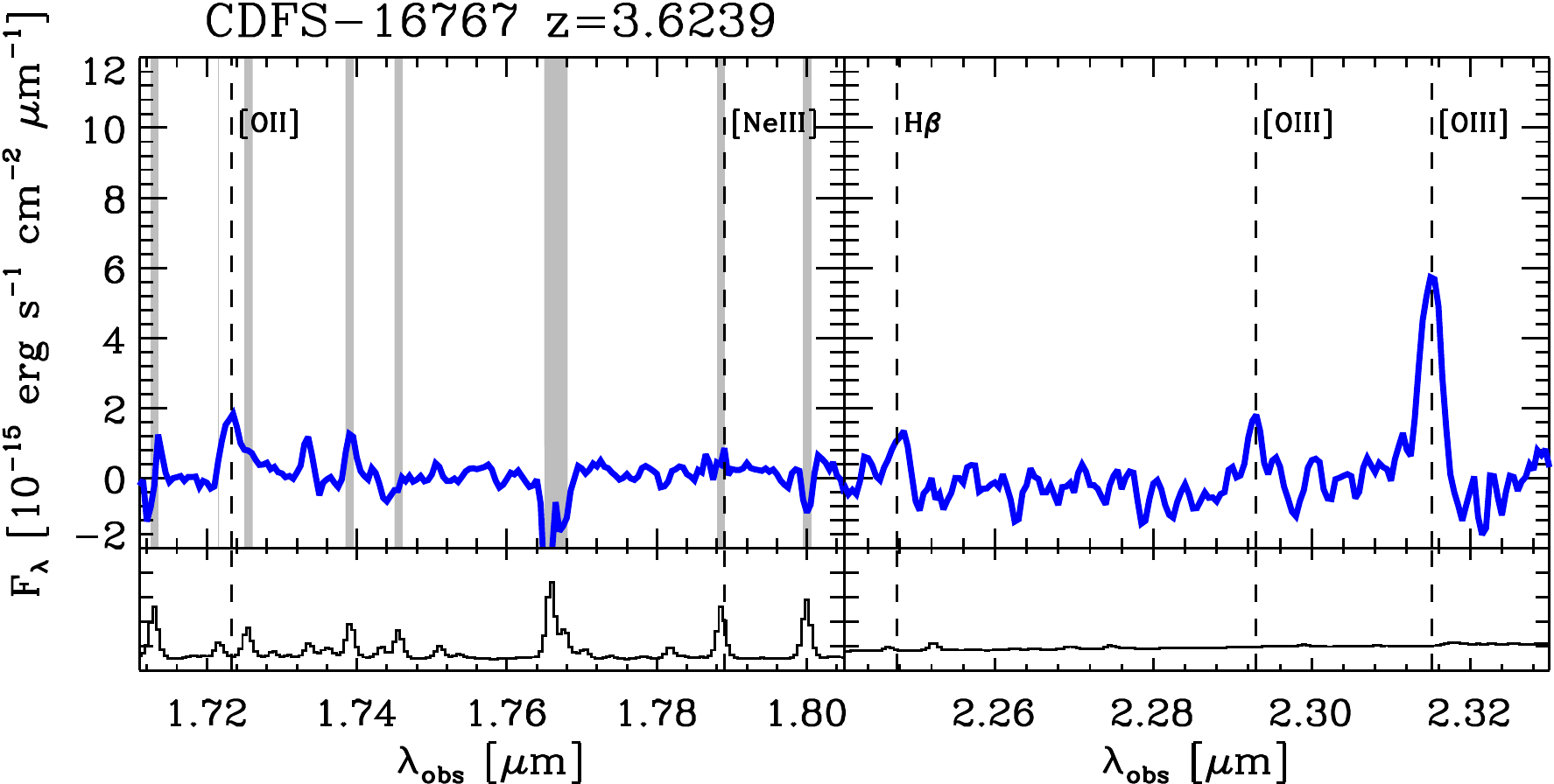}
  \includegraphics[width=0.49\linewidth]{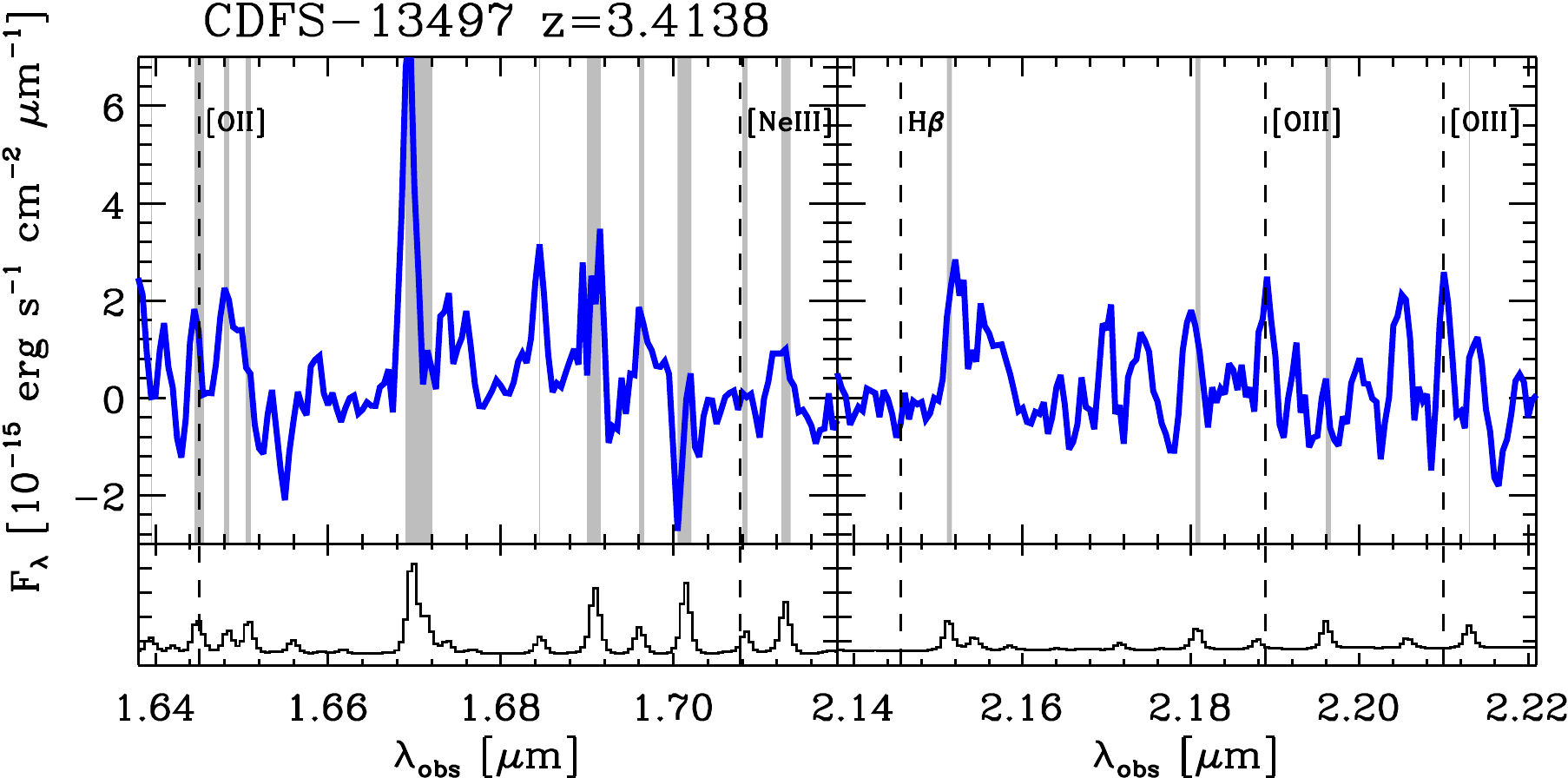}
  \includegraphics[width=0.49\linewidth]{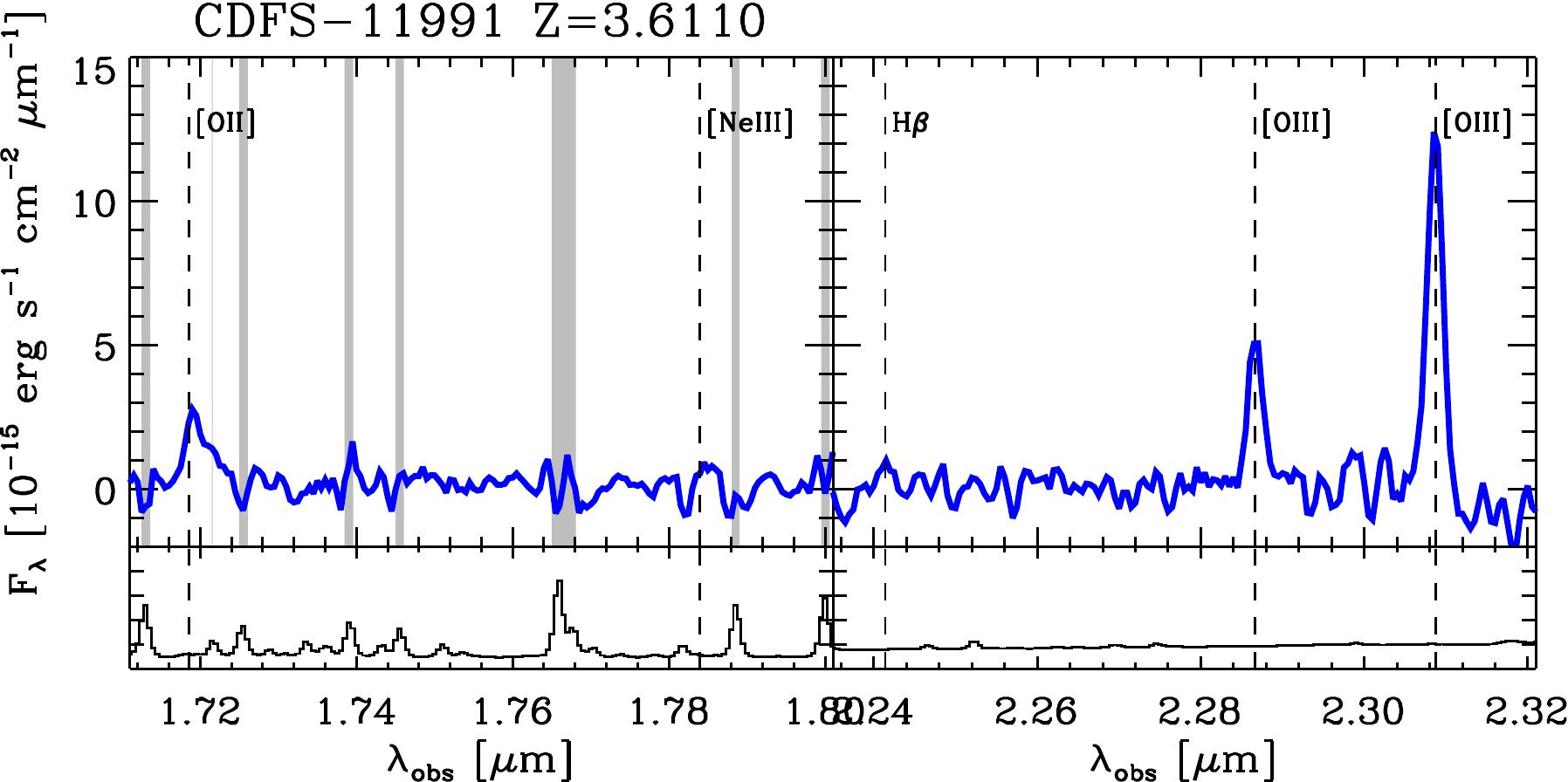}
  \includegraphics[width=0.49\linewidth]{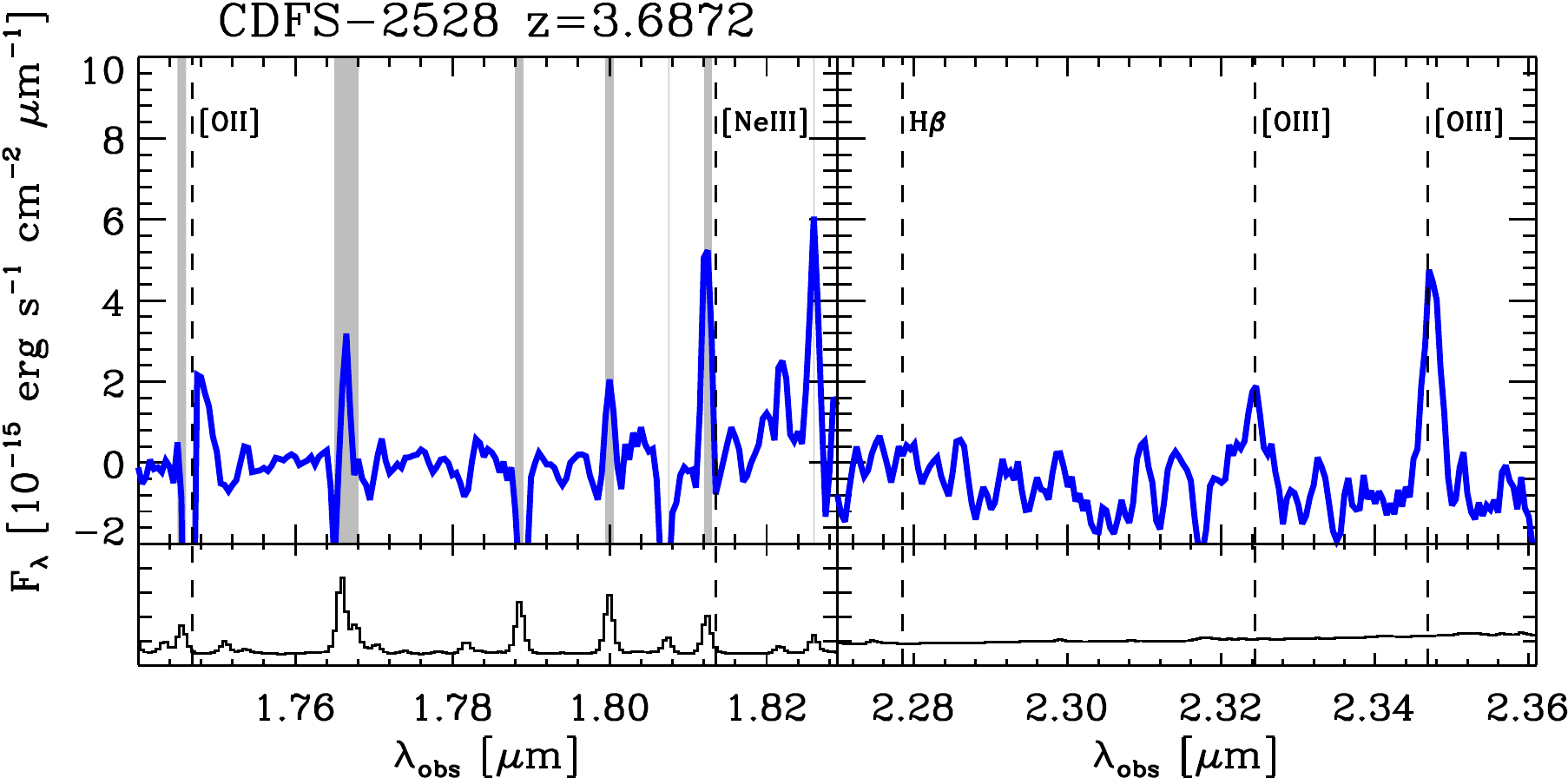}
  \includegraphics[width=0.49\linewidth]{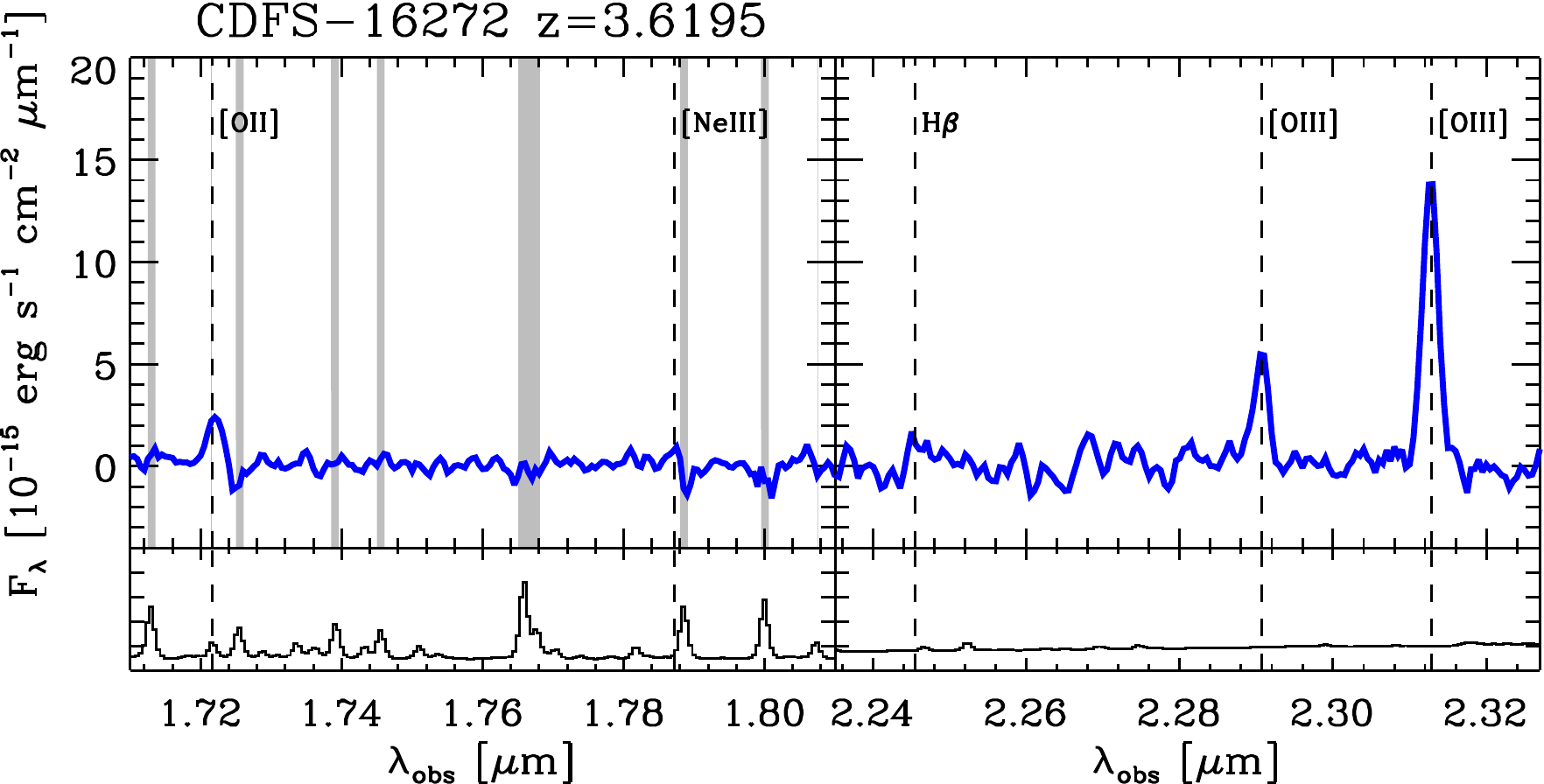}
  \includegraphics[width=0.49\linewidth]{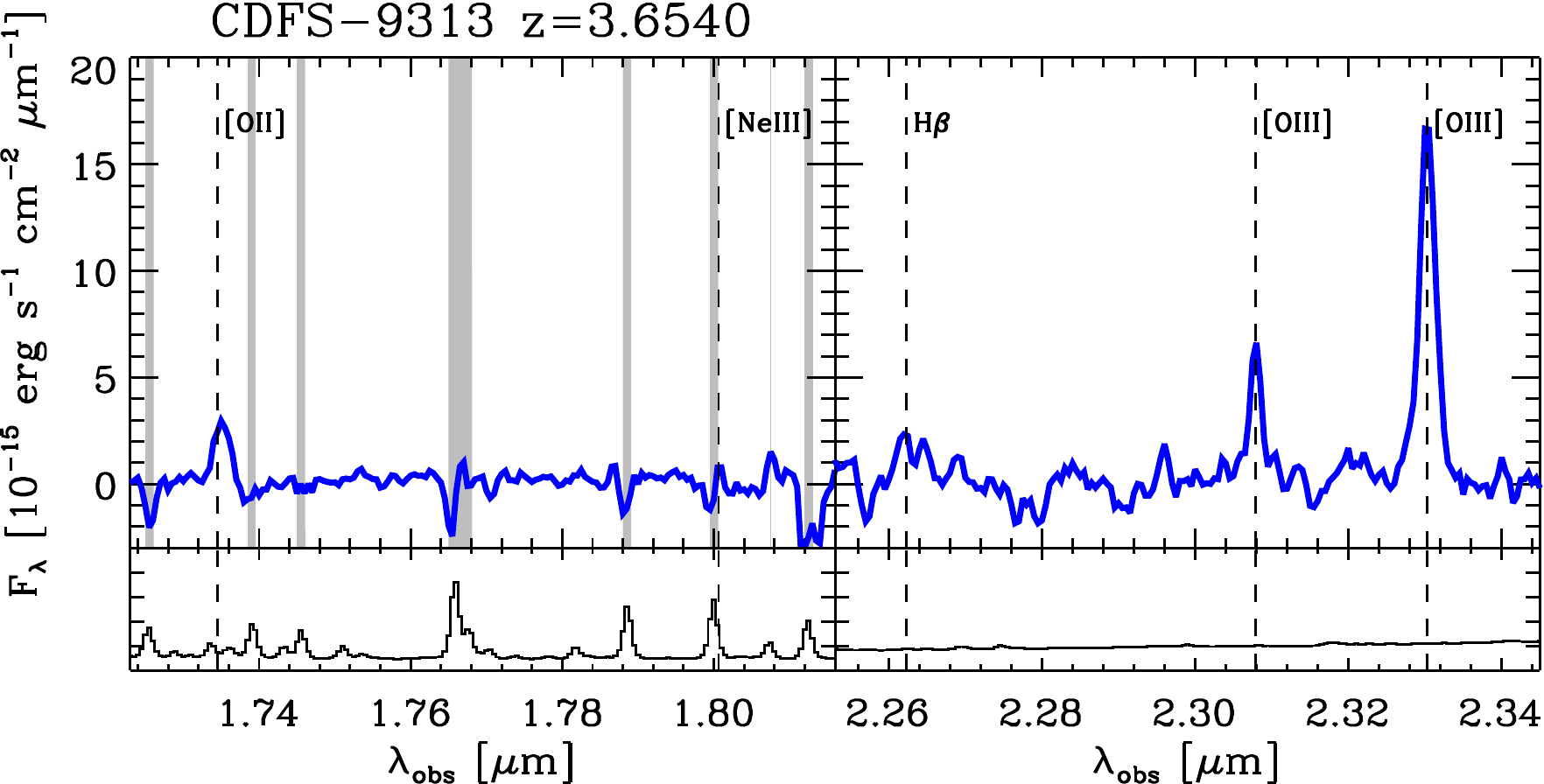}
  \caption{Near-IR spectra of the galaxies in the AMAZE sample.
  The vertical dotted lines indicate the expected location of nebular emission lines.
  The shaded vertical regions overlaid on each spectrum highlight spectral regions affected by strong sky emission lines.}
  \label{fig1sp}
  \end{figure*}

\begin{figure*}	  
  \centering
  \includegraphics[width=0.49\linewidth]{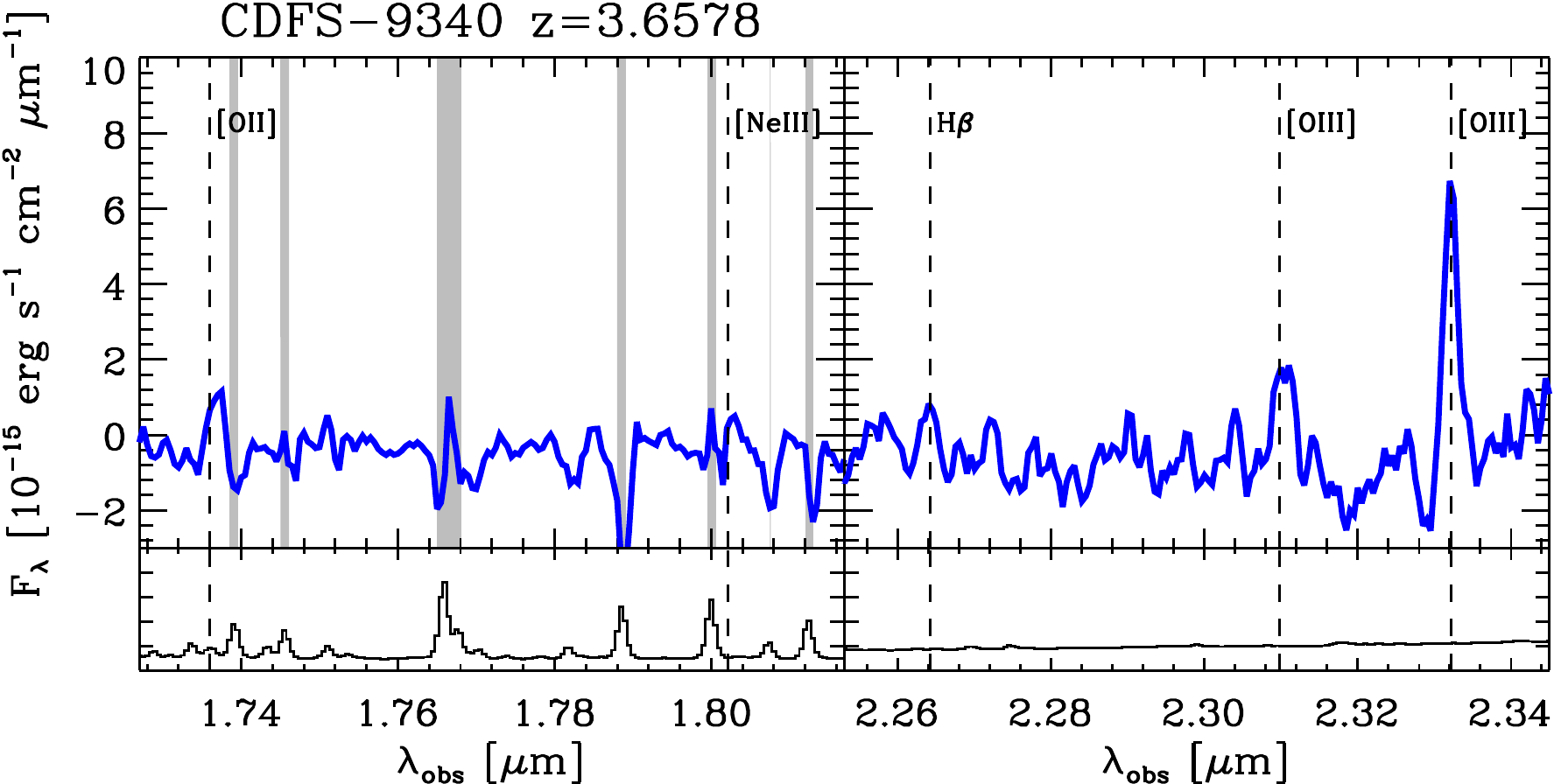}
  \includegraphics[width=0.49\linewidth]{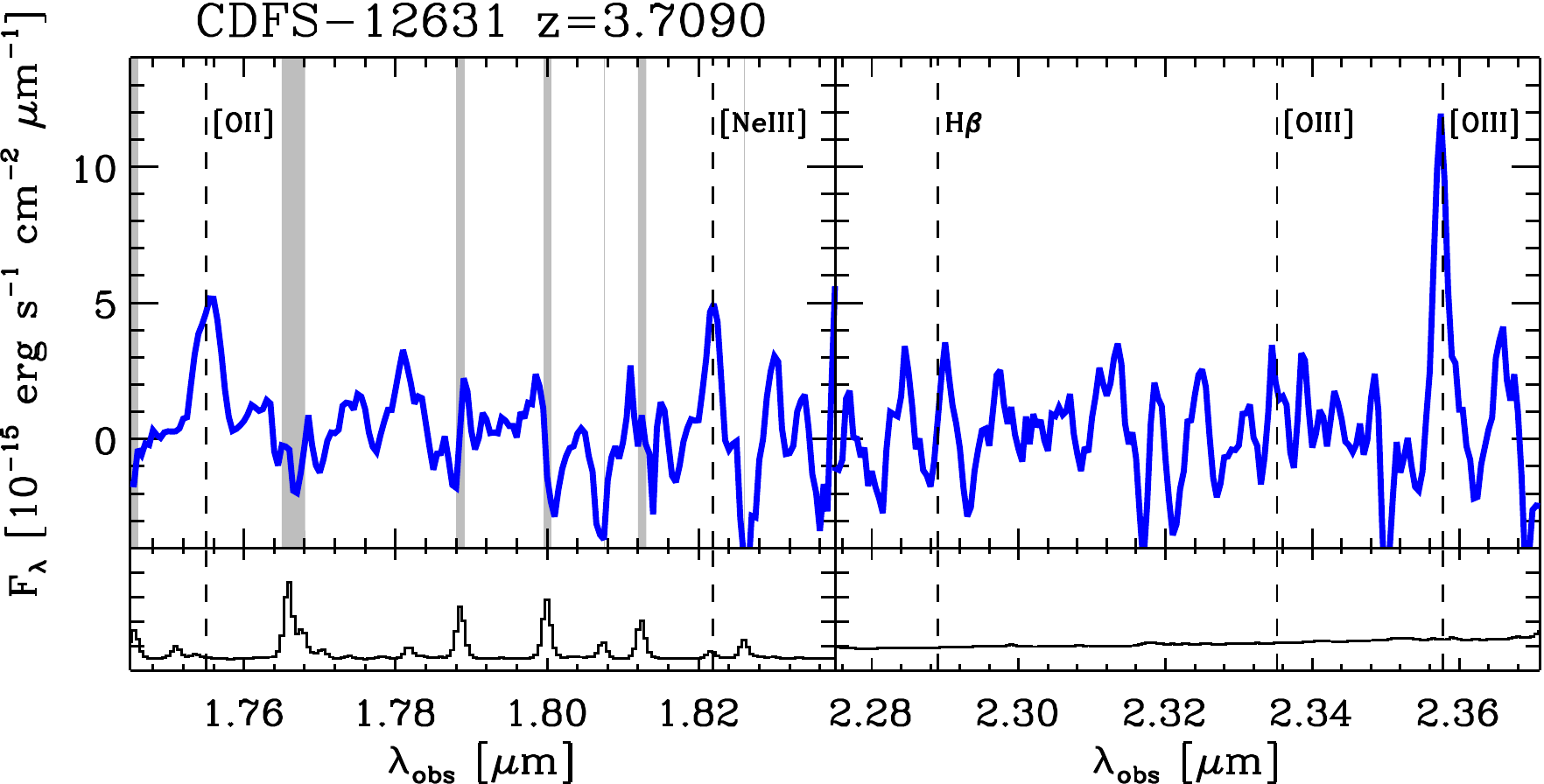}
  \includegraphics[width=0.49\linewidth]{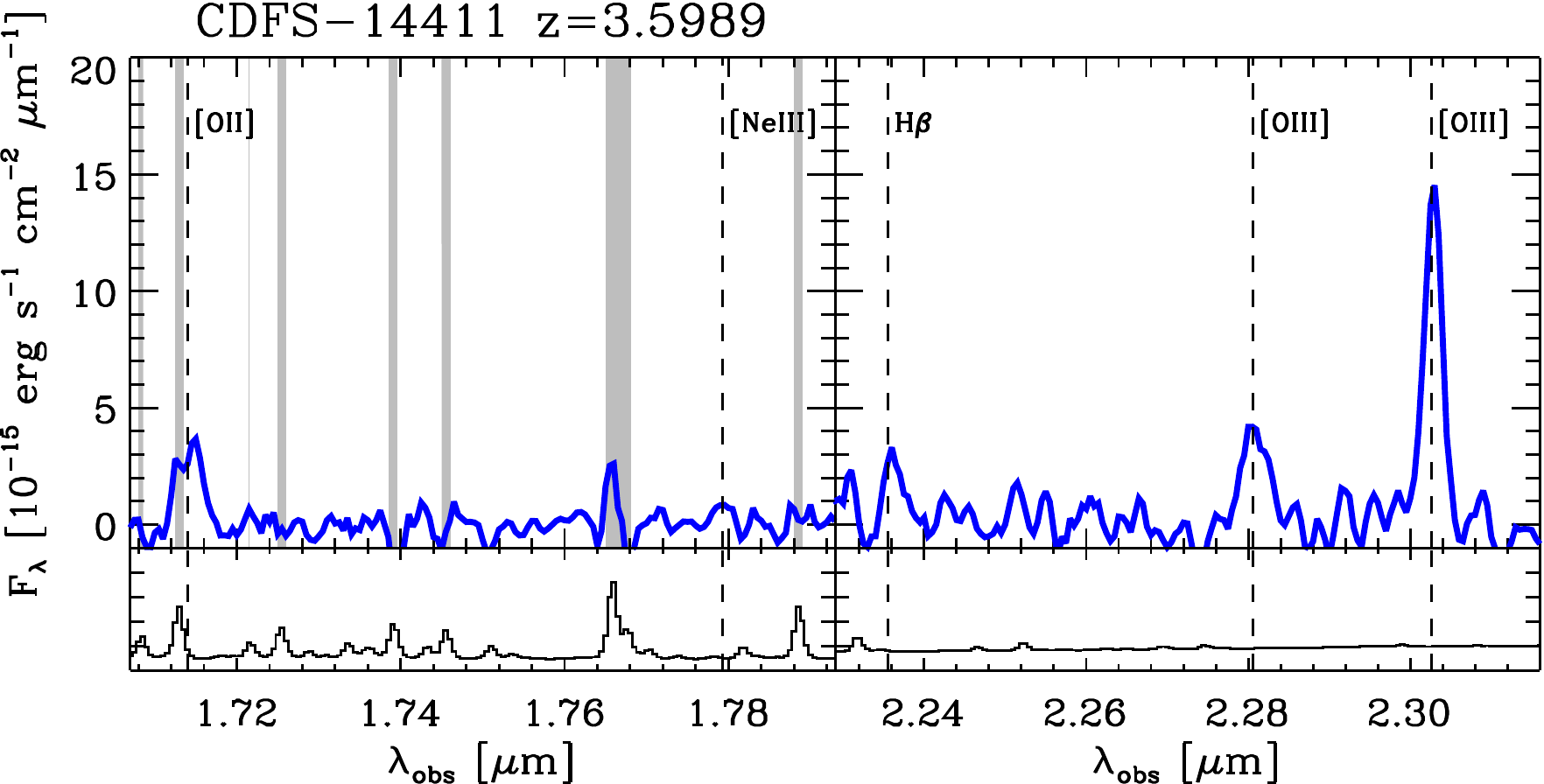}
  \includegraphics[width=0.49\linewidth]{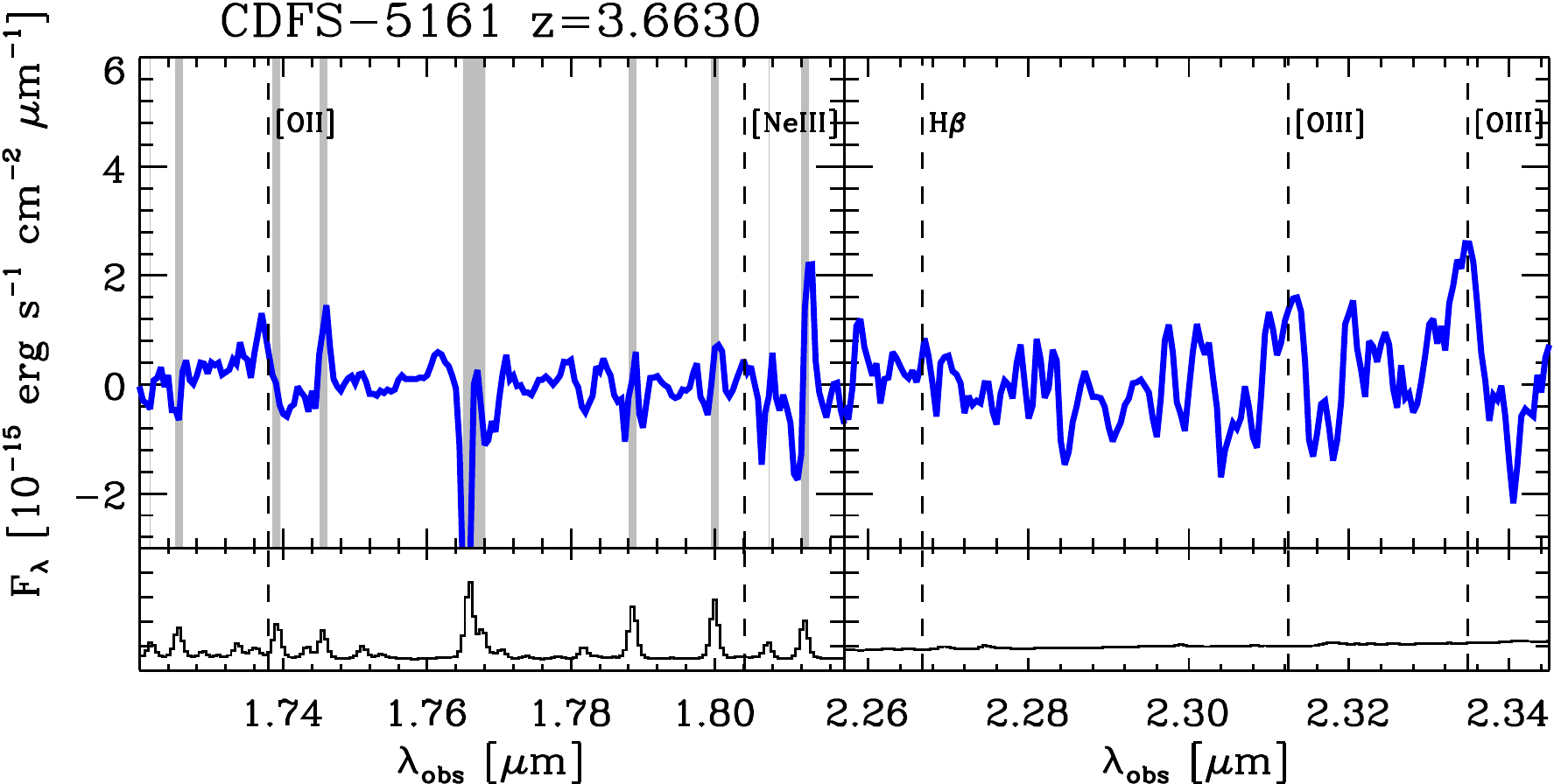}
  \includegraphics[width=0.49\linewidth]{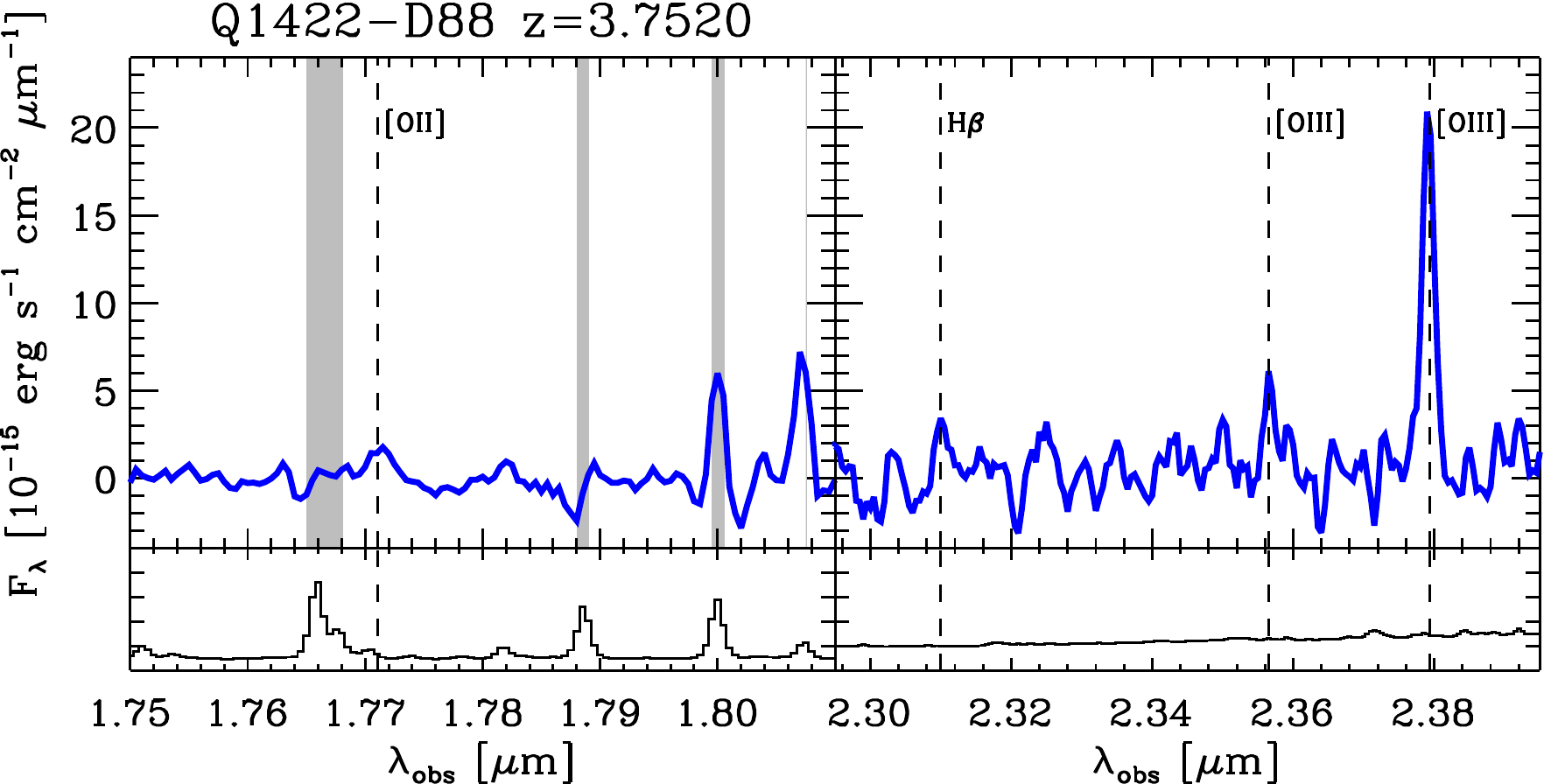}
  \includegraphics[width=0.49\linewidth]{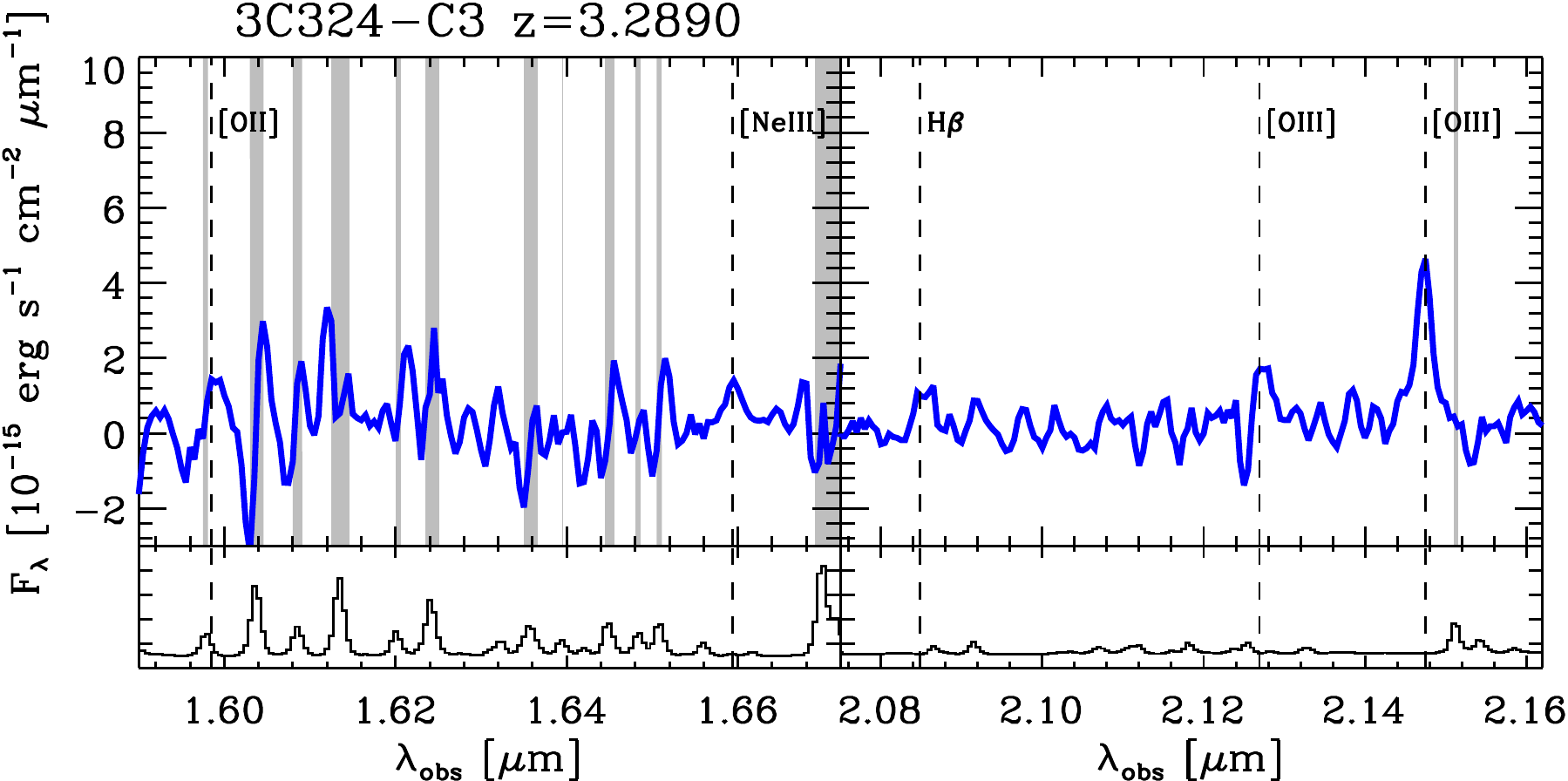}  
  \includegraphics[width=0.49\linewidth]{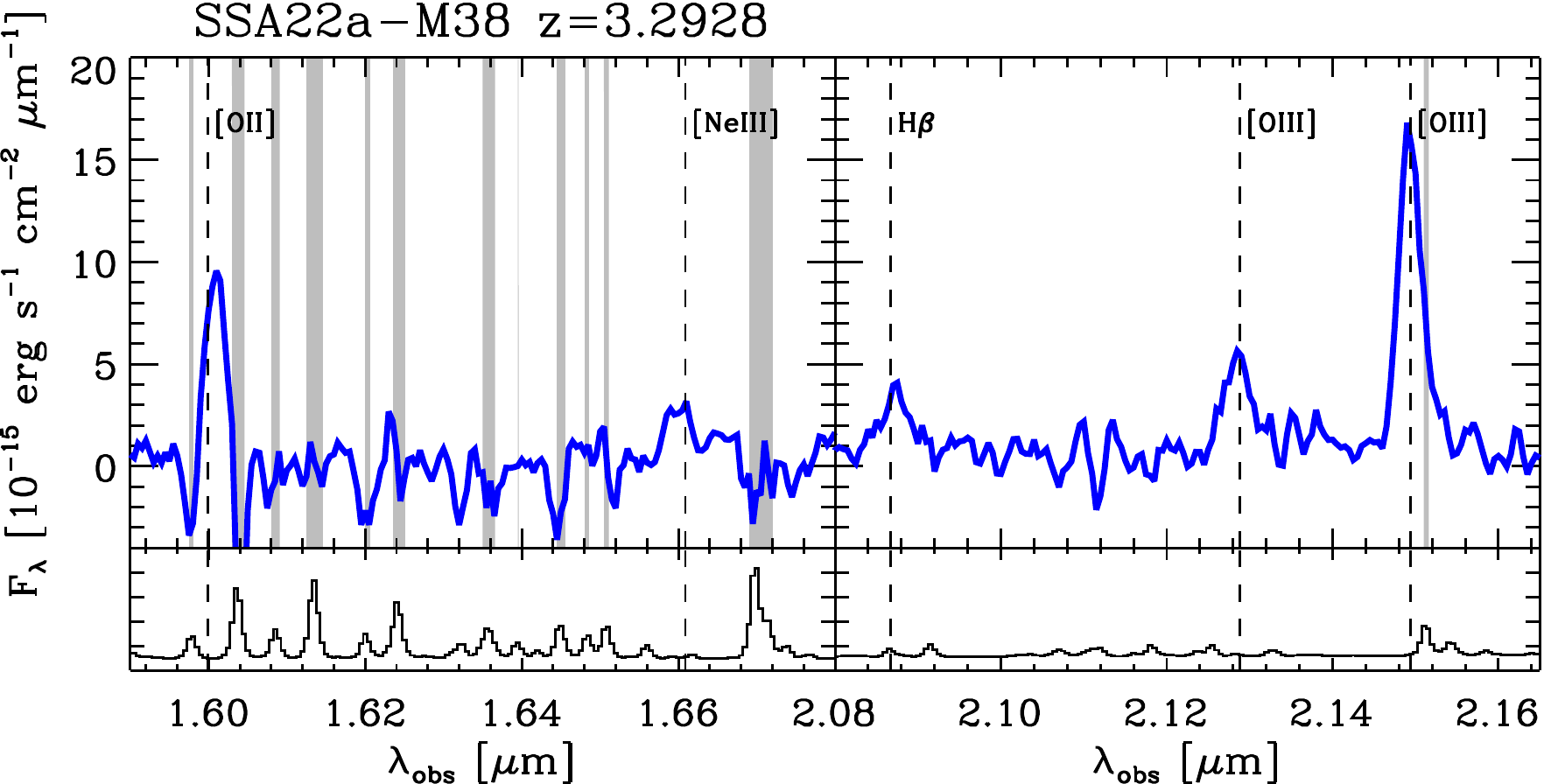}
  \includegraphics[width=0.49\linewidth]{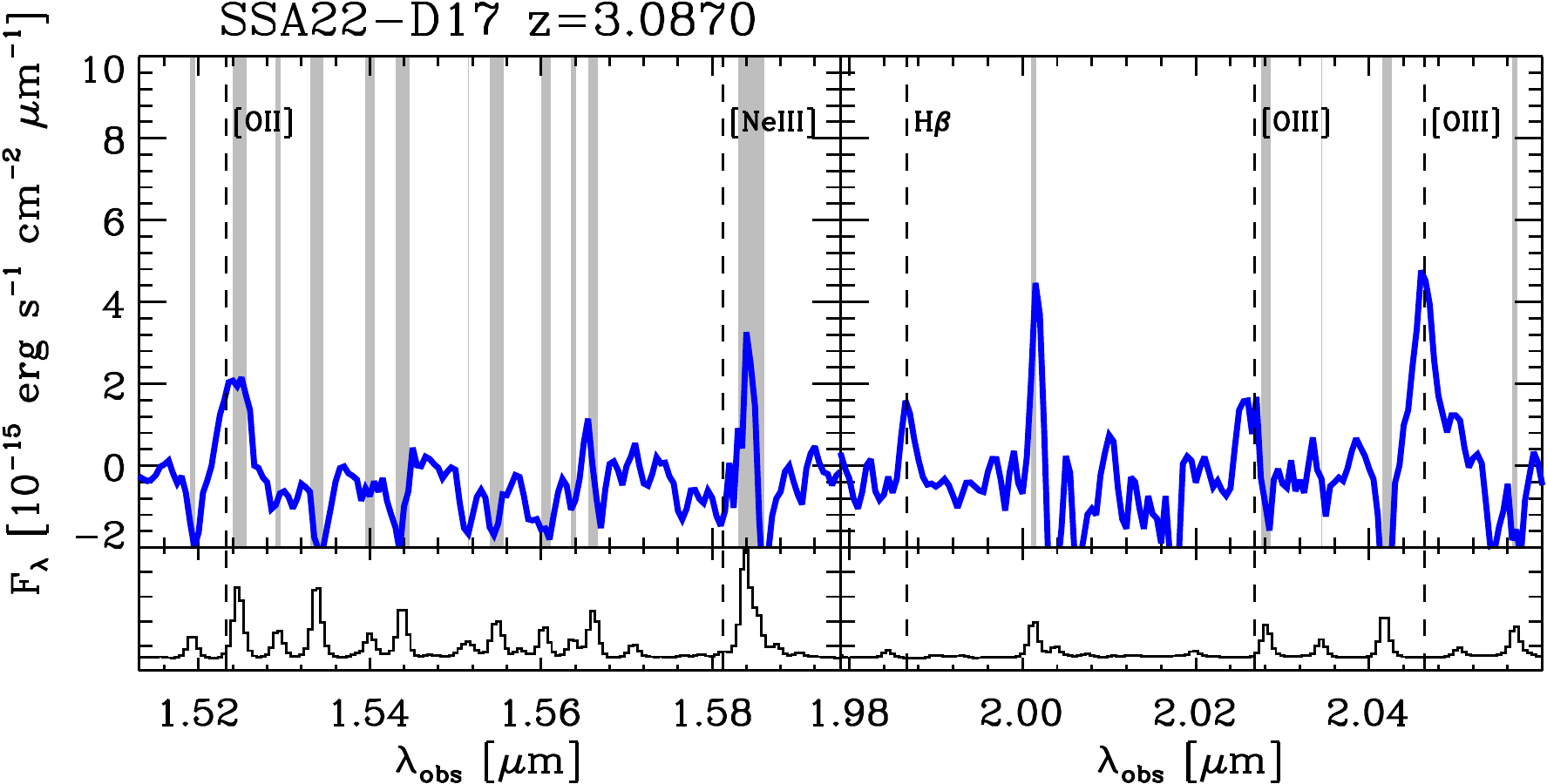}
  \includegraphics[width=0.49\linewidth]{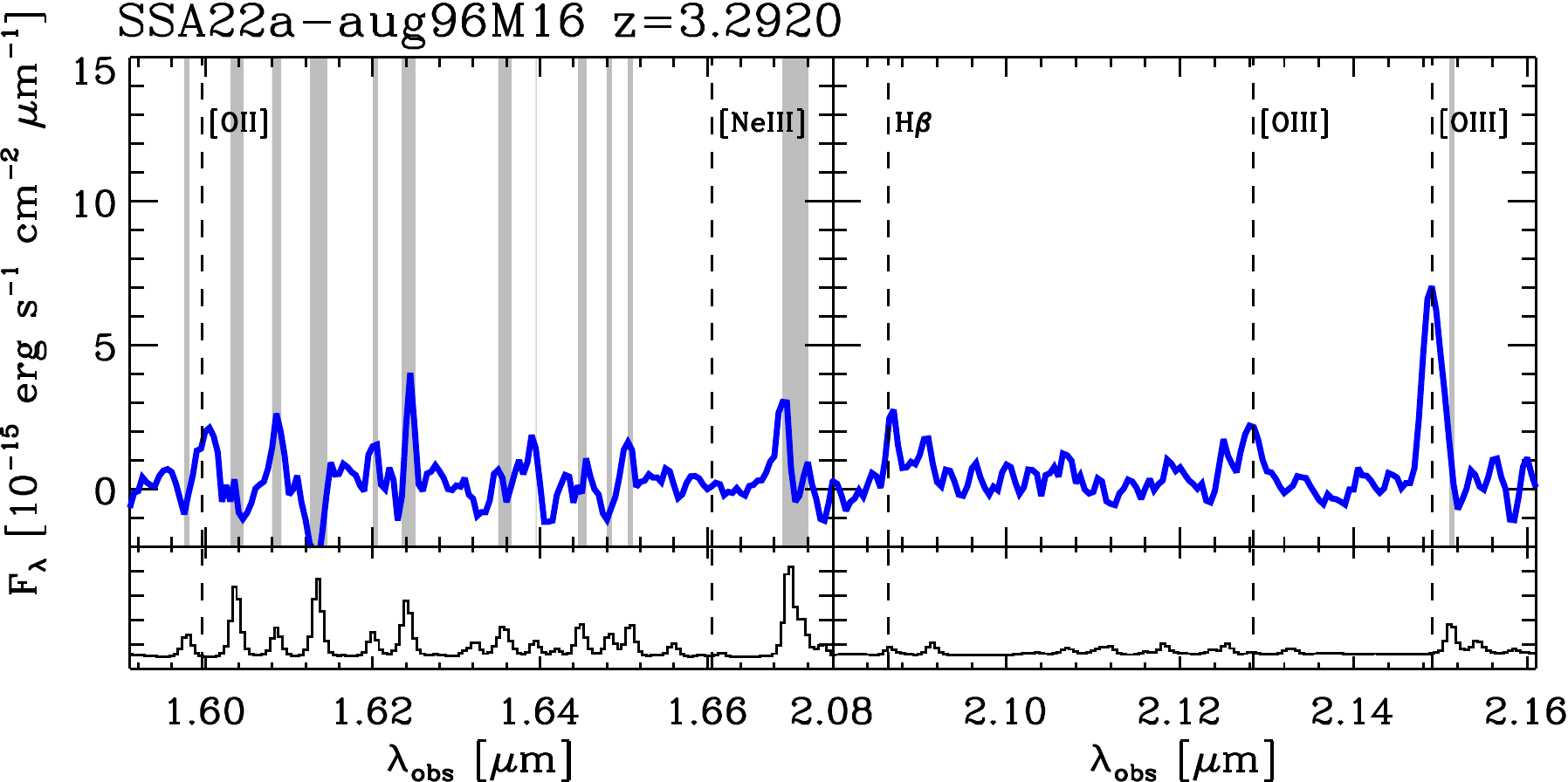}
  \includegraphics[width=0.49\linewidth]{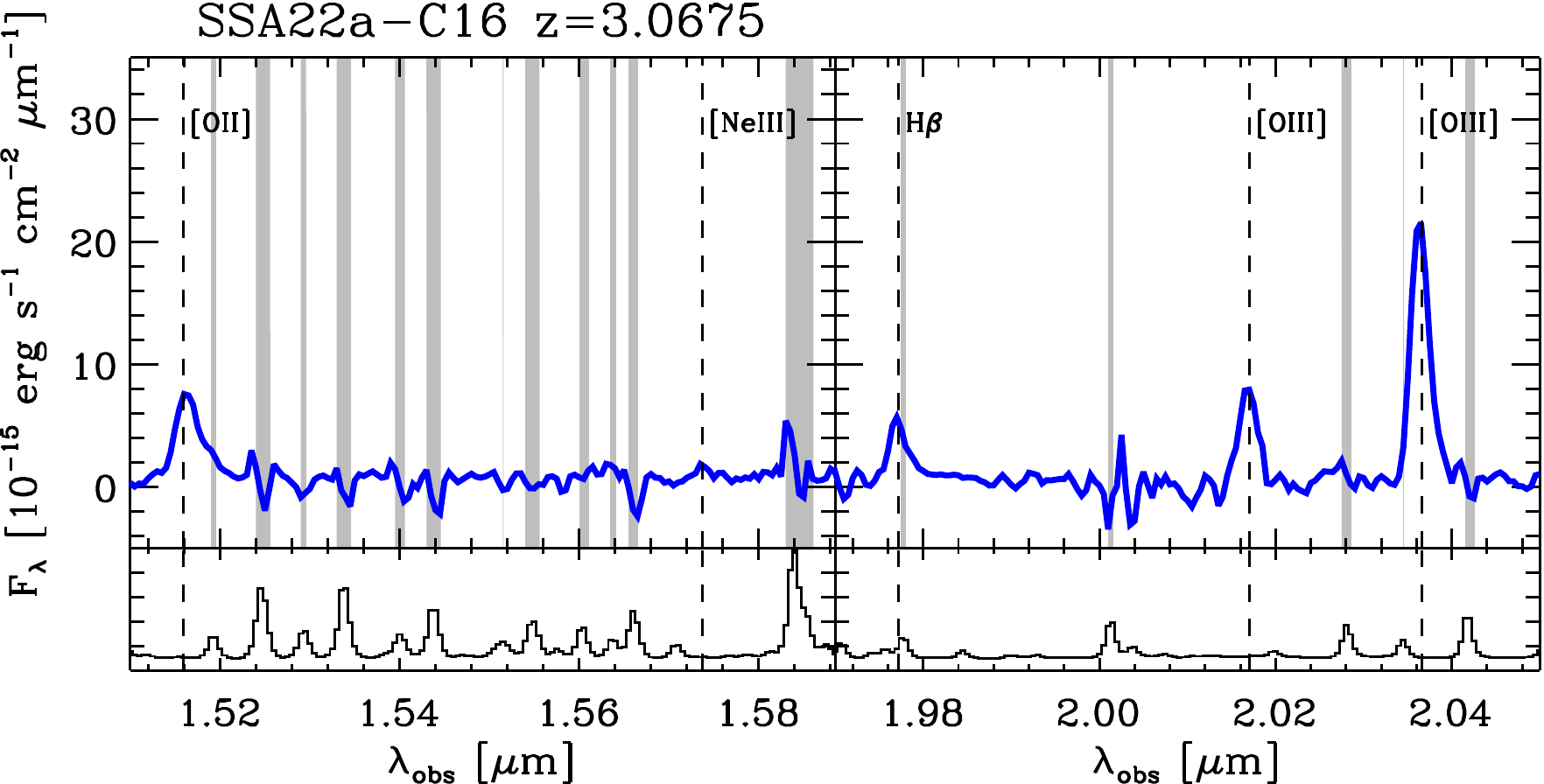}  
  \caption{Near-IR spectra of the galaxies in the AMAZE sample.
  The vertical dotted lines indicate the expected location of nebular emission lines.
  The shaded vertical regions overlaid on each spectrum highlight spectral regions affected by strong sky emission lines.}
\label{fig2sp}
  \end{figure*}

\begin{figure*}	  
  \centering
  \includegraphics[width=0.49\linewidth]{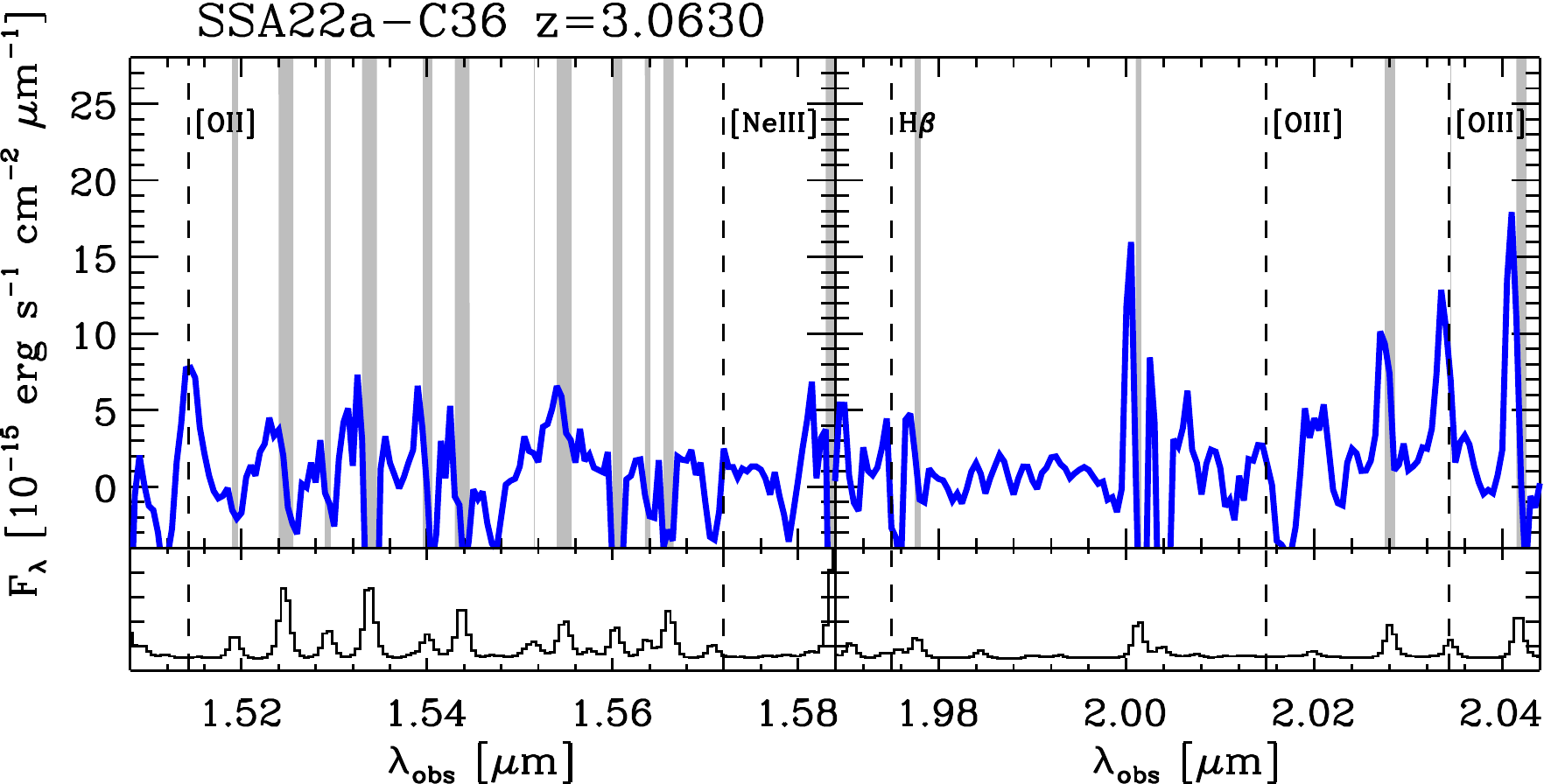}
   \includegraphics[width=0.49\linewidth]{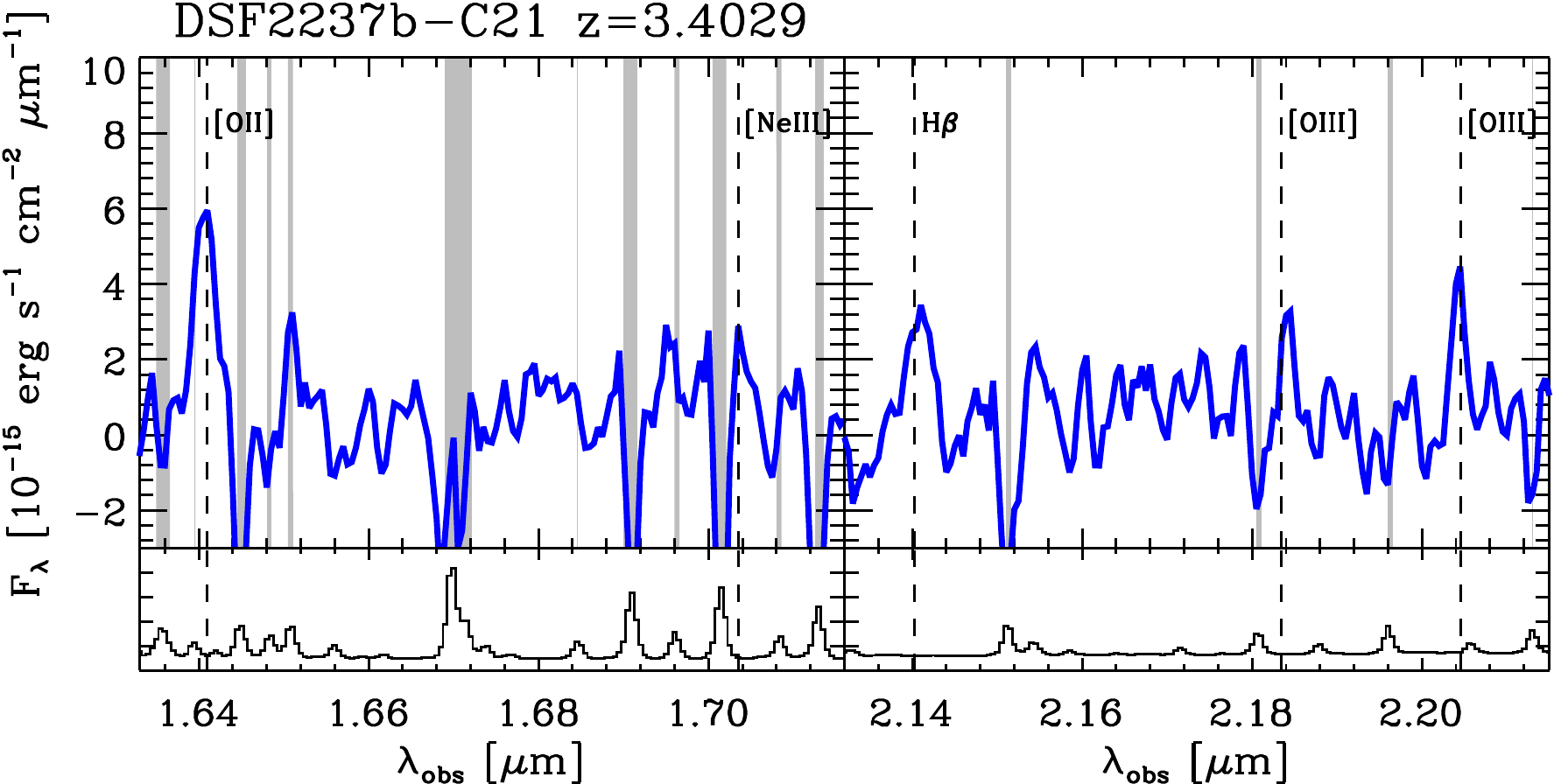}
  \includegraphics[width=0.49\linewidth]{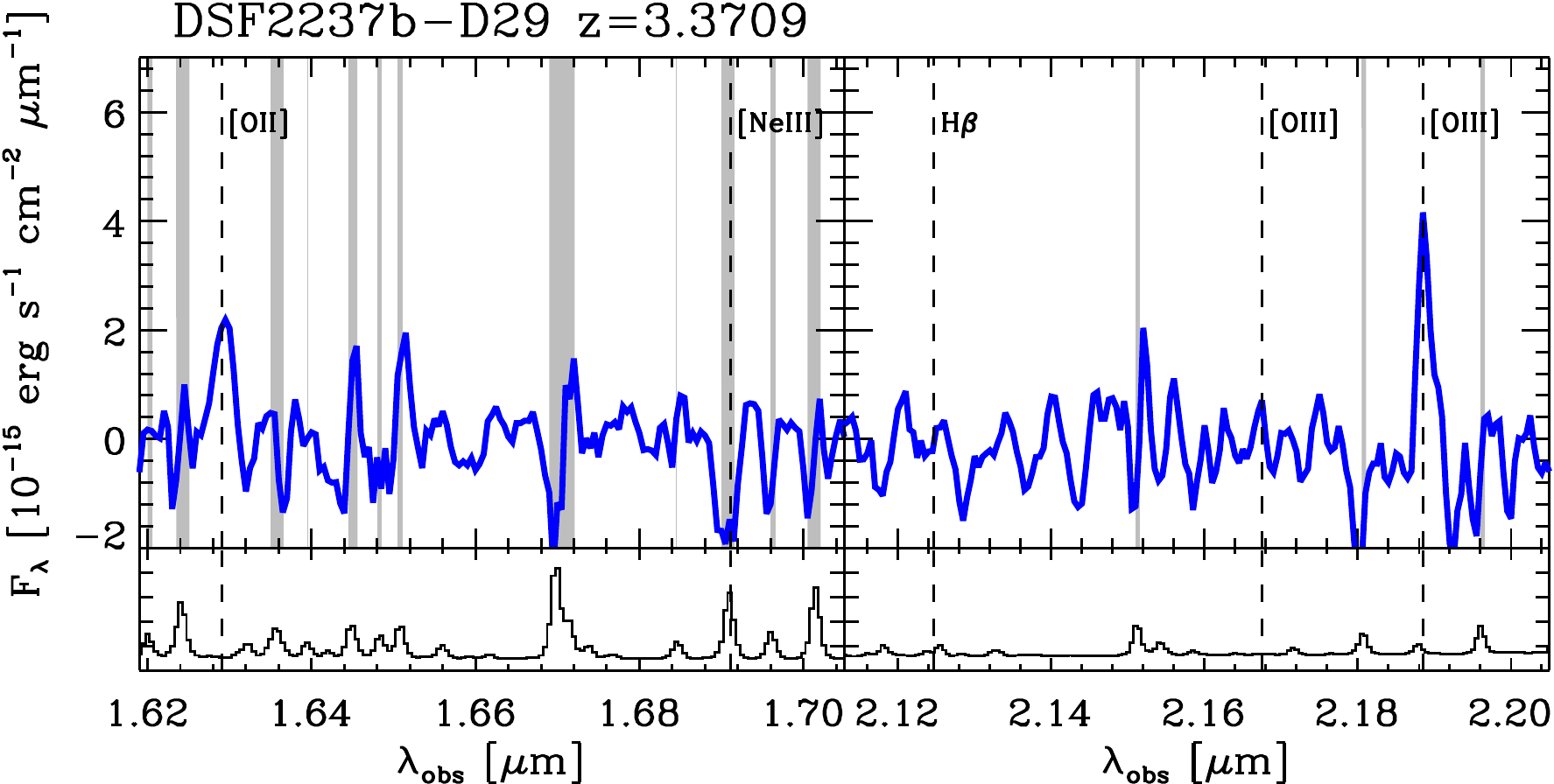}
  \caption{Near-IR spectra of the galaxies in the AMAZE sample.
  The vertical dotted lines indicate the expected location of nebular emission lines.
  The shaded vertical regions overlaid on each spectrum highlight spectral regions affected by strong sky emission lines.}
\label{fig3sp}
  \end{figure*}

\begin{figure*}
  \centering
  \includegraphics[width=0.49\linewidth]{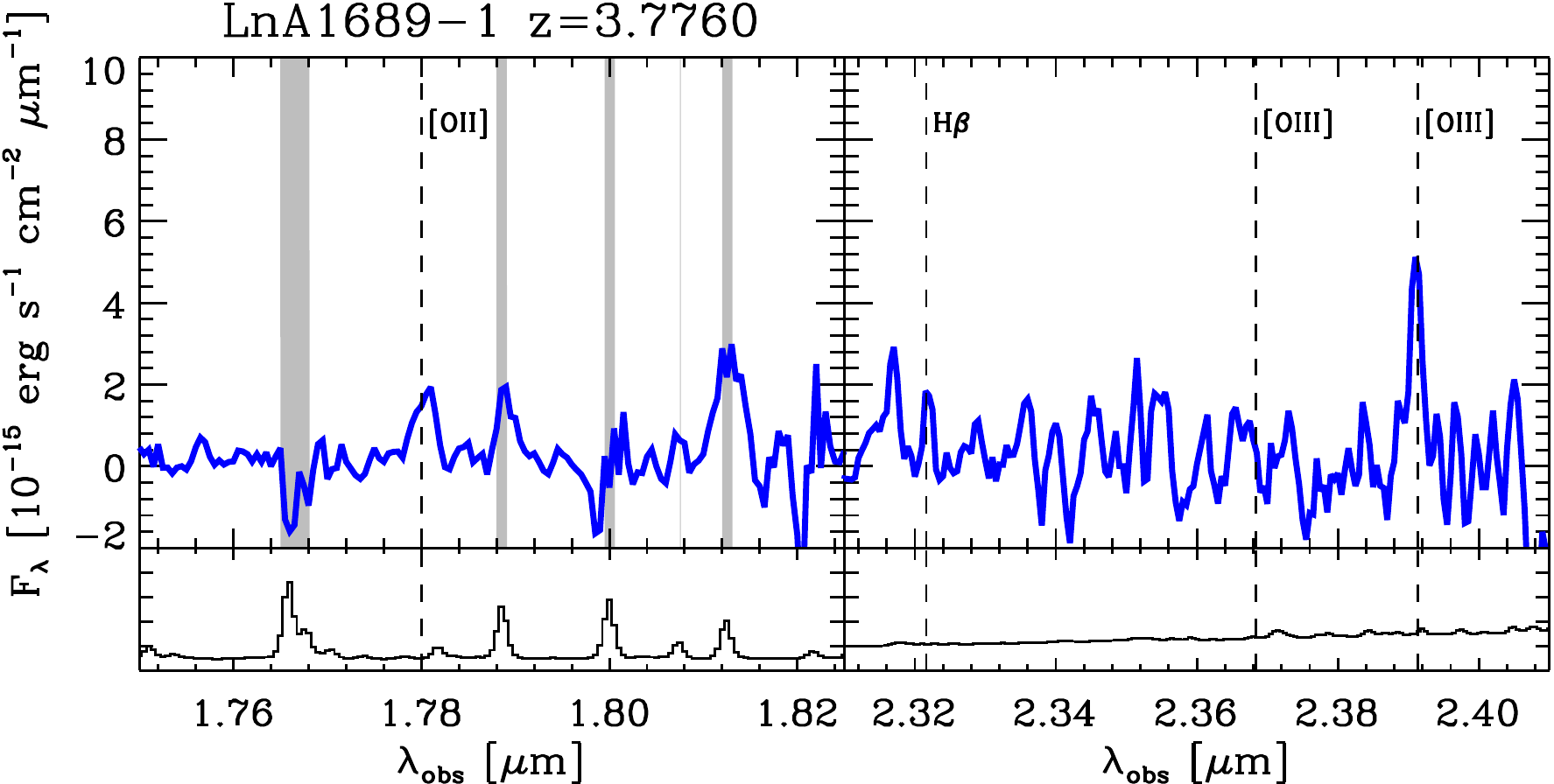}
  \includegraphics[width=0.49\linewidth]{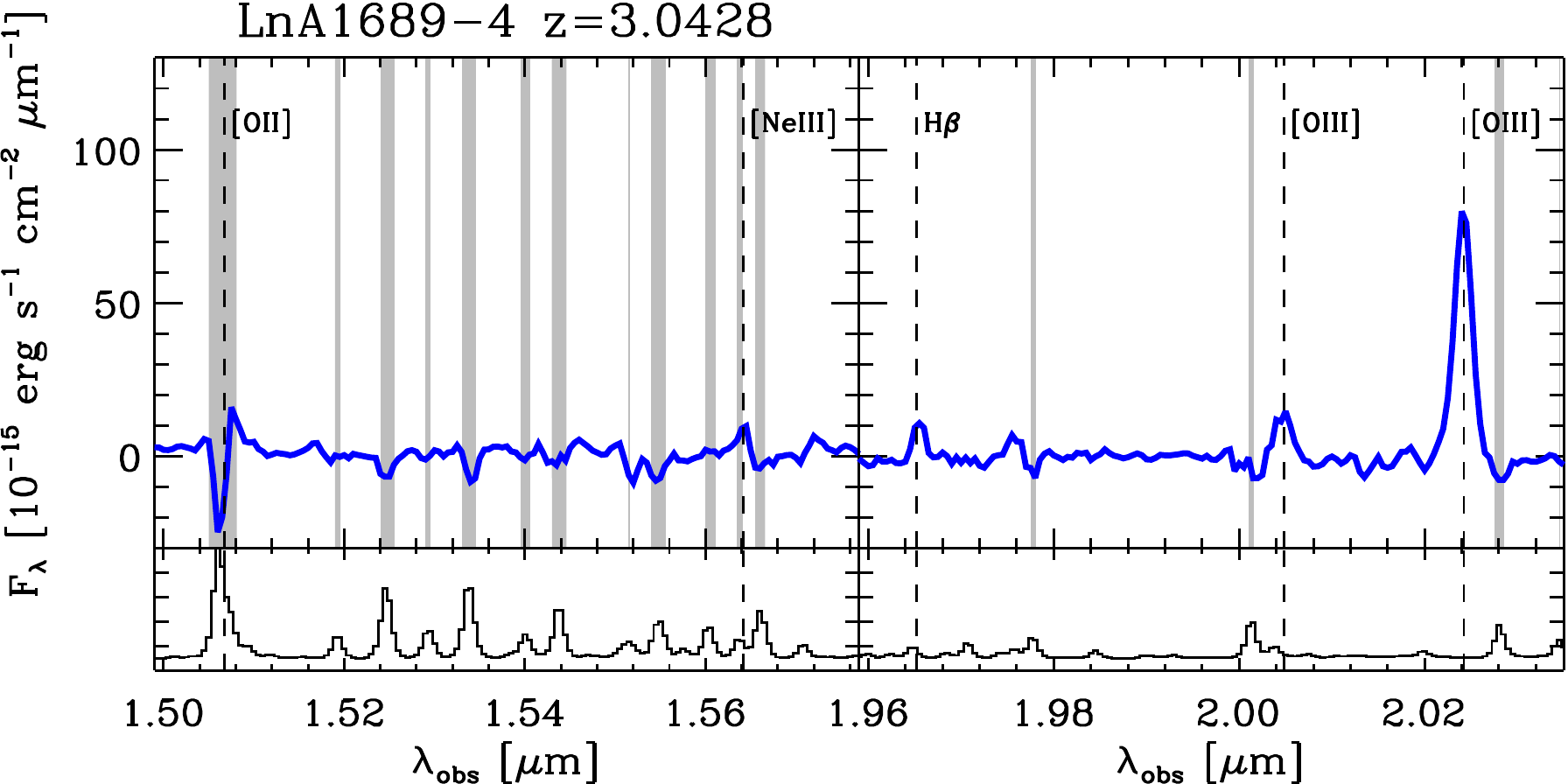}
  \includegraphics[width=0.49\linewidth]{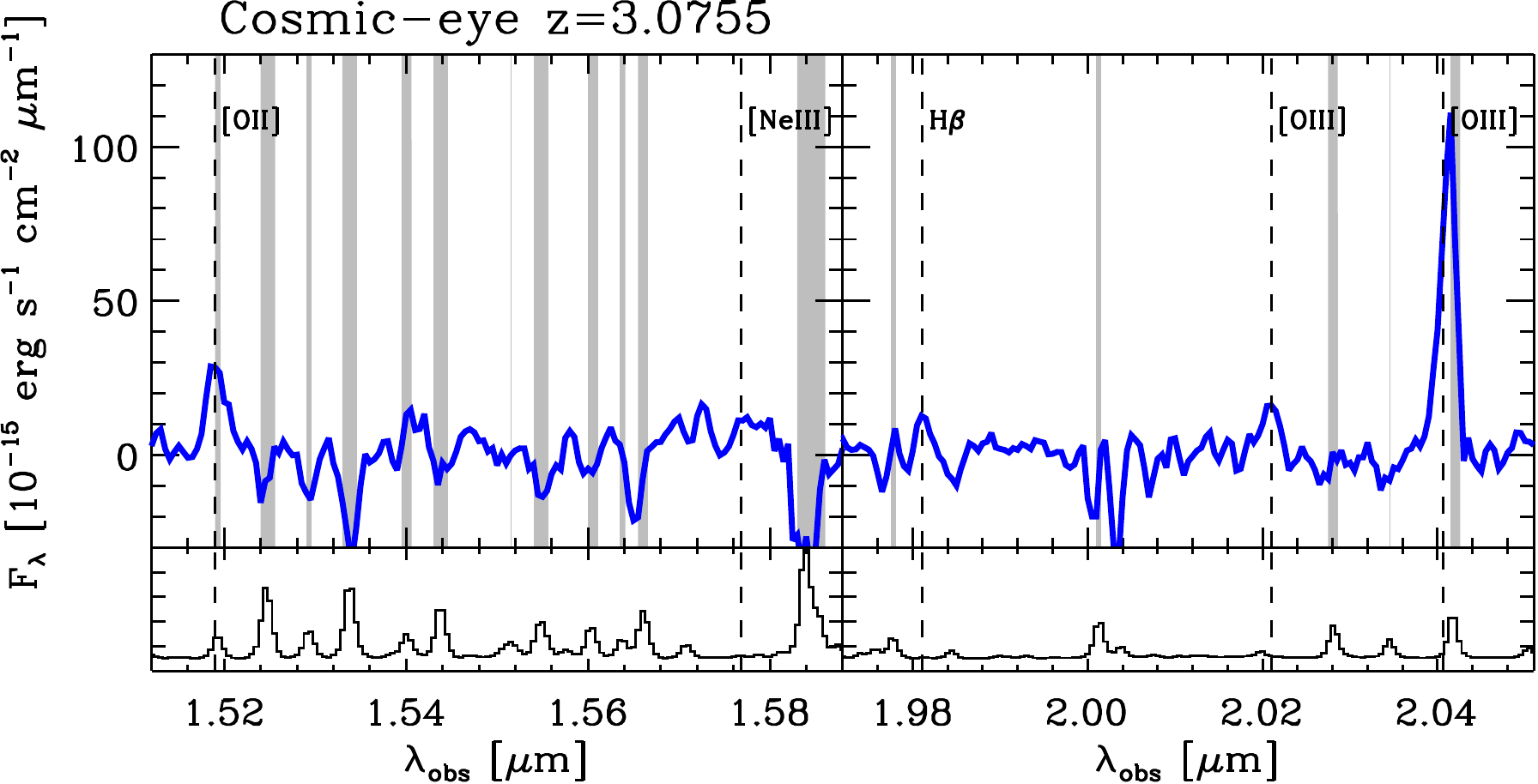}
  \includegraphics[width=0.27\linewidth]{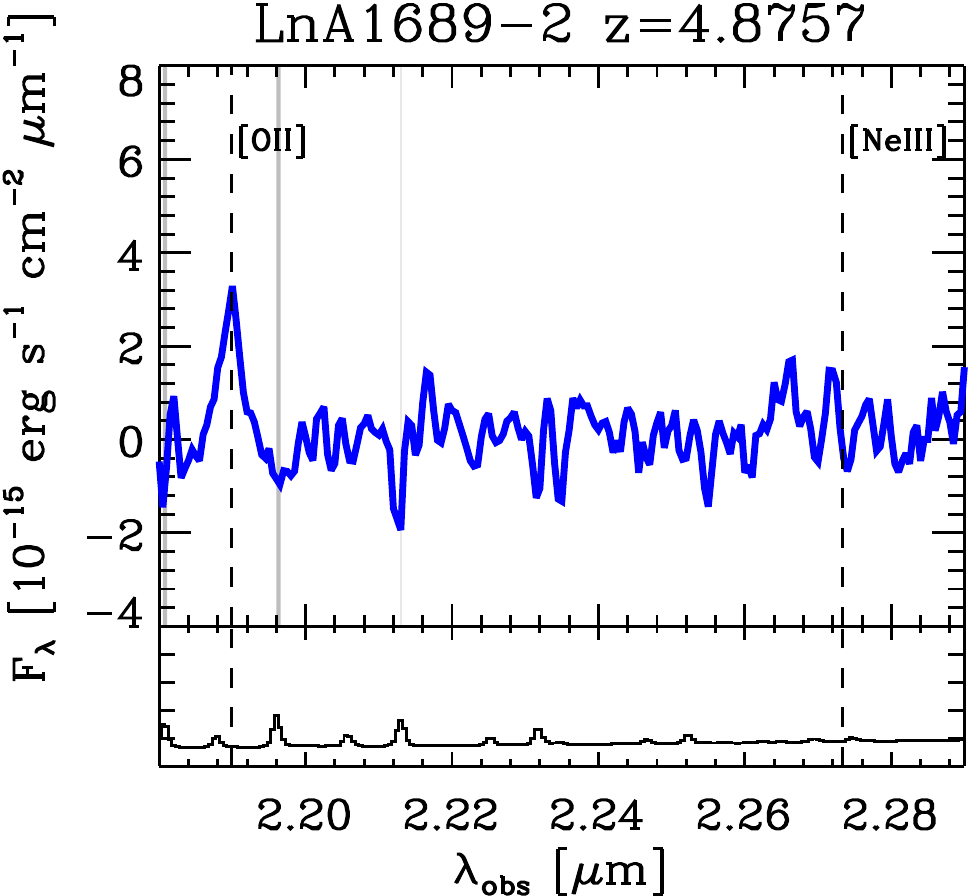}
  \caption{Near-IR spectra of the lensed galaxies in the AMAZE sample.
  The vertical dotted lines indicate the expected location of nebular emission lines.
  The shaded vertical regions overlaid on each spectrum highlight spectral regions affected by strong sky emission lines.}
\label{figlens}
  \end{figure*}

Figs.\ref{fig1sp}--\ref{figlens} show the integrated spectra of all galaxies 
in the AMAZE and LSD samples,
restricted to the spectral regions that cover the nebular lines used to measure the metallicities.

      \begin{figure*}	  
  \centering
  \includegraphics[width=0.49\linewidth]{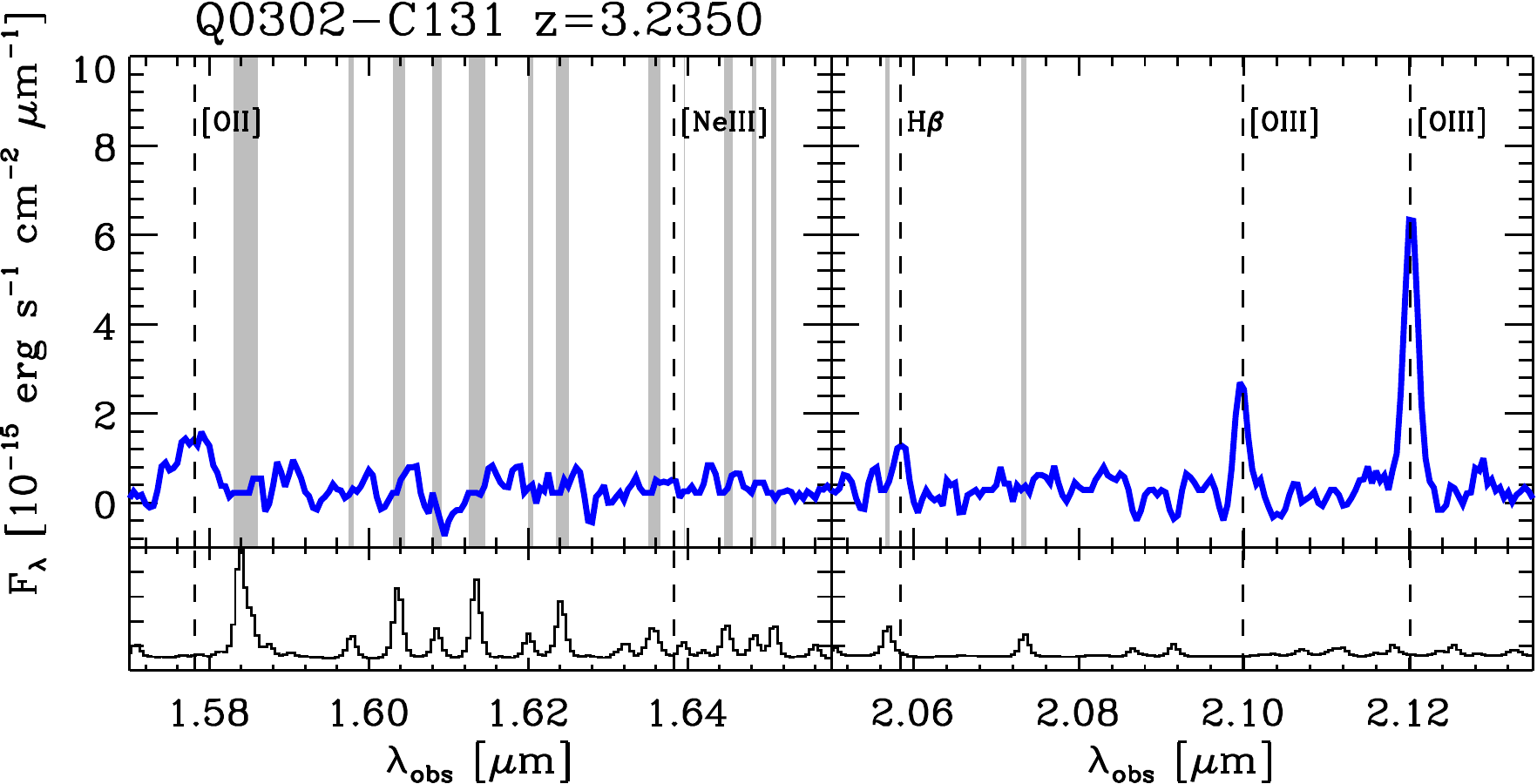}
  \includegraphics[width=0.49\linewidth]{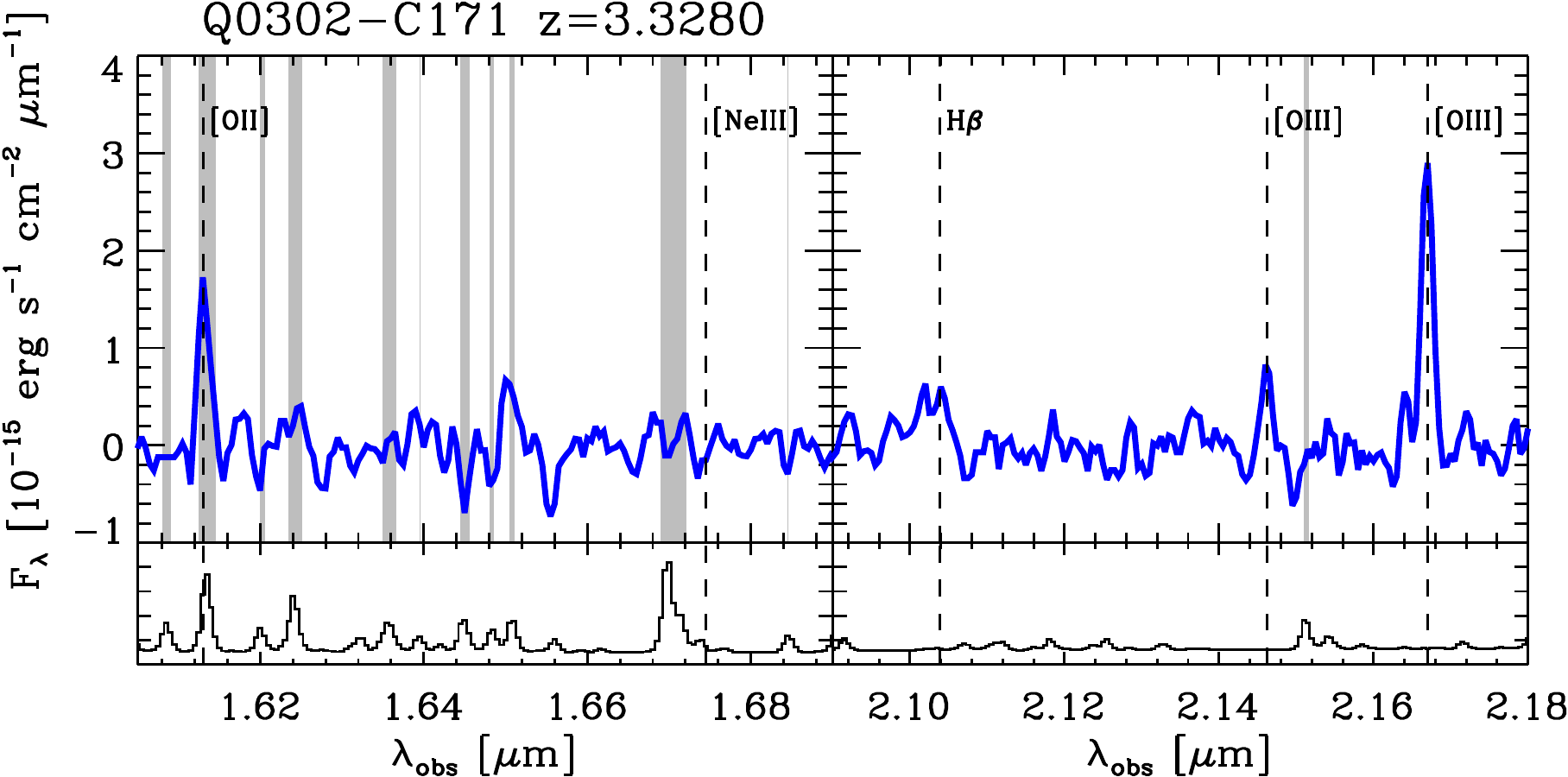}
  \includegraphics[width=0.49\linewidth]{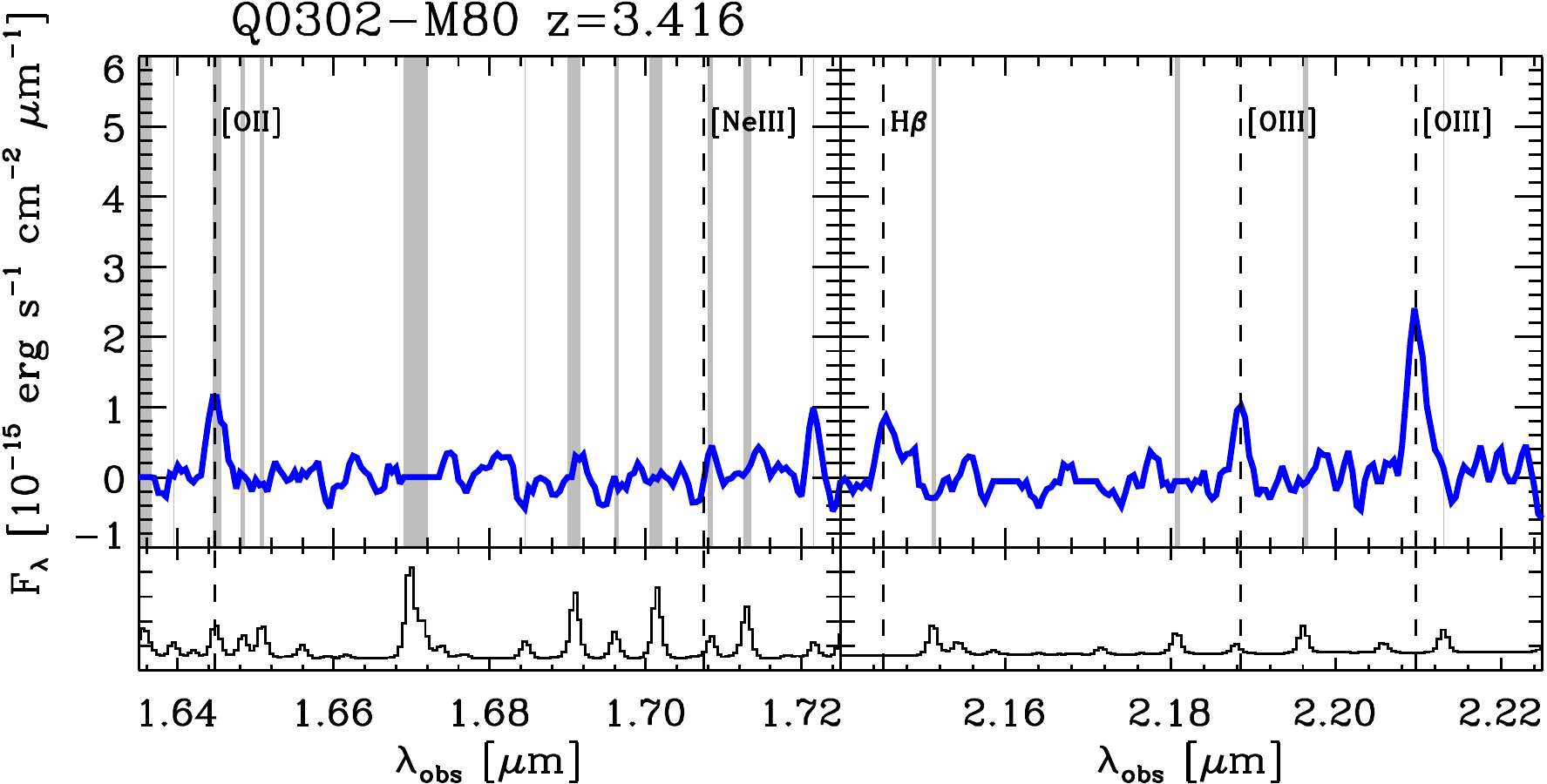}
  \includegraphics[width=0.49\linewidth]{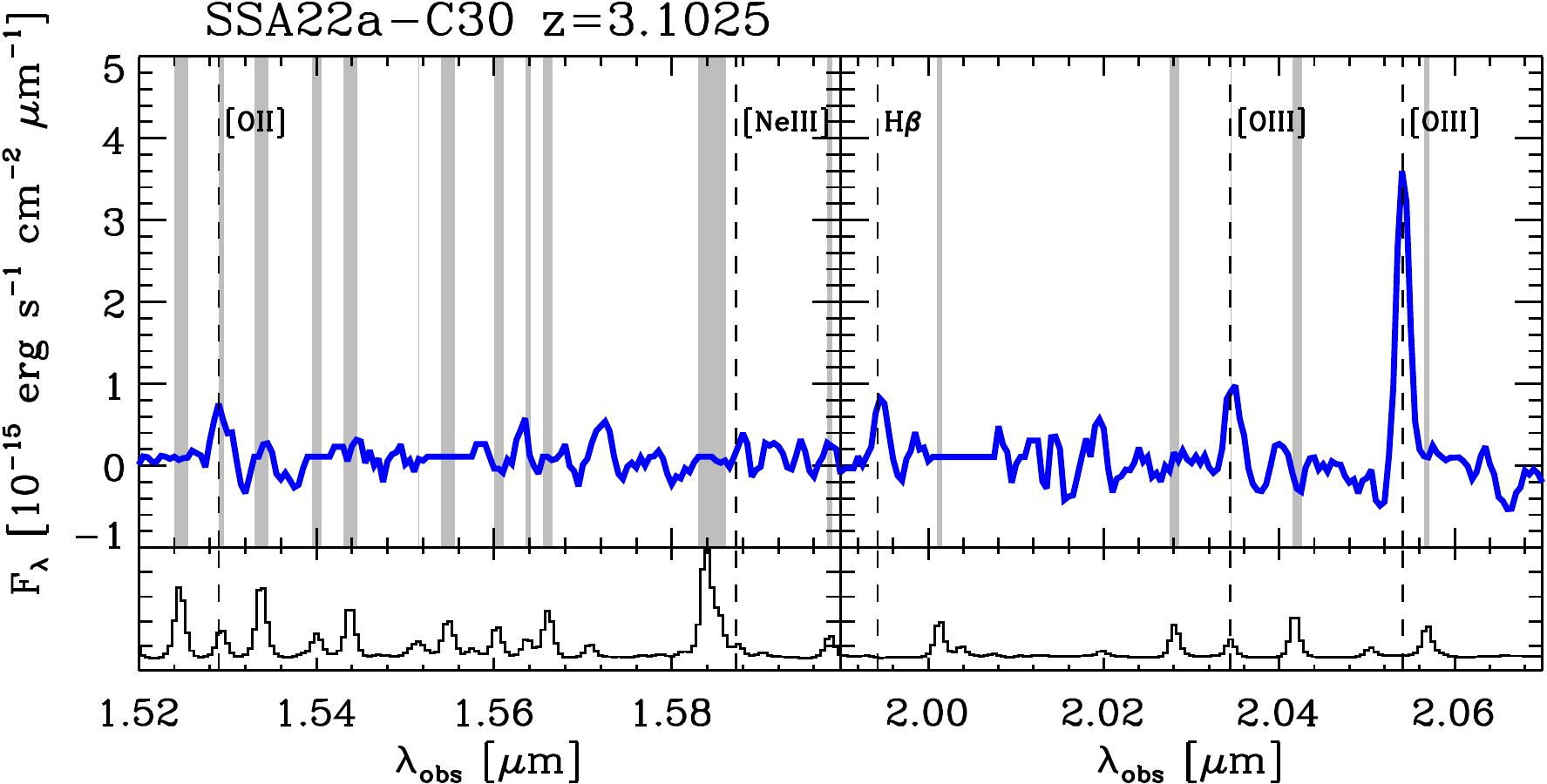}
  \includegraphics[width=0.49\linewidth]{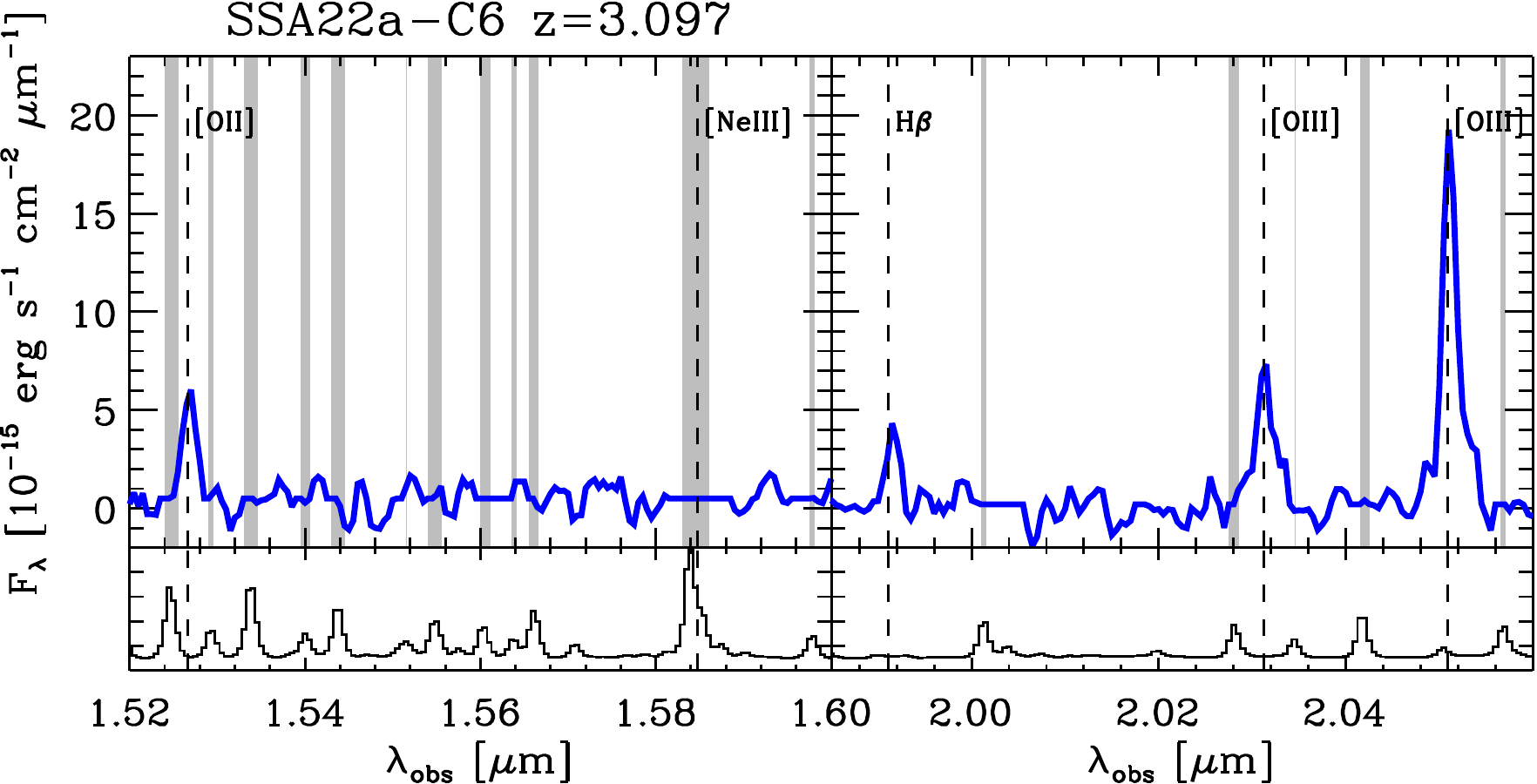}
  \includegraphics[width=0.49\linewidth]{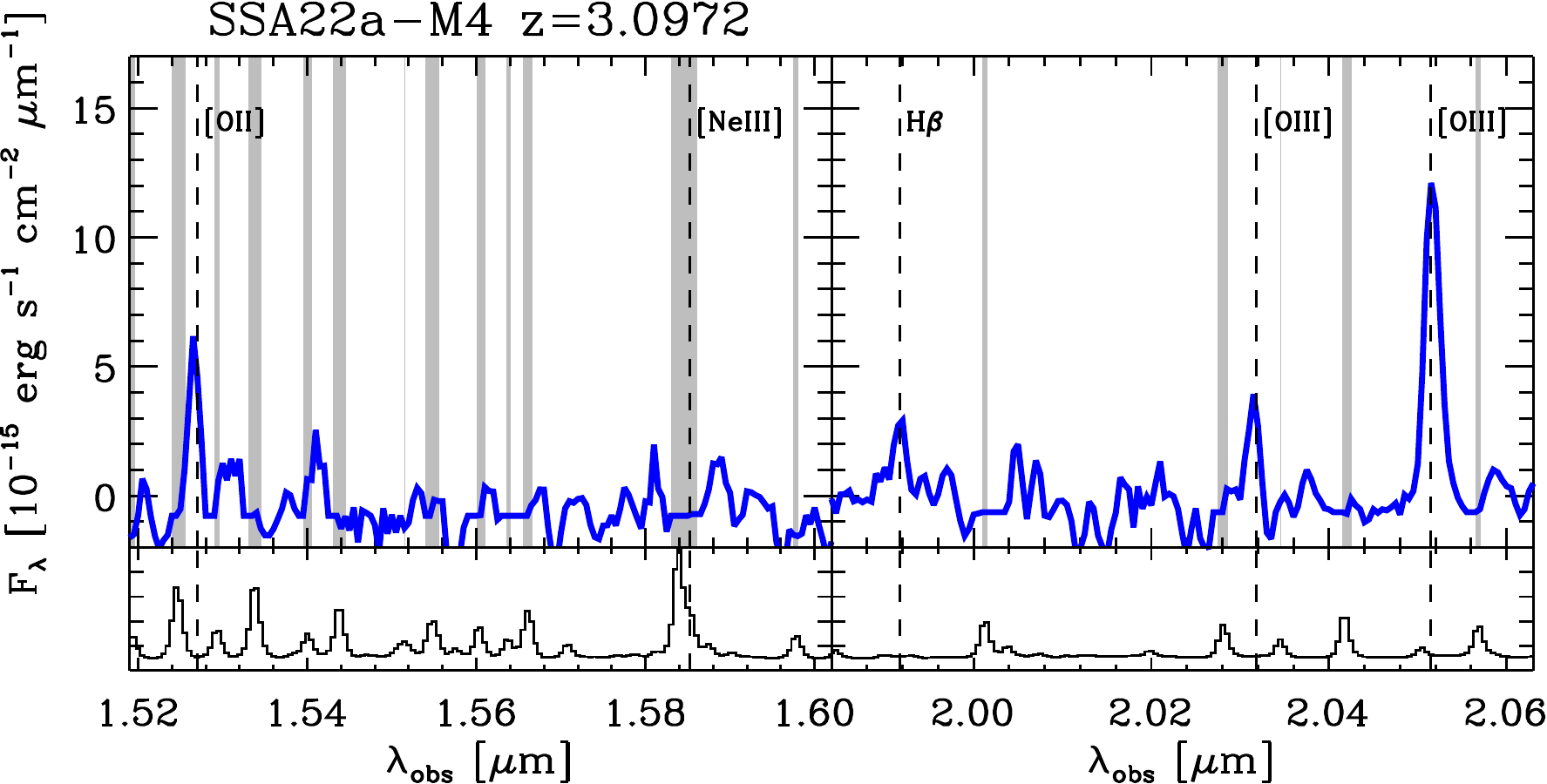}
  \includegraphics[width=0.49\linewidth]{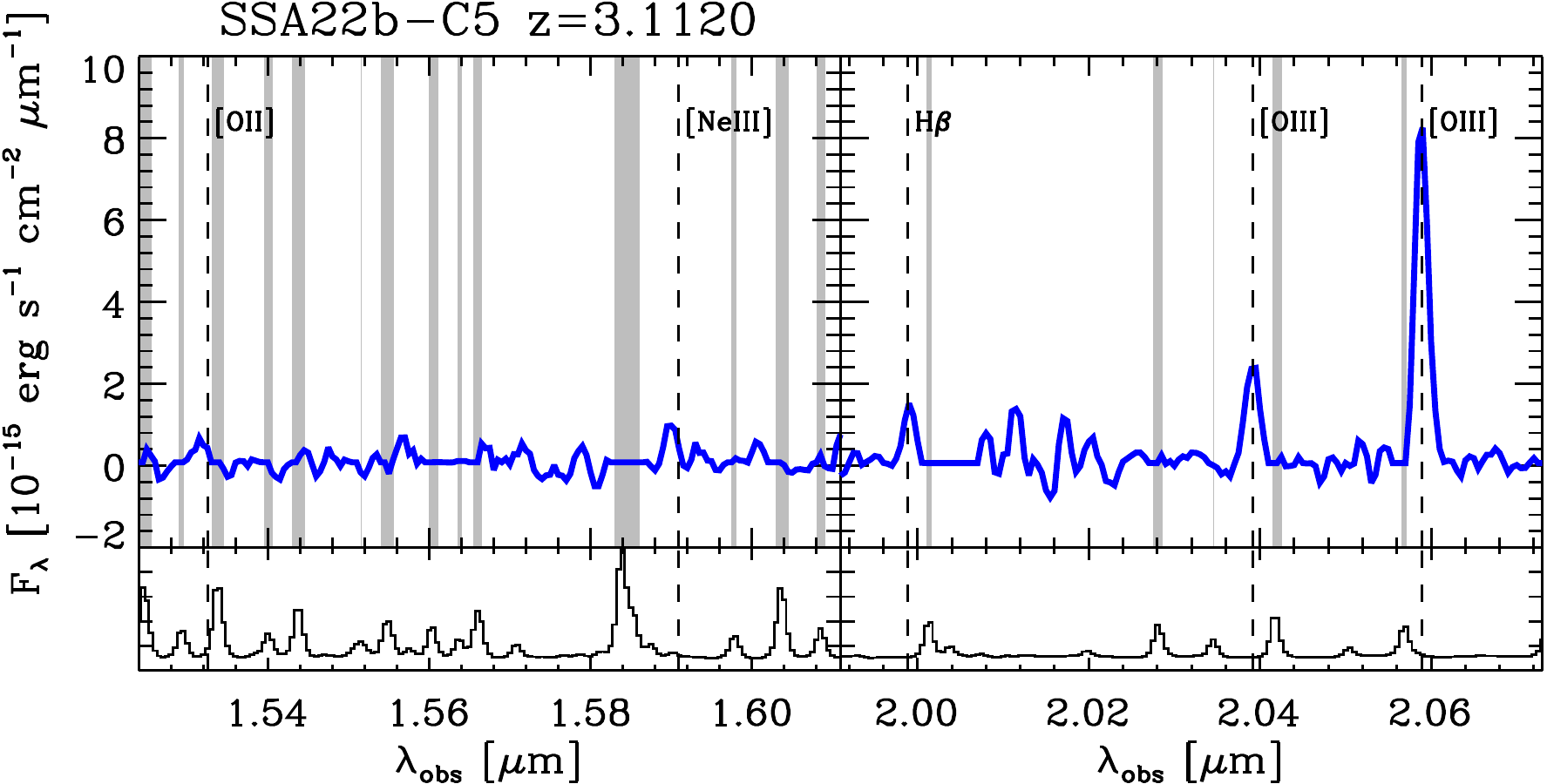}
  \includegraphics[width=0.49\linewidth]{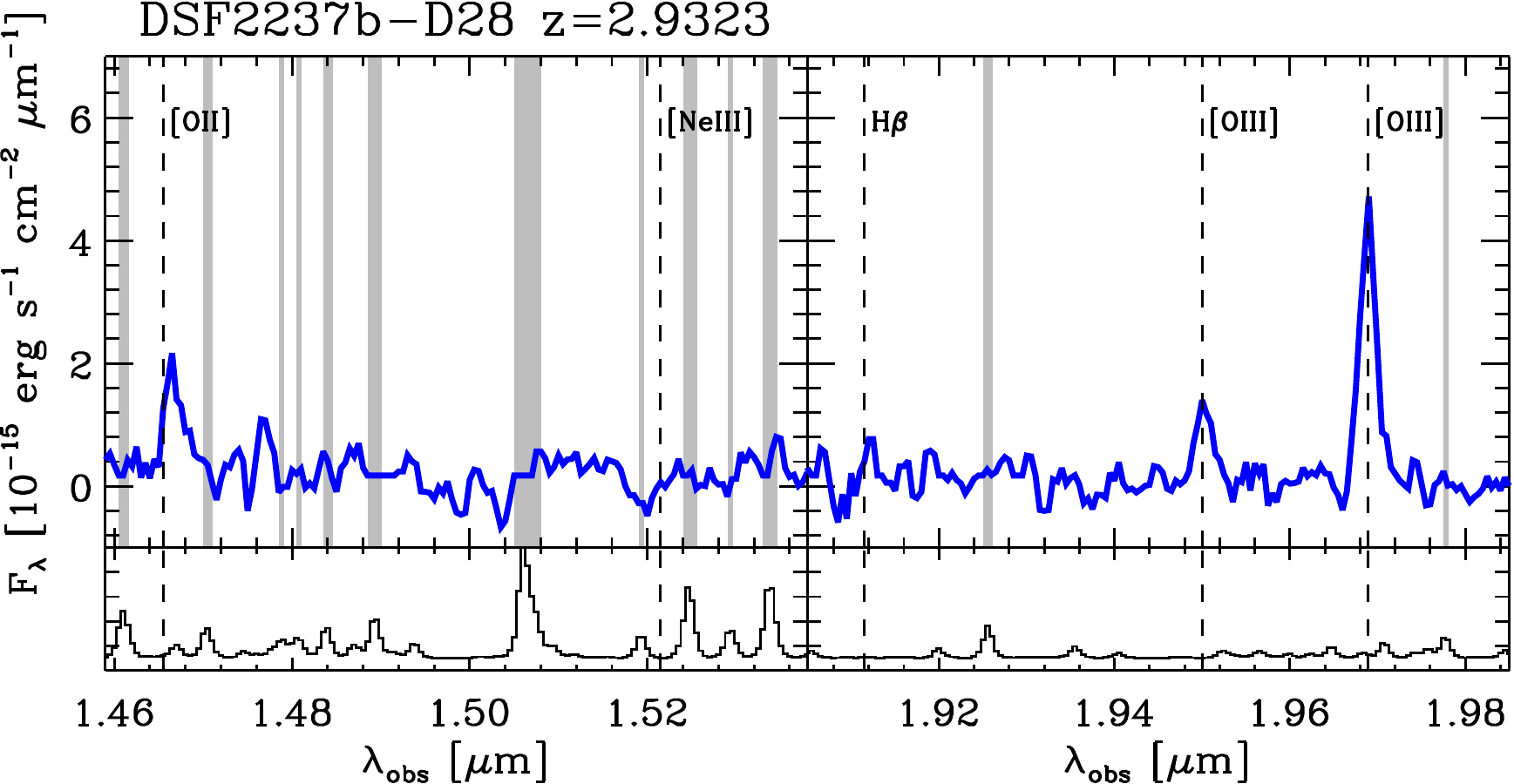}
  \caption{Near-IR spectra of the galaxies in the LSD sample.
  The vertical dotted lines indicate the expected location of nebular emission lines.
  The shaded vertical regions overlaid on each spectrum highlight spectral regions affected by strong sky emission lines.}
\label{fig4sp}
  \end{figure*}

\section{Metallicity measurements}

\begin{figure*}
  \centering
  \includegraphics[width=0.49\linewidth]{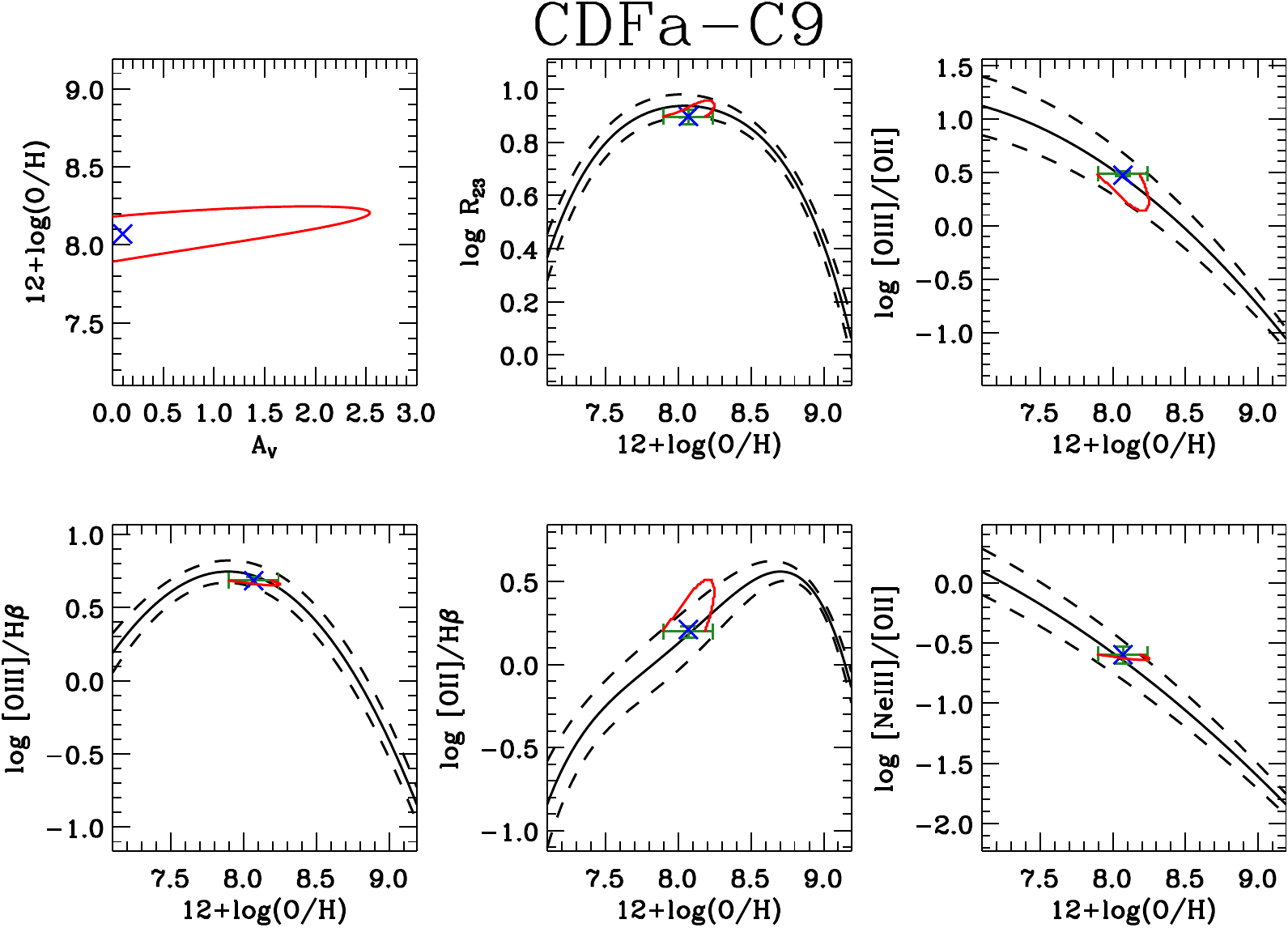}
  \includegraphics[width=0.49\linewidth]{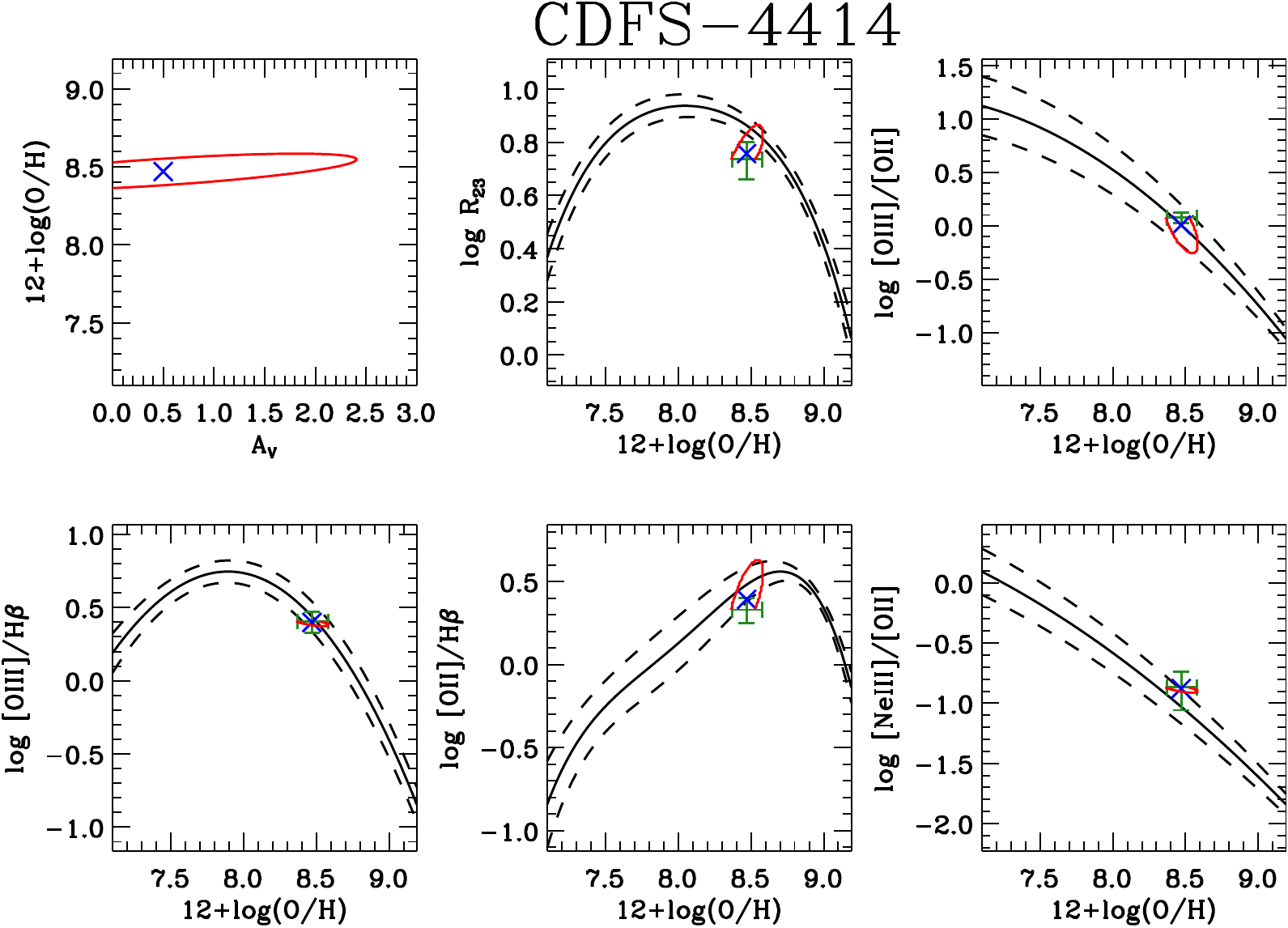}
  \includegraphics[width=0.49\linewidth]{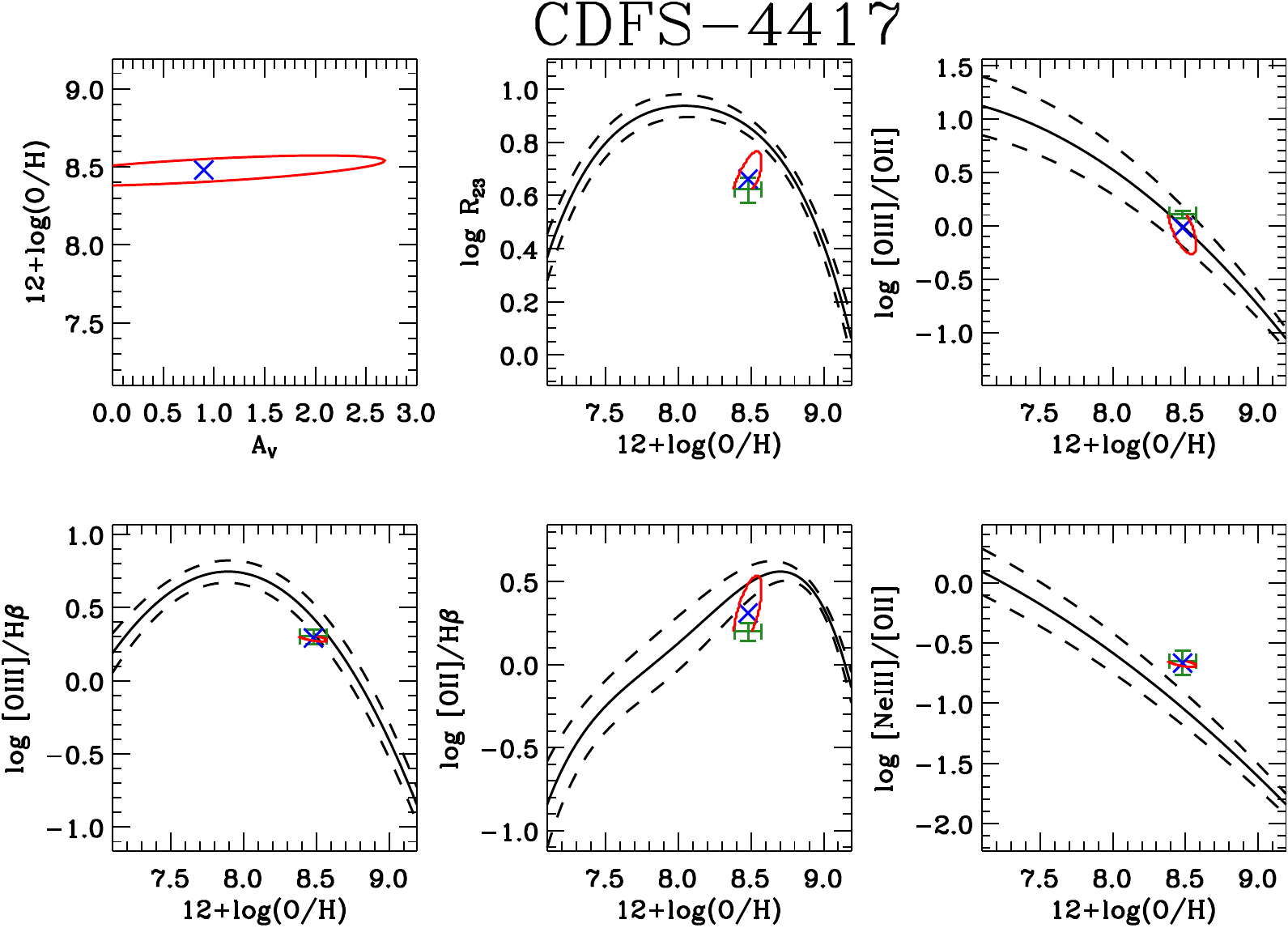}
  \includegraphics[width=0.49\linewidth]{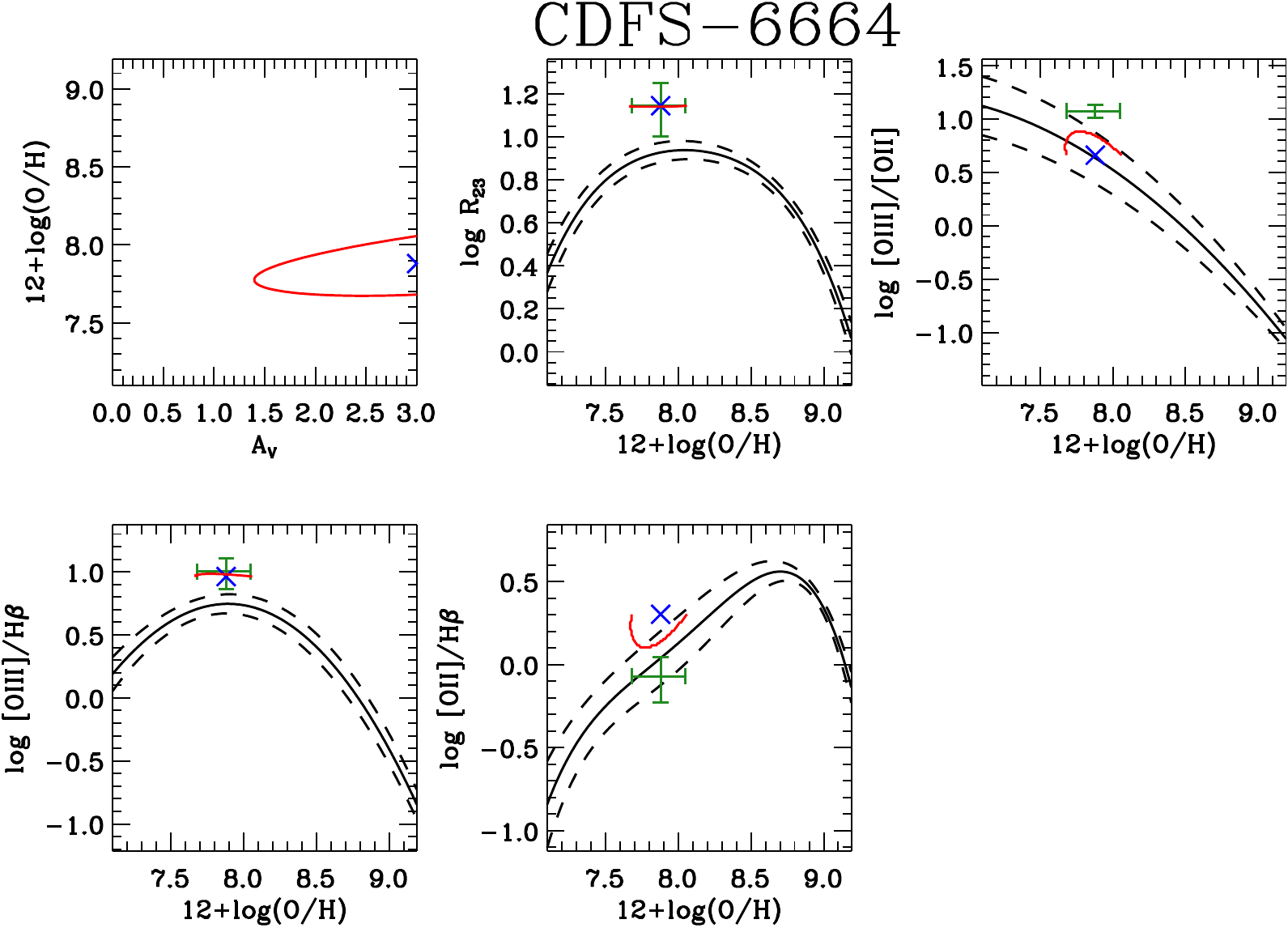}
  \includegraphics[width=0.49\linewidth]{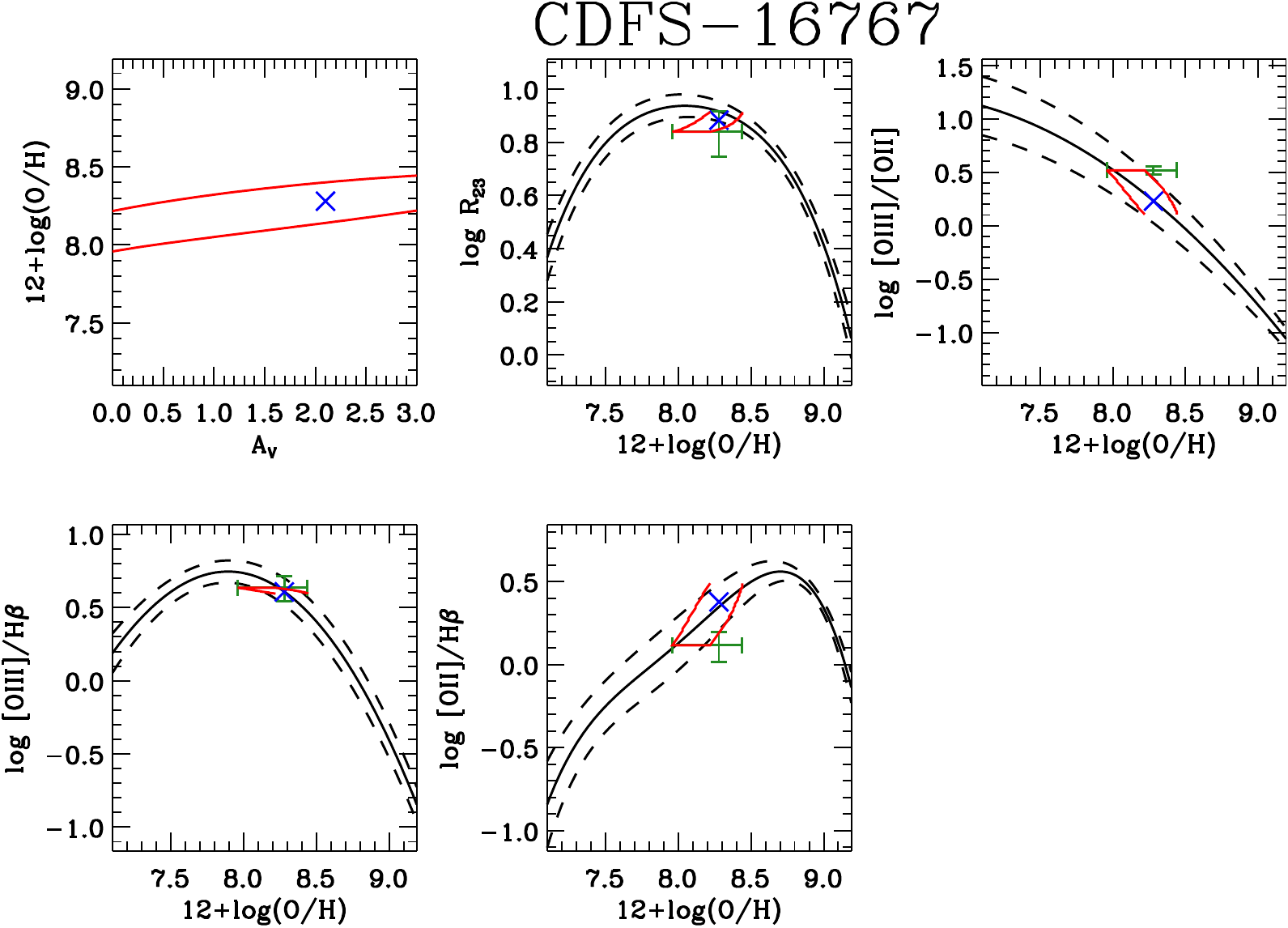}
  \includegraphics[width=0.49\linewidth]{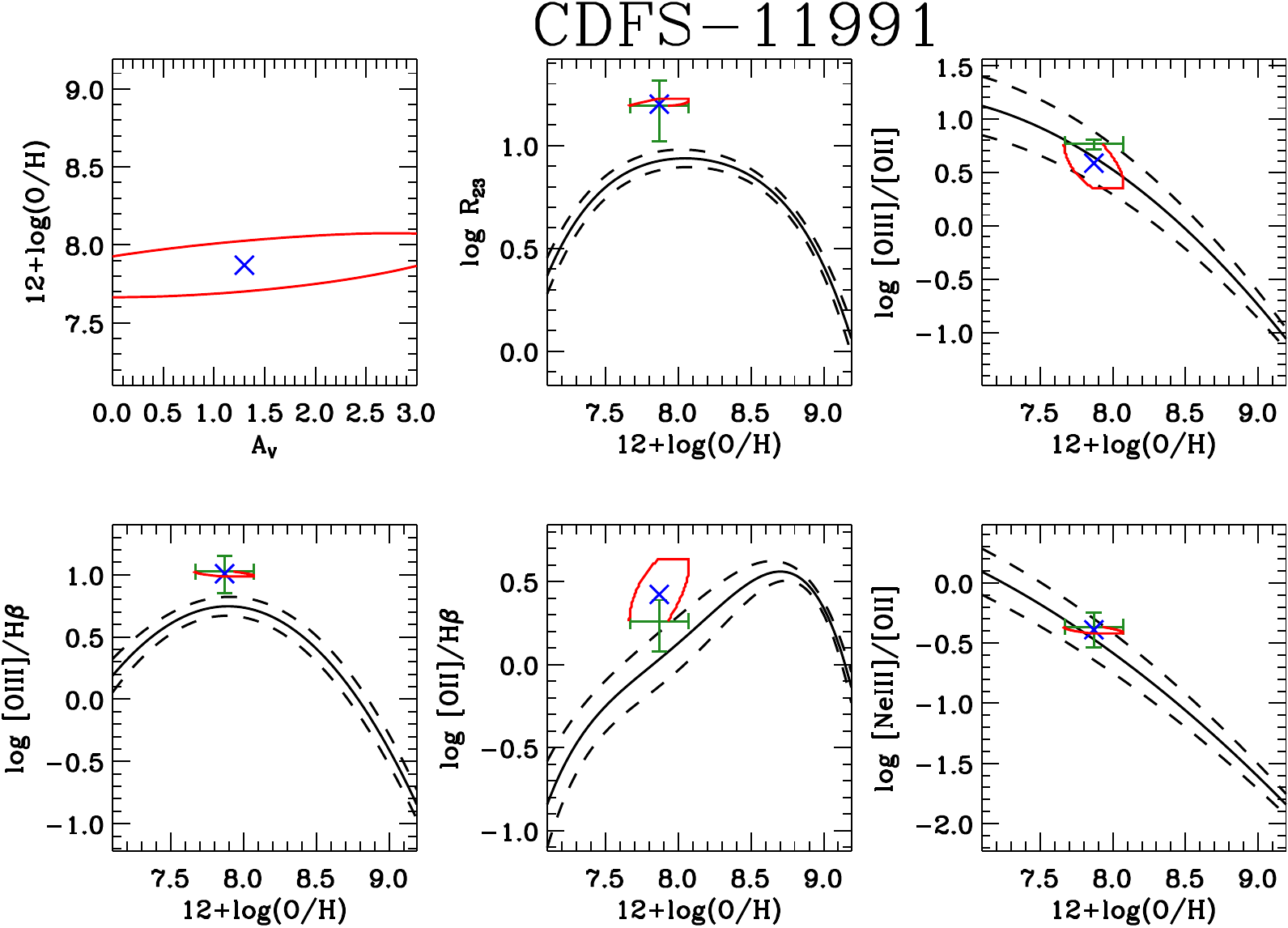}
 \caption{Diagnostic tools used to determine the metallicity of galaxies in the AMAZE sample.
In each plot, the upper left panel shows the best solution (blue cross) and the 1$\sigma$ confidence level in the AV-metallicity plane.
In the other panels the black solid line (best fit) and the dashed lines (dispersion)
show the empirical relations between various line ratios and the gas metallicity.
The green error bars show the observed ratios (along the y-axis) and the best-fit
metallicity with uncertainty (along the x-axis); the blue cross shows the de-reddened ratios
from adopting the best-fit extinction; 
the red line shows the projection of the 1$\sigma$ uncertainty of the fit in the top-left panel.
}
\label{figmet1}
  \end{figure*}

\begin{figure*}
  \centering
  \includegraphics[width=0.49\linewidth]{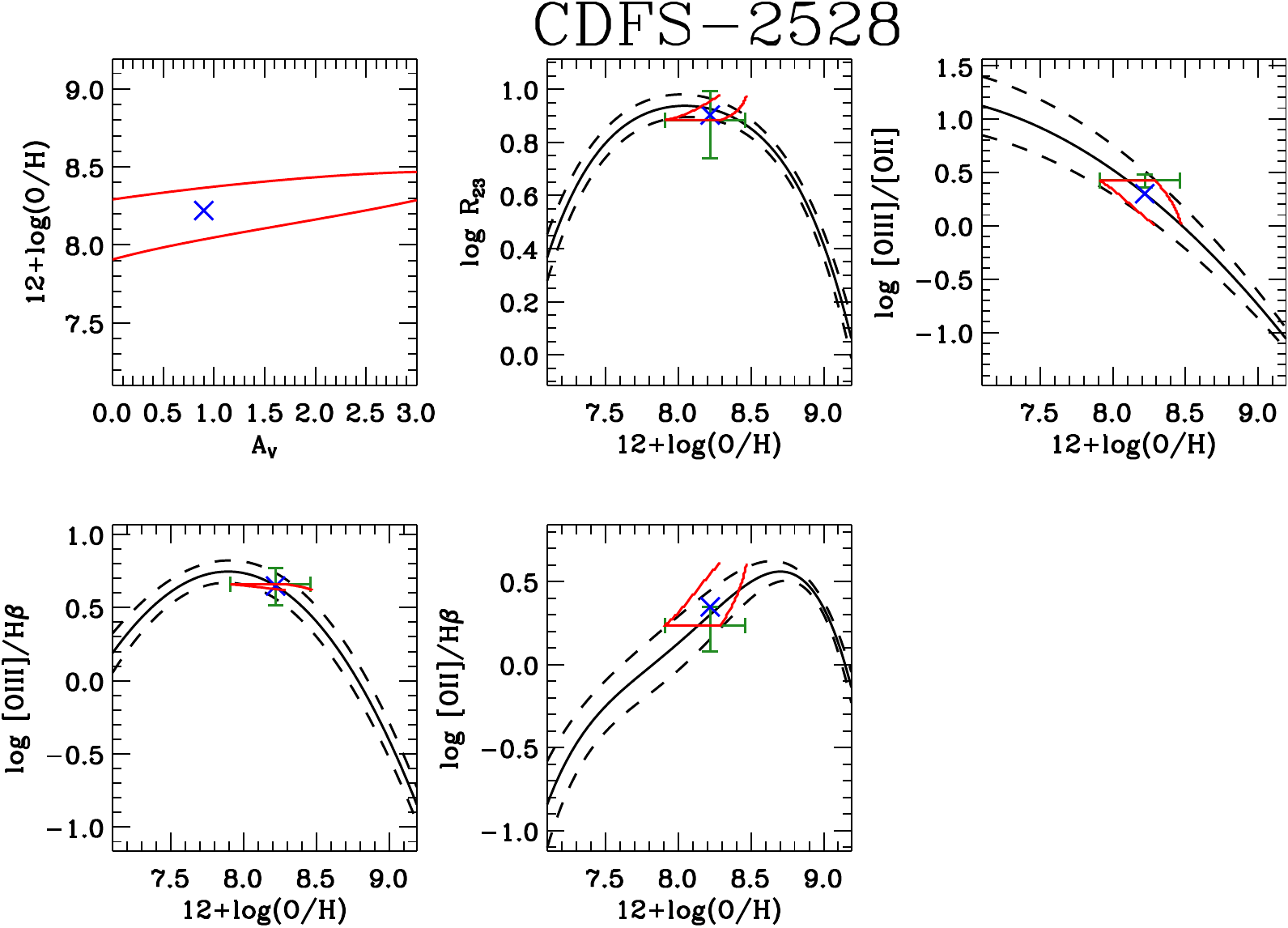}
  \includegraphics[width=0.49\linewidth]{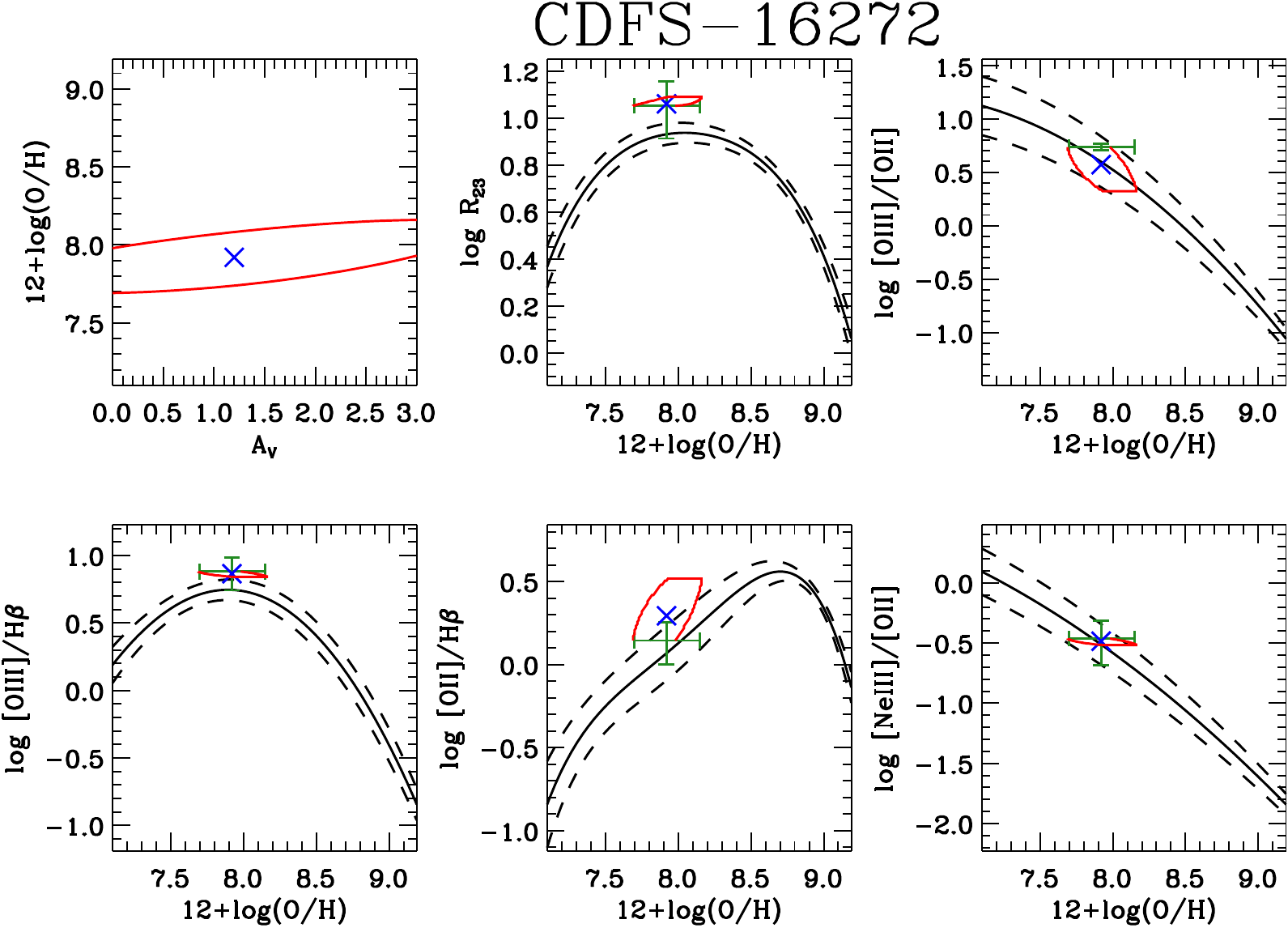}
  \includegraphics[width=0.49\linewidth]{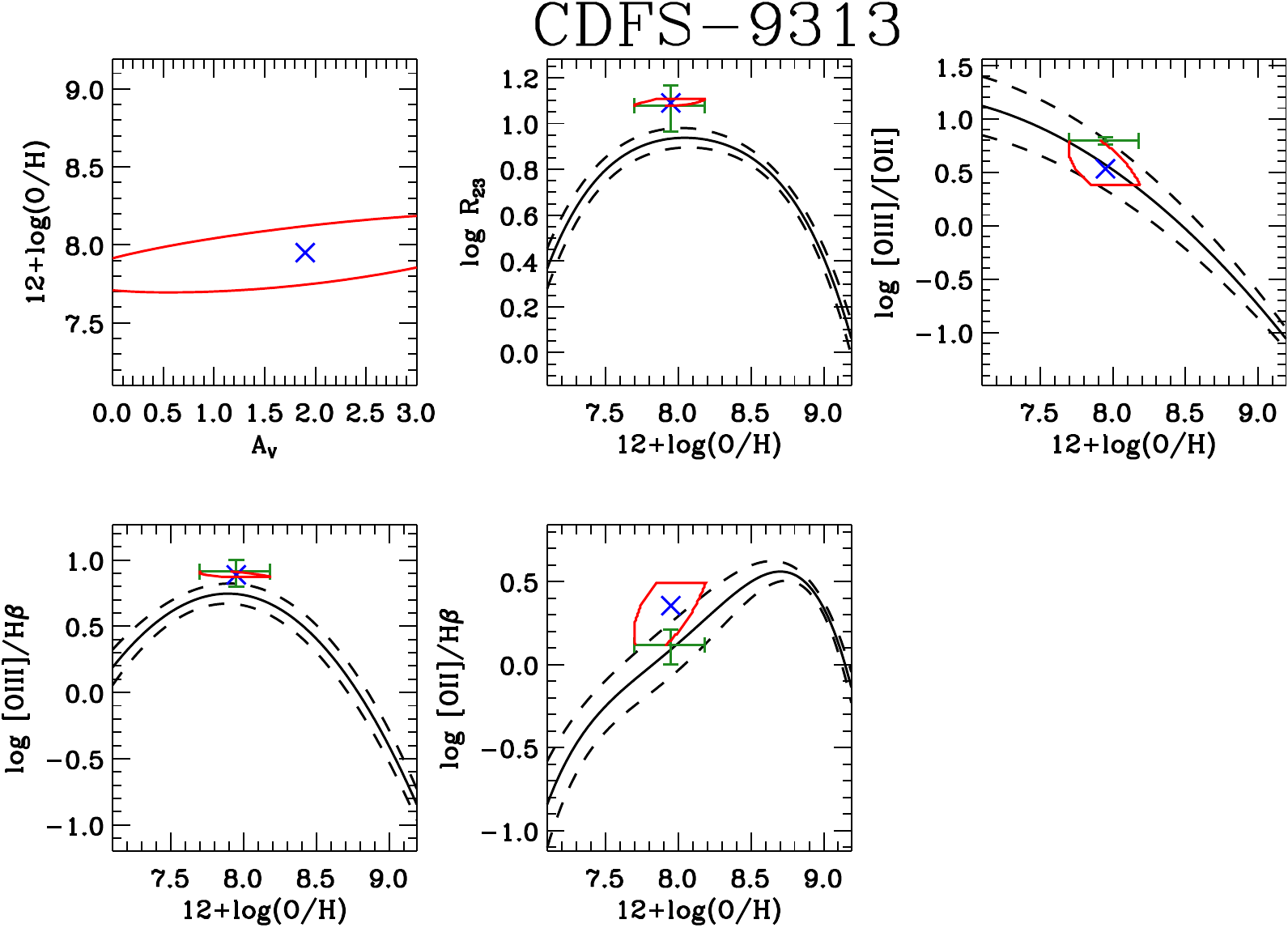}
  \includegraphics[width=0.49\linewidth]{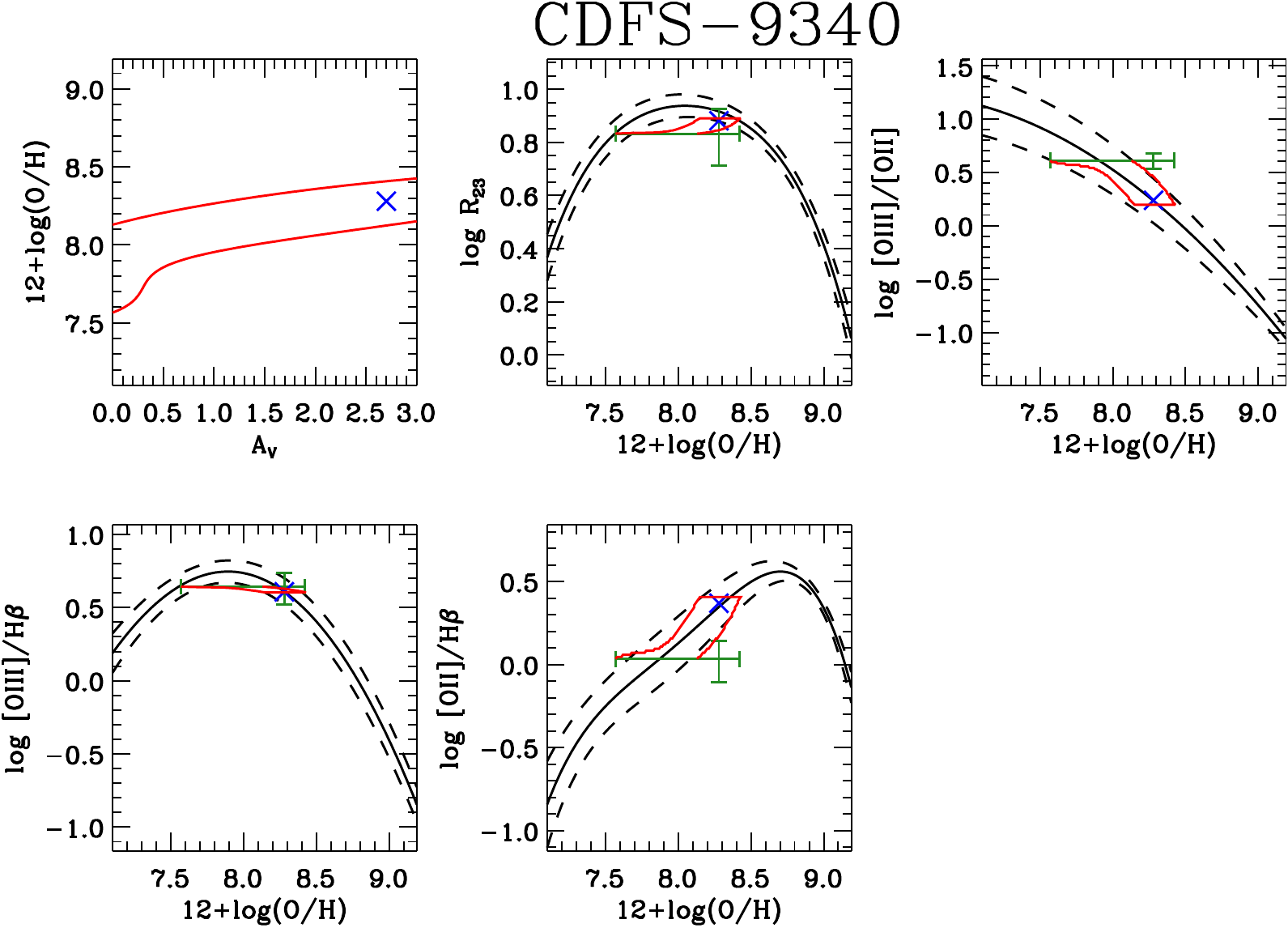}
  \includegraphics[width=0.49\linewidth]{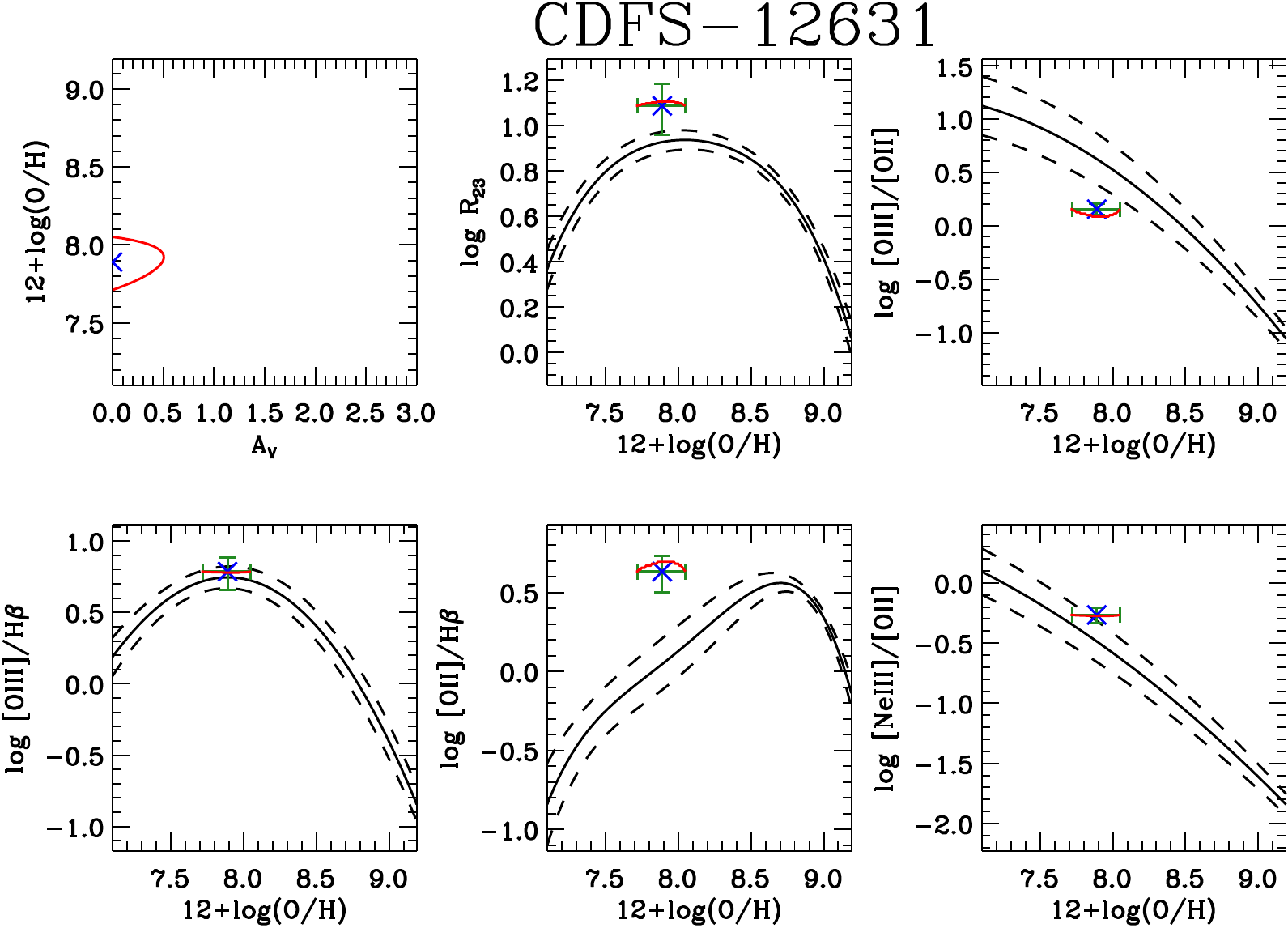}  
  \includegraphics[width=0.49\linewidth]{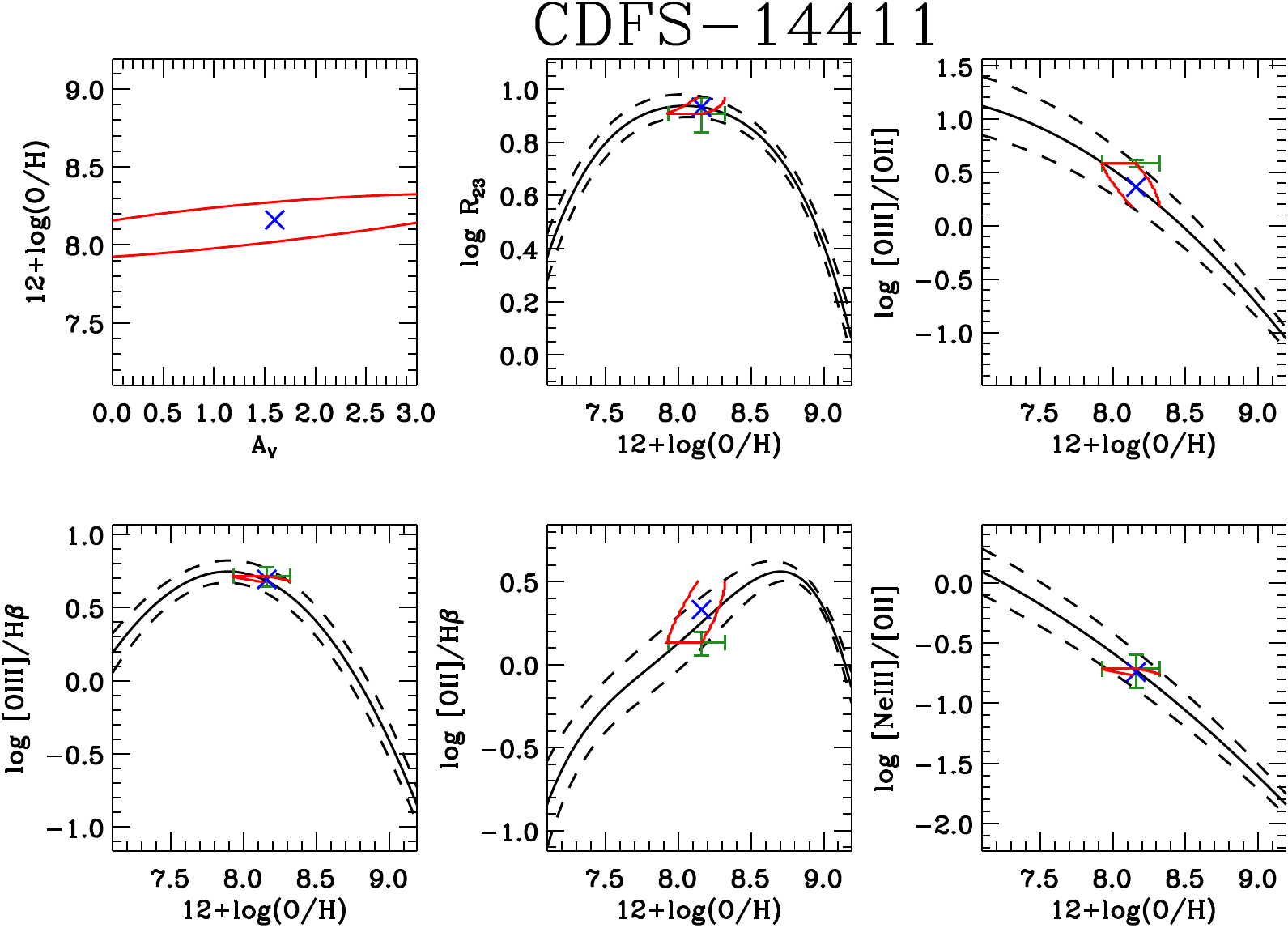}
 \caption{Diagnostic tools used to determine the metallicity of the galaxies in the AMAZE sample.
In each plot, the upper left panel shows the best solution (blue cross) and the 1$\sigma$ confidence level in the AV-metallicity plane.
In the other panels the black solid line (best fit) and the dashed lines (dispersion) 
show the empirical relations between various line ratios and the gas metallicity.
The green error bars show the observed ratios (along the y-axis) and the best-fit 
metallicity with uncertainty (along the x-axis); the blue cross shows the de-reddened ratios 
from adopting the best-fit extinction; the red line shows the projection of the 1$\sigma$ uncertainty of the fit in the top-left panel. 
}
\label{figmet2}
  \end{figure*}

\begin{figure*}
  \centering
  \includegraphics[width=0.49\linewidth]{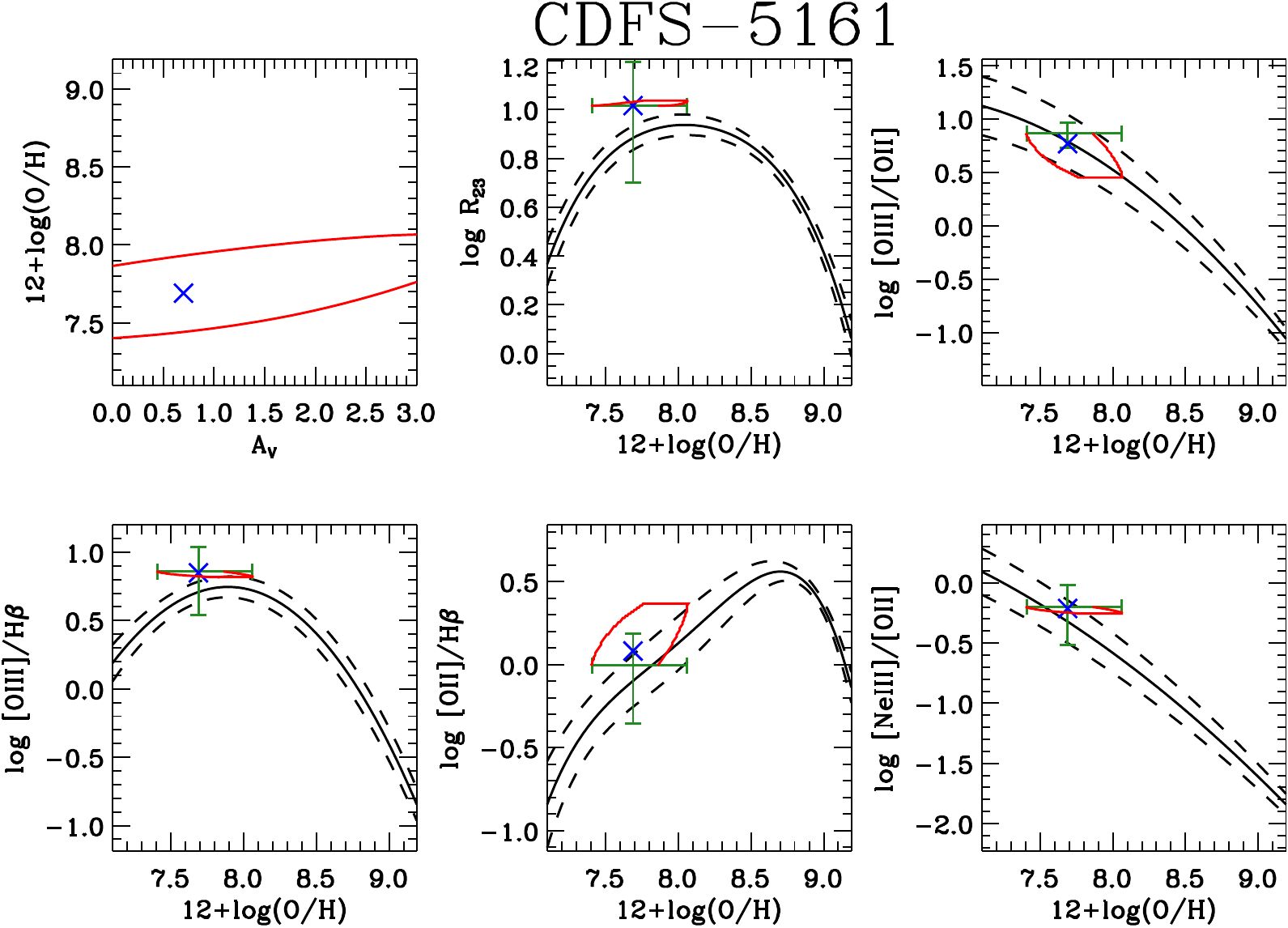}
  \includegraphics[width=0.49\linewidth]{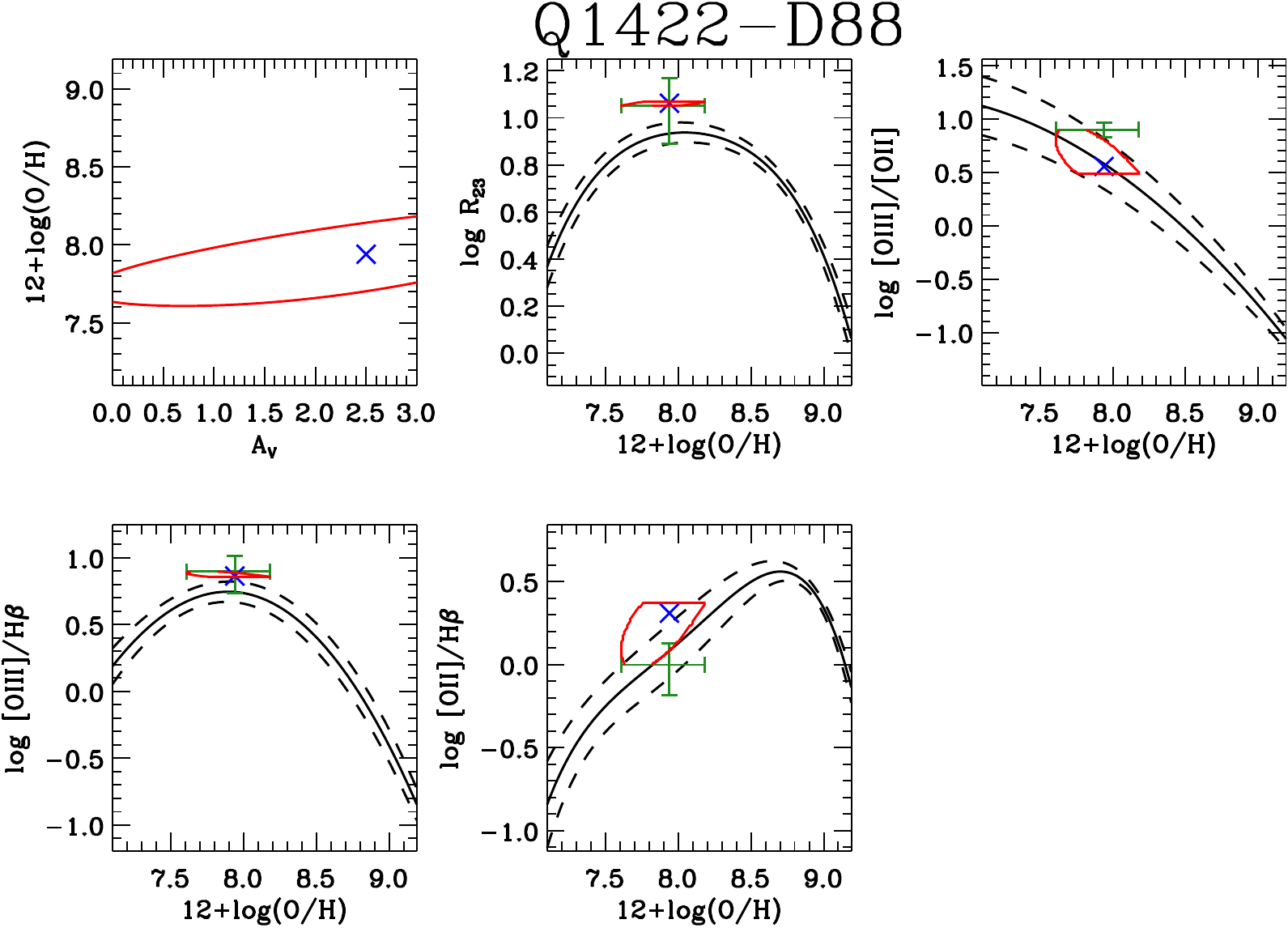}
  \includegraphics[width=0.49\linewidth]{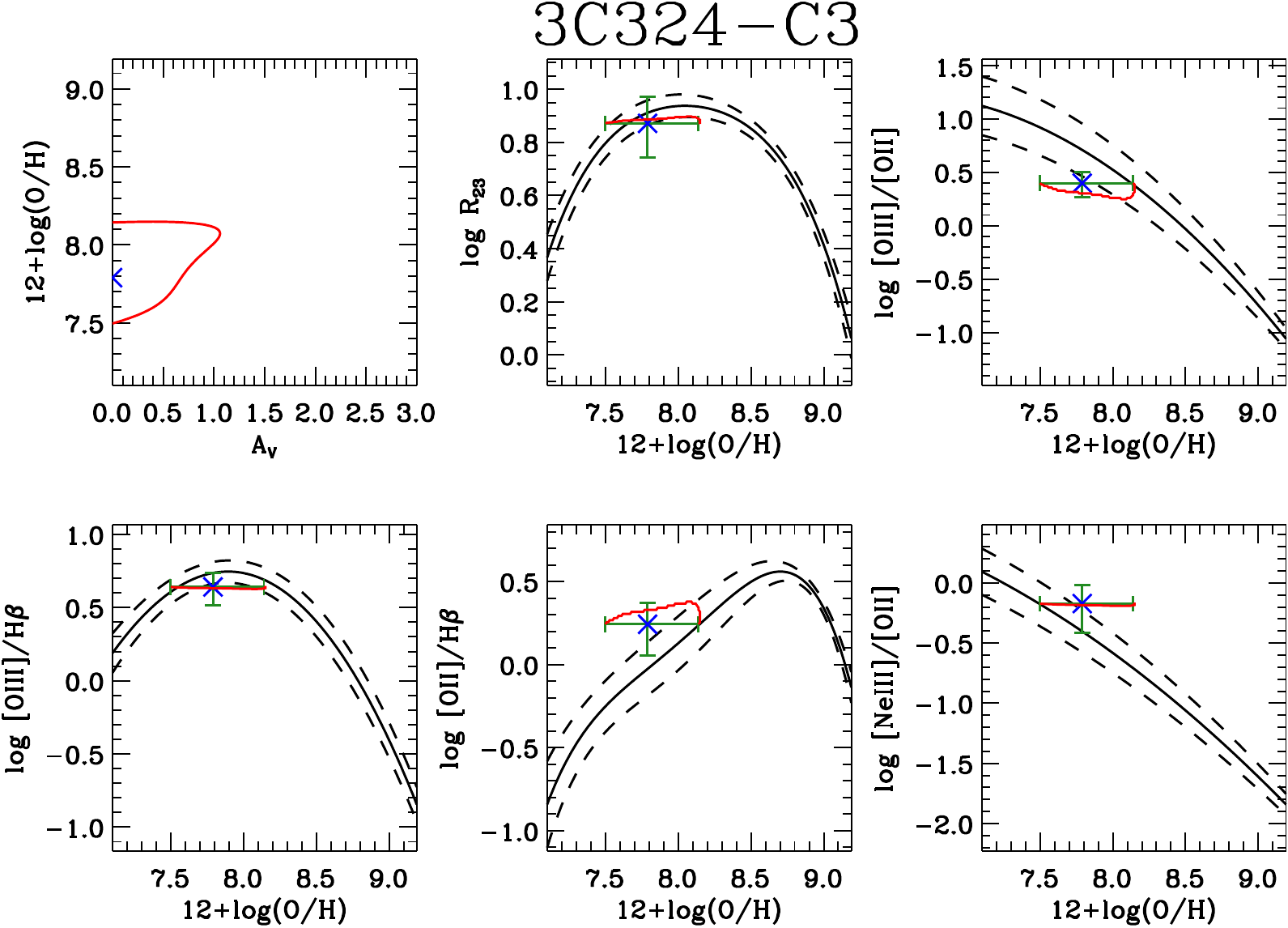}
  \includegraphics[width=0.49\linewidth]{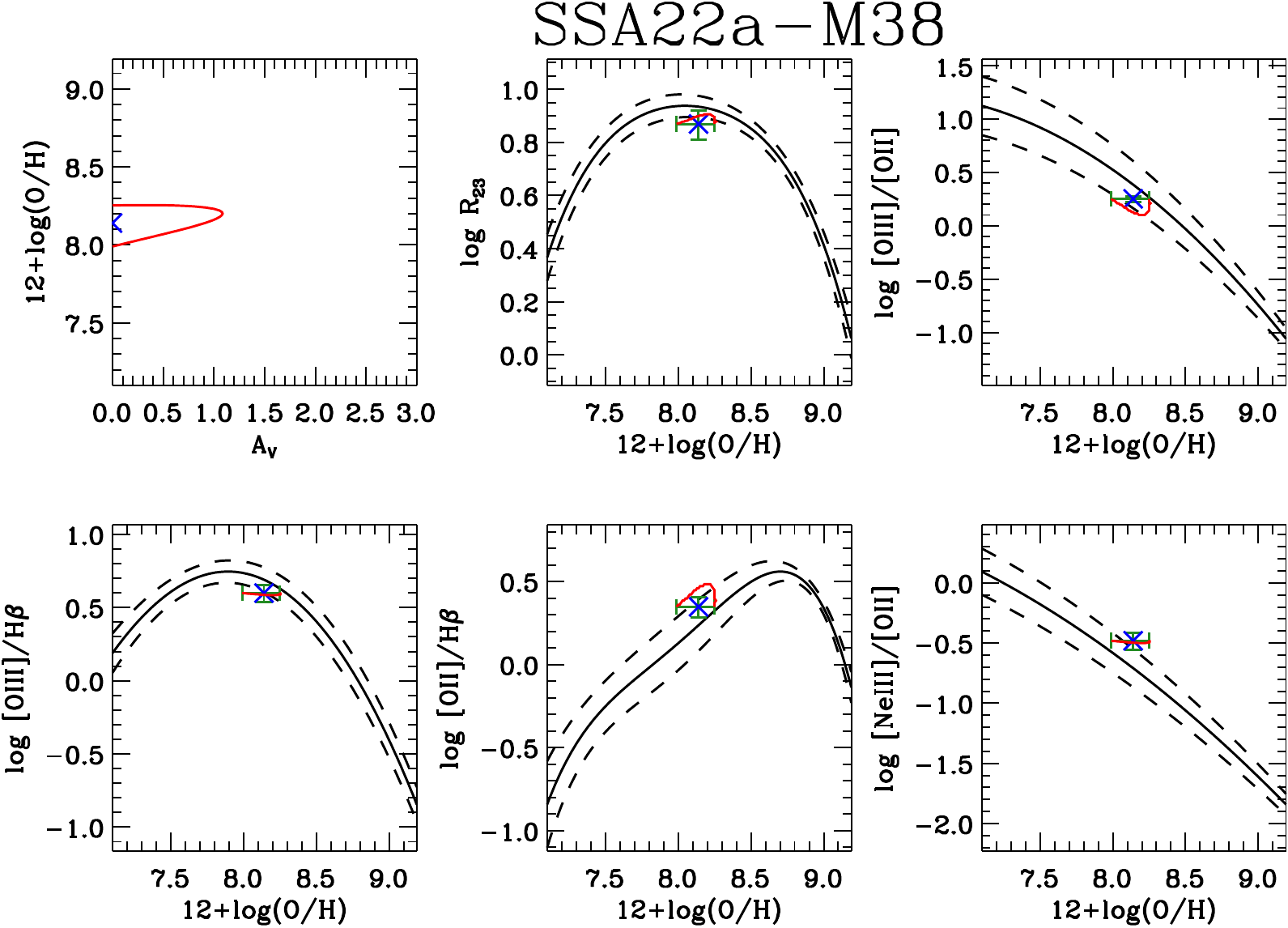}
   \includegraphics[width=0.49\linewidth]{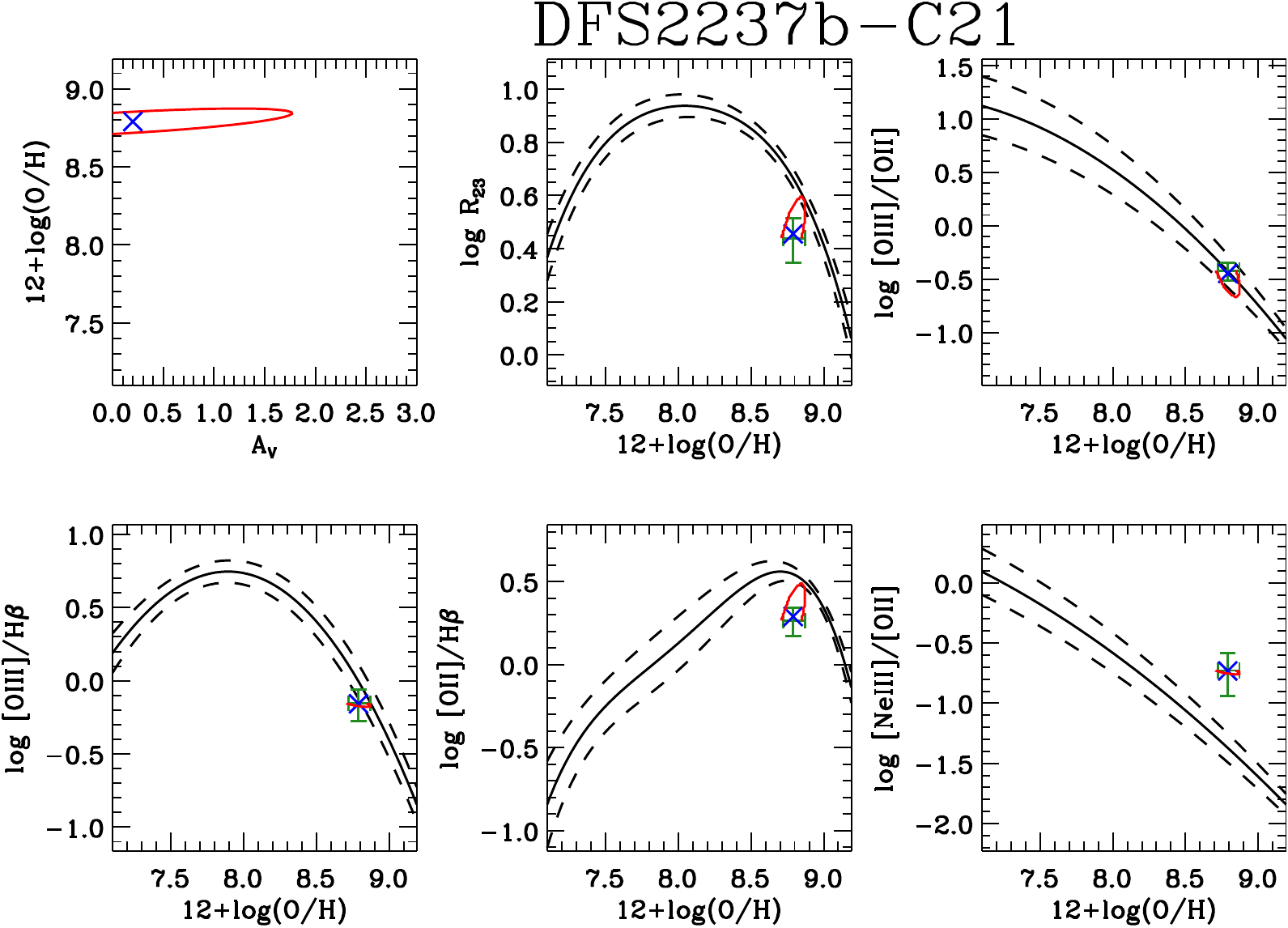}
  \caption{Diagnostic tools used to determine the metallicity of the galaxies in the AMAZE sample.
In each plot, the upper left panel shows the best solution (blue cross) and the 1$\sigma$ confidence level in the AV-metallicity plane.
In the other panels the black solid line (best fit) and the dashed lines (dispersion)
show the empirical relations between various line ratios and the gas metallicity.
The green error bars show the observed ratios (along the y-axis) and the best-fit 
metallicity with uncertainty (along the x-axis); the blue cross shows the de-reddened ratios
from adopting the best-fit extinction; the red line shows the projection of the 1$\sigma$ uncertainty of the fit in the top-left panel. 
}
\label{figmet3}
  \end{figure*}

\begin{figure*}
  \centering

  \includegraphics[width=0.49\linewidth]{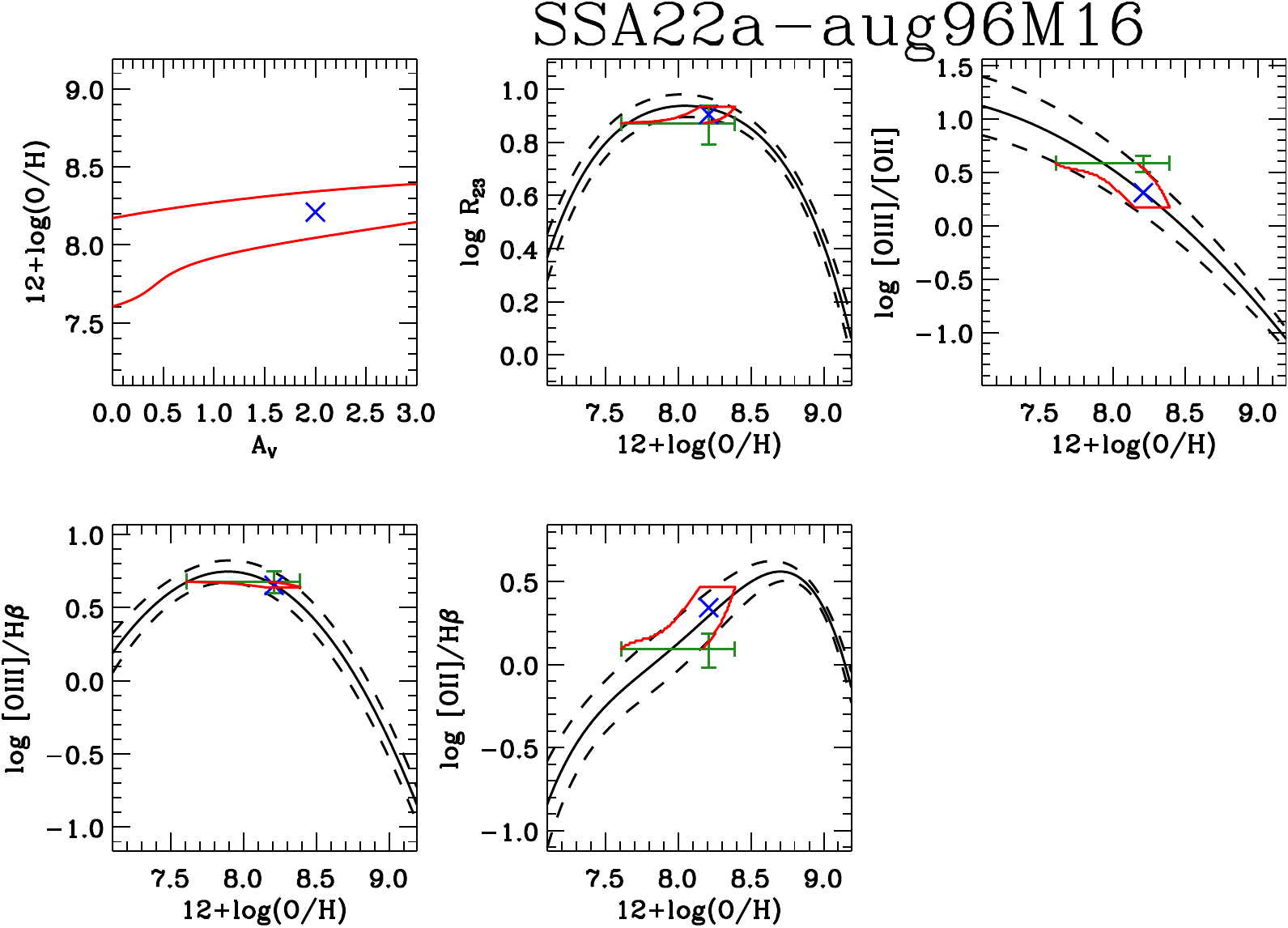}
  \includegraphics[width=0.49\linewidth]{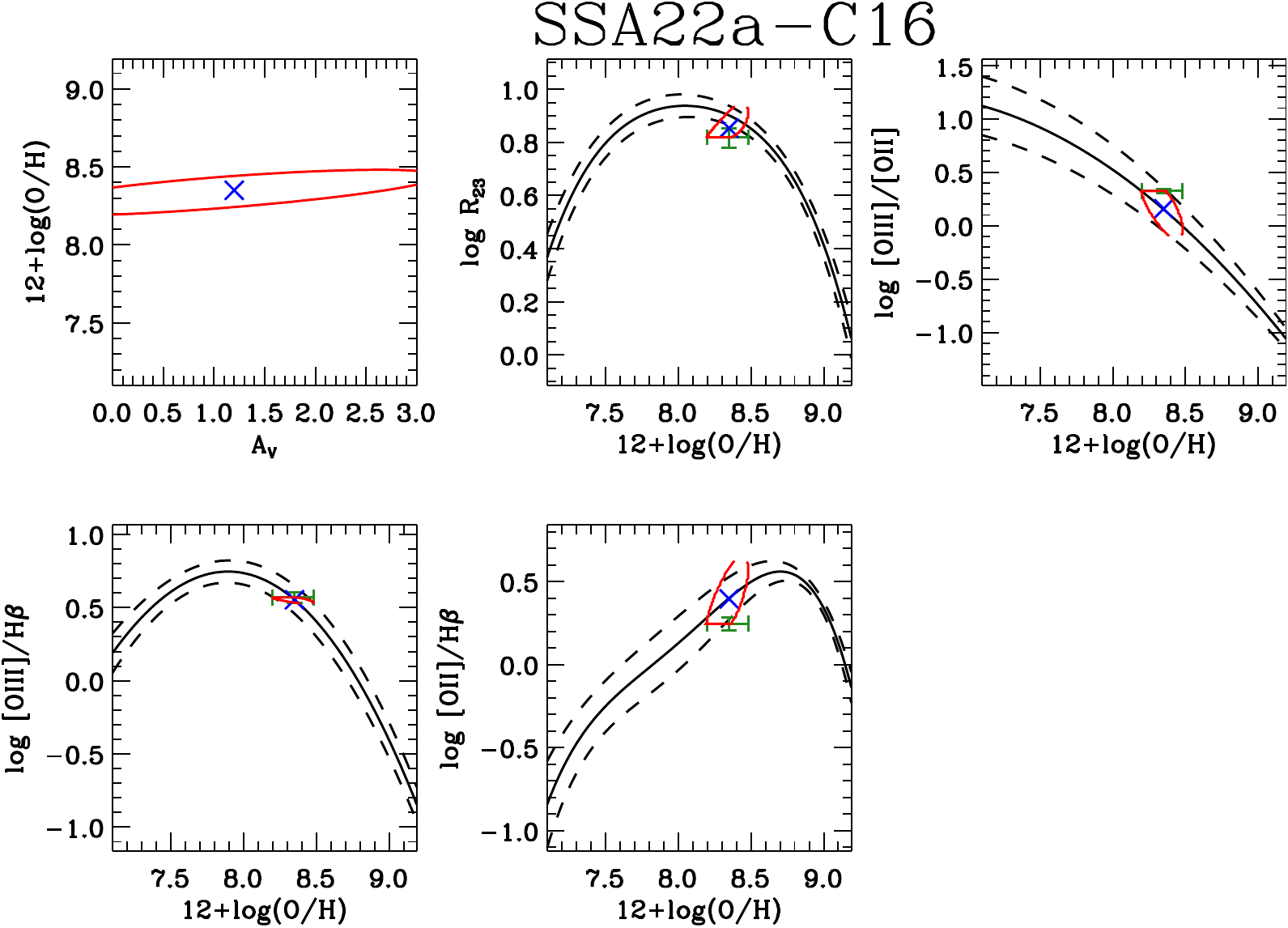}
  \includegraphics[width=0.49\linewidth]{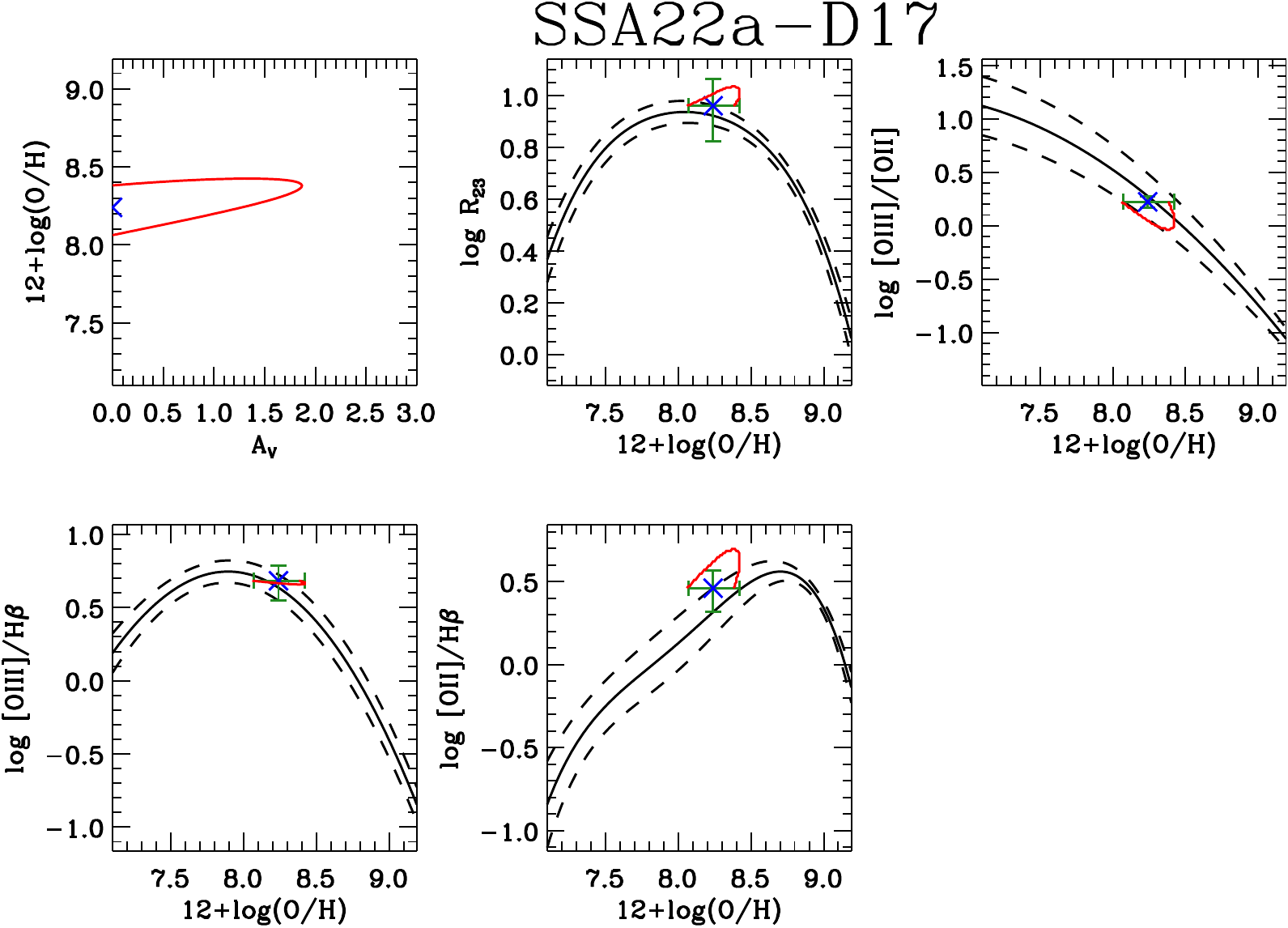}
  \includegraphics[width=0.49\linewidth]{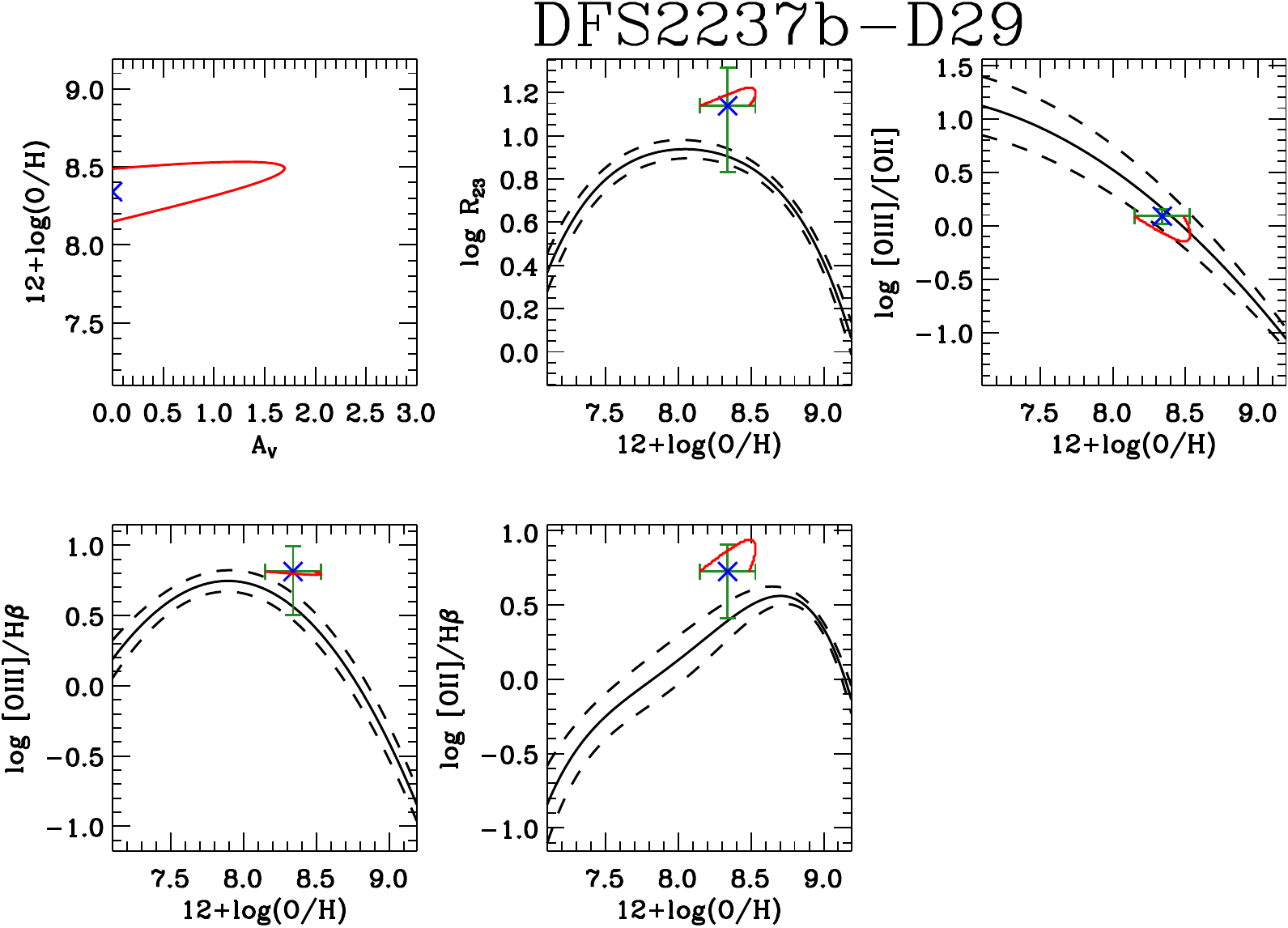}
  \includegraphics[width=0.49\linewidth]{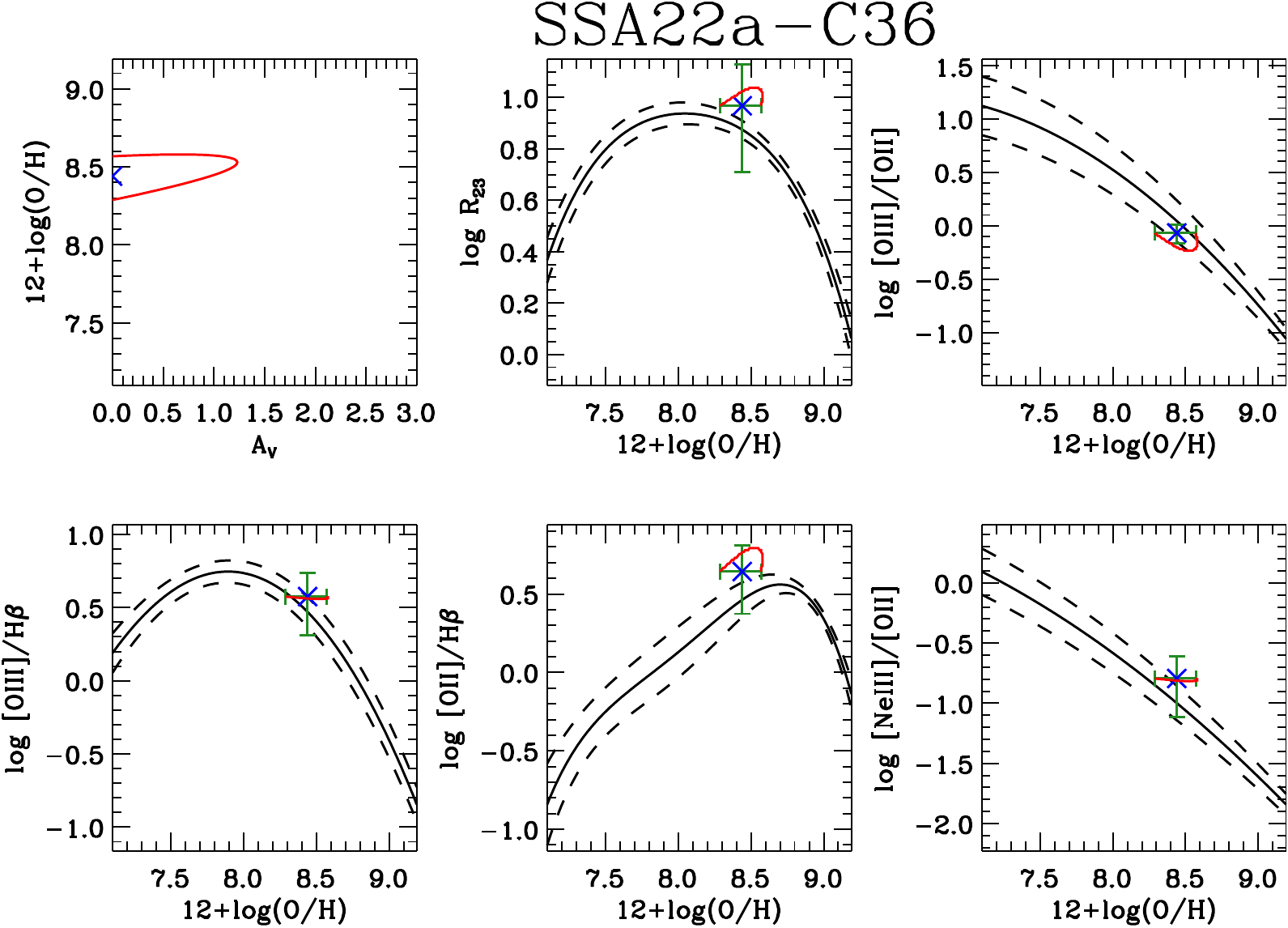}
  \caption{Diagnostic tools used to determine the metallicity of the galaxies in the AMAZE sample.
In each plot, the upper left panel shows the best solution (blue cross) and the 1$\sigma$ confidence level in the AV-metallicity plane.
In the other panels the black solid line (best fit) and the dashed lines (dispersion) 
show the empirical relations between various line ratios and the gas metallicity.
The green error bars show the observed ratios (along the y-axis) and the best-fit 
metallicity with uncertainty (along the x-axis); 
the blue cross shows the de-reddened ratios from adopting the best-fit extinction; 
the red line shows the projection of the 1$\sigma$ uncertainty of the fit in the top-left panel. 
}
 \label{figmet4}
  \end{figure*}

\begin{figure*}
  \centering
  \includegraphics[width=0.49\linewidth]{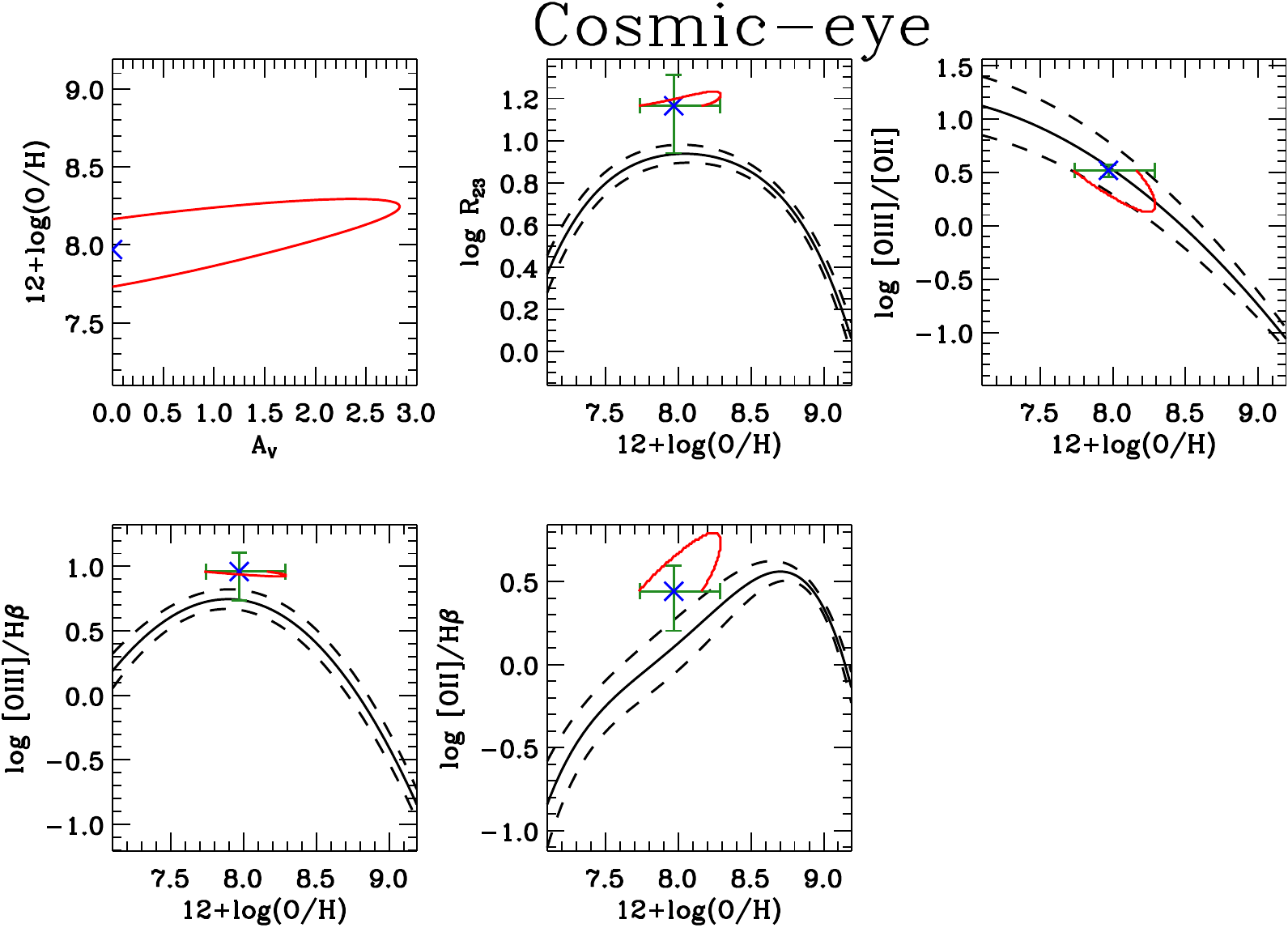}
  \includegraphics[width=0.32\linewidth]{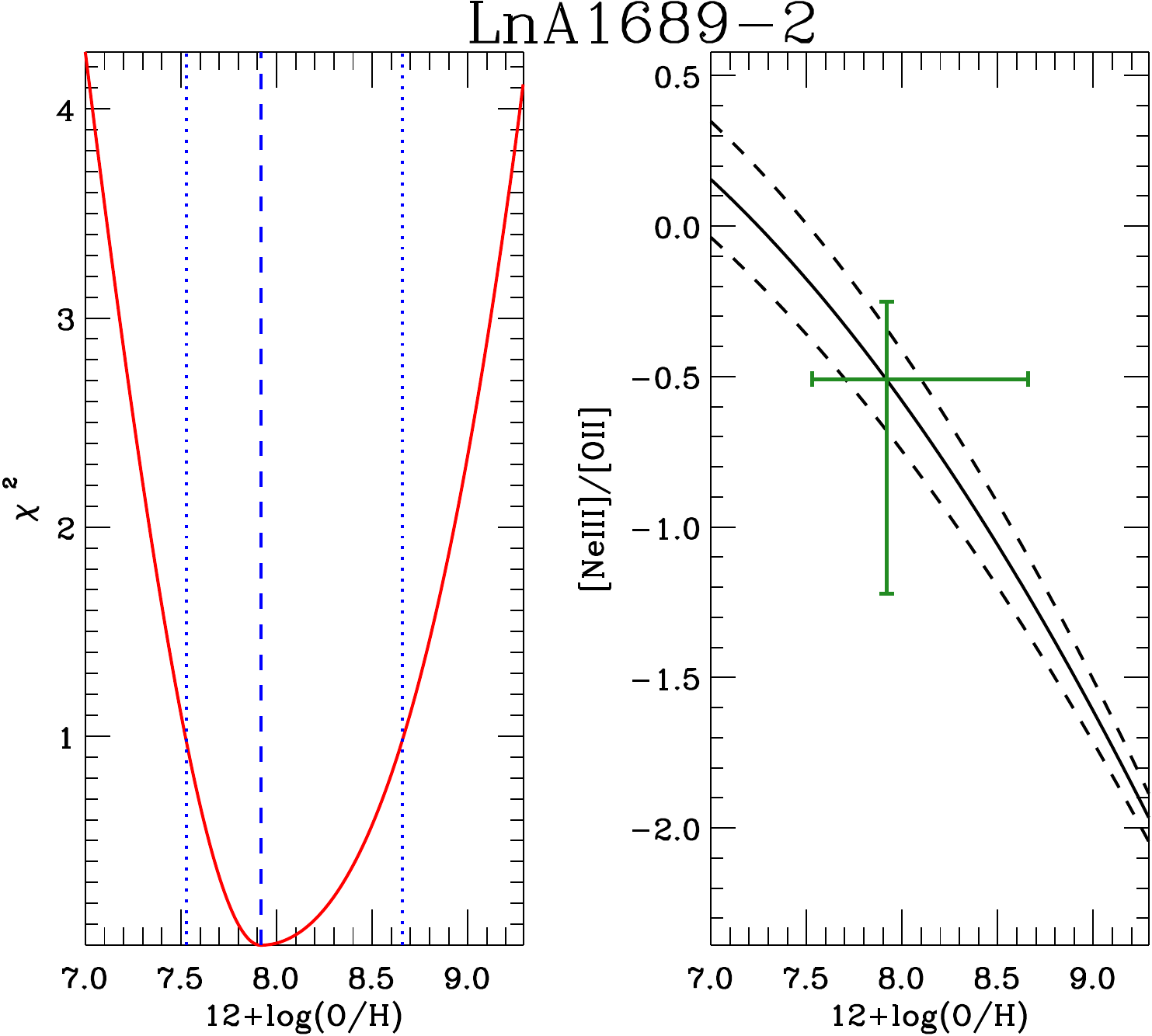}
  \includegraphics[width=0.49\linewidth]{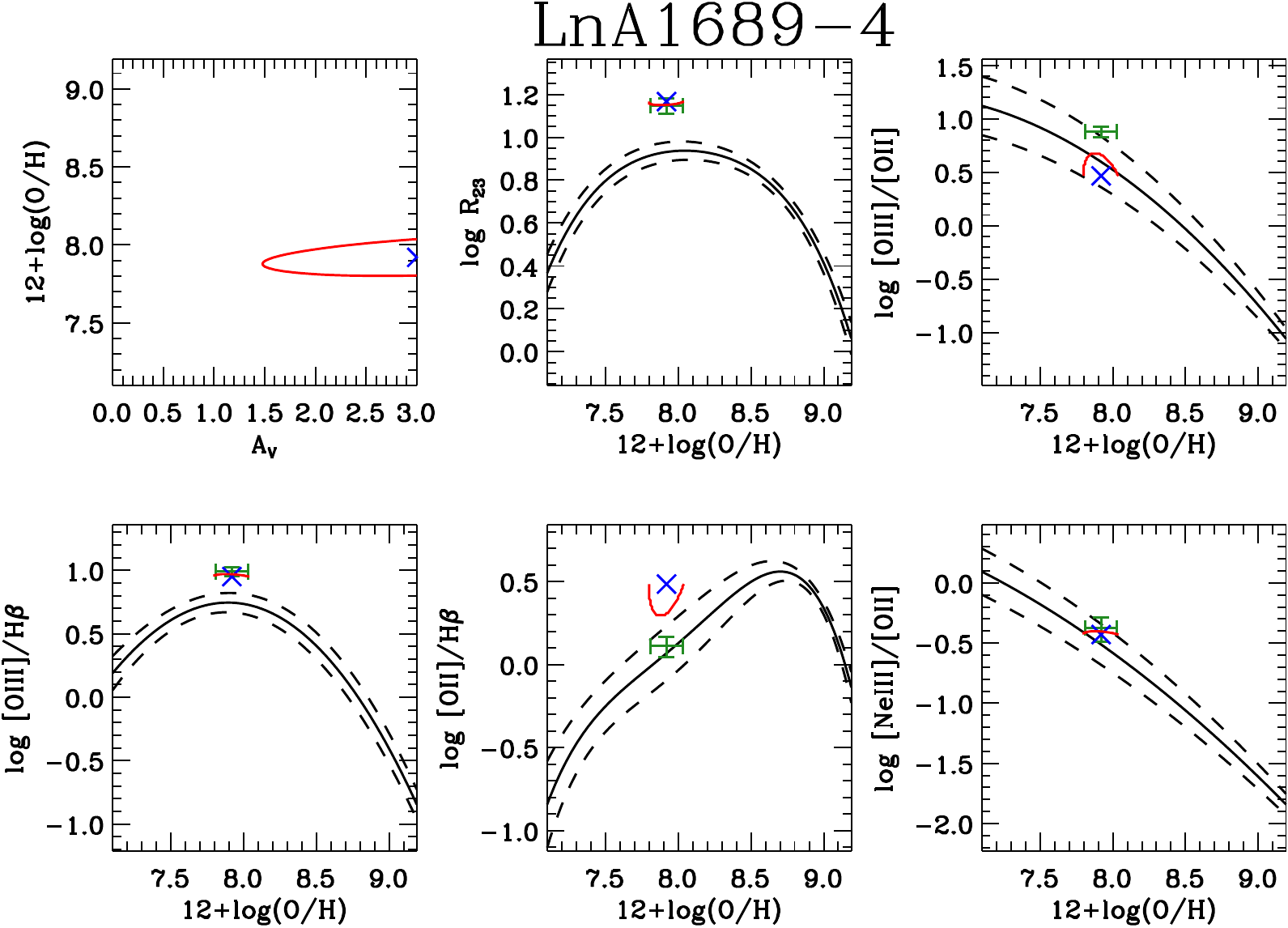}
  \includegraphics[width=0.49\linewidth]{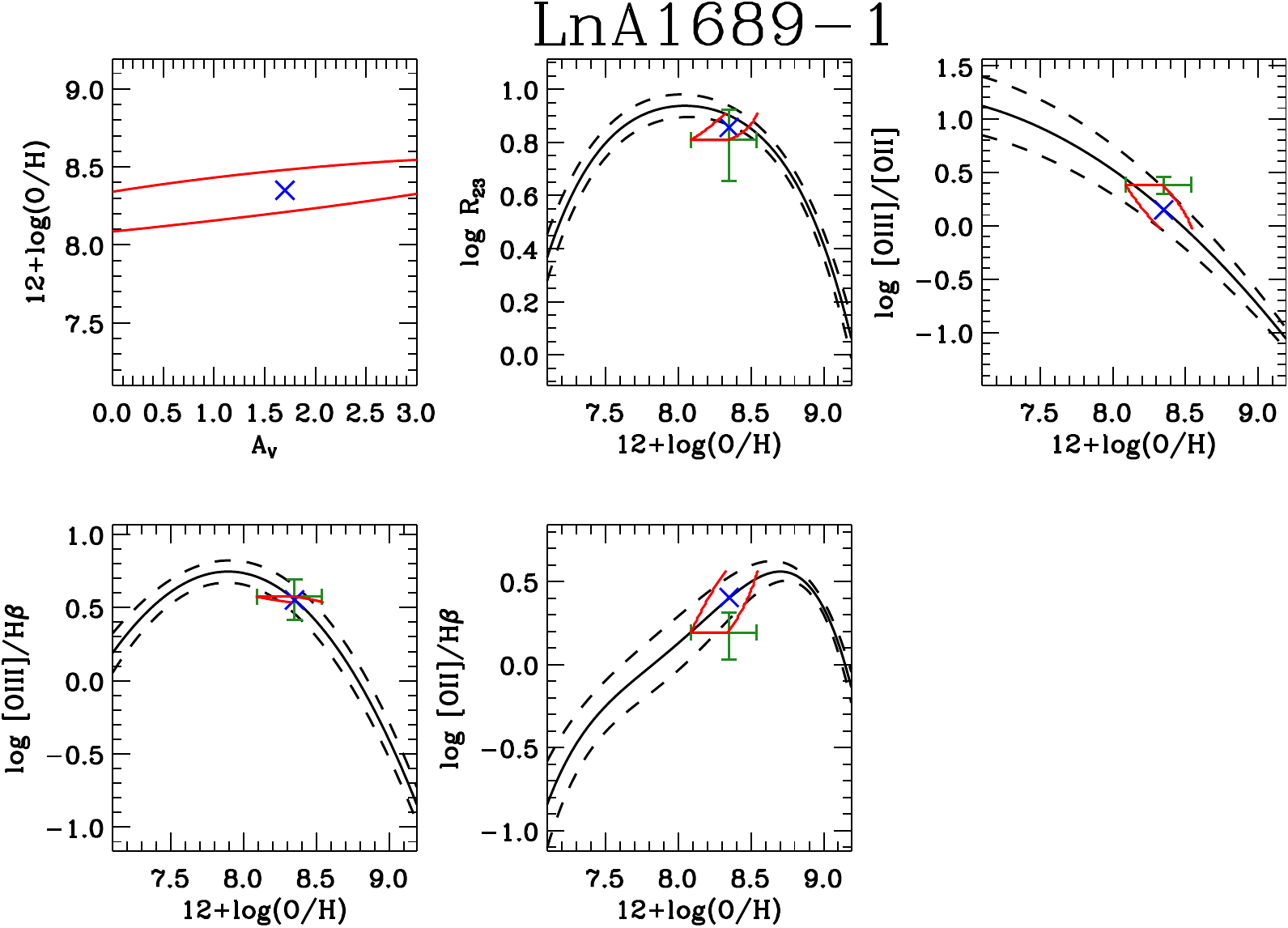}
 \caption{Diagnostic tools used to determine the metallicity of lensed galaxies in the AMAZE sample.
In each plot, the upper left panel shows the best solution (blue cross) and 
the 1$\sigma$ confidence level in the A$_V$--metallicity plane.
In the other panels the black solid line show the average relations between various line ratios 
and the gas metallicity as inferred by \cite{maiolino08}. 
The dashed lines show the associated dispersions.
The green error bars show the observed ratios (along the y-axis) and the best-fit metallicity with
uncertainty (along the x-axis);
the blue cross shows the best-fit metallicity and the
de-reddened ratios from adopting the best-fit extinction;
the red line shows the projection of the 1$\sigma$ uncertainty of the fit in the top-left panel. 
For the lensed AMAZE galaxy LnA1689-2 the metallicity was inferred from the [NeIII]/[OII] line ratio 
exploiting the anticorrelation (though with high dispersion) between this line ratio and 
the metallicity found in \cite{maiolino08}.
}
\label{figmetlens}
  \end{figure*}

In this appendix we provide some additional details on the measurements of the gas metallicity.
As mentioned in section \S \ref{met_z3}, the metallicities were inferred following the calibrations and
the method described in detail in \cite{maiolino08}. 
Here we briefly summarize the method
and show the application to our full sample of galaxies. 
\cite{maiolino08} inferred the relationship between metallicity and
various nebular line ratios by exploiting a combination of direct
measurements (based on the T$_e$ method, mostly at low metallicities) and photoionization
modeling (mostly at high metallicities).
It has recently been argued that the excitation
conditions in high-z galaxies are different from the local ones and that,
therefore, the local calibrations may not apply to high redshift \citep{kewley13}.
However, we found in a parallel work \citep{maiolino13} that
local calibrations are indeed appropriate for high-z galaxies,
after selection effects are properly taken in to account.
These relations are shown in Fig \ref{figmet1}, where
solid black lines show the average of the galaxy
distribution and the dashed black lines show the dispersion.
Each of these diagnostics has advantages and disadvantages.
Some of them are practically unaffected by extinction
(e.g., [OIII]/H$\beta$), but have a double-degenerate metallicity solution;
others have a monotonic dependence on metallicity (e.g., [OIII]/[OII], [NeIII]/[OII]), 
but have high dispersion and are affected by reddening.
However, when these diagnostics are used simultaneously,
only some combinations of metallicity and reddening are allowed by the data.
Obviously, not all of the diagnostics shown in Figs. \ref{figmet1} -- \ref{figmetlens} are independent.

We decided to use the following independent metallicity diagnostics:
R$_{23}$ (which has the tightest relationship with
metallicity, although with double solution), [OIII]/[OII] (which has a monotonic dependence on
metallicity, although with large scatter, which allowed us to remove the degeneracy on R$_{23}$),
and, when available, [NeIII]/[OII] (which also has a monotonic dependence on metallicity).
In practice, we determined the $\chi^2$ for each combination of metallicity and dust extinction for
each of these relations, and found the best combination as the minimum of $\chi^2$. 

The result of this method is shown, for each object in our sample, 
in Figs. \ref{figmet1} \ref{figmet2}, \ref{figmet3}, \ref{figmet4}, and \ref{figmetlens}.
The upper-left panel of the figure associated with each object shows in the
metallicity-extinction plane the best-fitting value (blue cross) 
and the 1$\sigma$ confidence level.
Clearly, the metallicity is generally well constrained.
The extinction is instead poorly constrained by this method;
but this figure shows us that the uncertainties on extinction do not significantly affect
the metallicity determination.
In the other panels the vertical green error bar indicates the measured ratio with the associated
uncertainties, while the blue cross gives the best-fit value of the metallicity and the dereddened
ratio assuming the best-fit reddening value; the red lines show
the projection of the 1$\sigma$ confidence levels.
The horizontal green error bar shows the resulting uncertainty on the metallicity.

While Figs. \ref{figmet1} \ref{figmet2}, \ref{figmet3}, \ref{figmet4}, and \ref{figmetlens}
only show the results for the line ratios obtained from the integrated fluxes in each galaxy,
the same method was applied pixel-to-pixel for resolved galaxies to infer the metallicity
maps shown in Figs. \ref{fig_metmap1}, \ref{fig_metmap2}, and \ref{fig_metmap3} .

\end{document}